\documentclass[aps,prl,reprint,10pt,superscriptaddress,nofootinbib,floatfix]{revtex4-2}

\usepackage{graphicx,amsmath,amssymb,amsthm,mathtools,bm} \usepackage{microtype}
\usepackage{hyperref}
\usepackage{xcolor}
 \usepackage{tikz,pgfplots}

\usepgfplotslibrary{groupplots,fillbetween}
\pgfplotsset{compat=1.18}
\hypersetup{colorlinks=true,citecolor=blue!55!black,linkcolor=blue!55!black,urlcolor=blue!55!black,pdftitle={Quantum ensembles as waves on state space}}

\providecommand{\openone}{\mathbb I}
\newcommand{\Tr}{\operatorname{Tr}}
\newcommand{\rank}{\operatorname{rank}}
\newcommand{\ket}[1]{|#1\rangle}
\newcommand{\bra}[1]{\langle #1|}
\newcommand{\ketbra}[1]{\ket{#1}\bra{#1}}
\newcommand{\doilink}[2]{\href{https://doi.org/#1}{#2}}
\newcommand{\arxivlink}[2]{\href{https://arxiv.org/abs/#1}{#2}}
\newcommand{\backmatterhead}[1]{\par\vspace*{1.1\baselineskip}\noindent{\normalfont\bfseries\MakeUppercase{#1}}\par\nobreak\vspace{0.3\baselineskip}\nobreak\noindent\ignorespaces}

\begin{document}
\raggedbottom

\title{Quantum ensembles as waves on state space}

\author{Yu-Xuan Zhang} \affiliation{School of Physics, Nankai University, Tianjin 300071, People's Republic of China} \affiliation{Centre for Quantum Technologies, National University of Singapore, Singapore 117543, Singapore}

\author{Leong-Chuan Kwek} \email{kwekleongchuan@nus.edu.sg} \affiliation{Centre for Quantum Technologies, National University of Singapore, Singapore 117543, Singapore} \affiliation{National Institute of Education, Nanyang Technological University, Singapore 637616, Singapore}

\author{Jing-Ling Chen} \email{chenjl@nankai.edu.cn} \affiliation{Theoretical Physics Division, Chern Institute of Mathematics, Nankai University, Tianjin 300071, People's Republic of China}

\date{\today}
\begin{abstract}
A quantum ensemble contains preparation weights that are lost in its average density operator. We show how optimal decisions read these weights as a wave on projective state space. Varying the input prior sweeps a surface of fixed Born overlap through the ensemble; the resulting response propagates according to the geometry of pure states and reconstructs their distribution in every finite dimension. Conditional preparations can fix this wave through contact, selecting the ensemble at a steering boundary. Positive propagation gives exact biased-qubit solutions and explicit parent measurements for noisy channels. When measurement records are combined, their state directions are averaged and the wave acquires a source whose growth measures the lost discrimination advantage. Preparation, inference and noise can thus be understood as operations on the same recoverable distribution, linking quantum correlations to measurement resolution.
\end{abstract}
\maketitle

\section*{Introduction}

A quantum state can be prepared by many different distributions of pure states. Its density operator gives their common average; the preparation labels retain which states contributed. Keeping those labels allows subsets of the distribution to be selected as conditional preparations. Forgetting them can erase distinctions that a measurement would otherwise resolve. Such operations depend on how the weight is distributed among state directions, information that the density operator has averaged away. We show that optimal binary decisions read this distribution as a wave on projective state space.

Remote preparation brings these weights into quantum correlations. A measurement on a purifying system selects an ensemble of the reduced state~\cite{Schrodinger1935,HJW1993}. Steering tests whether one fixed local ensemble can supply all the conditional states, with the remote measurement changing only how its weights are divided~\cite{EPR1935,Wiseman2007,UolaReview}. Normalizing the trusted mean converts that ensemble into a parent measurement: dividing its weights becomes classical processing of its outcomes~\cite{Uola2014,Quintino2014,Uola2015,Guhne2023}. The same distribution thereby connects a preparation problem to the resolution of quantum observables~\cite{AubrunLancien2016}.

Finding a common ensemble usually requires controlling a continuum of measurement choices. Semidefinite programming treats specified assemblages~\cite{Skrzypczyk2014,CavalcantiReview}; finite constructions and analytic criteria supply models for continuous families~\cite{Hirsch2016,Cavalcanti2016,BowlesCriterion}. For qubits, steering ellipsoids describe conditional states~\cite{Jevtic2014}, and critical radii compare their demands with the response available from an ensemble~\cite{NguyenVuCritical,NguyenGeometry2019}. Exact unbiased solutions~\cite{Werner1989,ZhangZhangBell,ZhangChitambar2024,Renner2024,ZhangPOVM2026} already depend on how weight is distributed over state directions. Polarization makes that dependence directional: exchanging the trusted party can change the boundary of the same state~\cite{Handchen2012,BowlesOneWay,Tischler2018}.

We investigate the ensemble distribution through its response to a moving decision threshold. Each outcome changes its preferred input when the threshold crosses its Born overlap. These switches trace the ensemble over pure-state space. The geometry of quantum overlaps~\cite{ProvostVallee1980} makes their accumulated response obey a wave equation in every finite dimension. At a preparation boundary, saturated demands can provide the data that determine this wave. Once the ensemble is recovered, its weights specify both the conditional preparations it can supply and the information lost when its outcomes are mixed.

\section*{Decision waves on state space}

\begin{figure*}[tbp]
\centering \includegraphics[width=\textwidth]{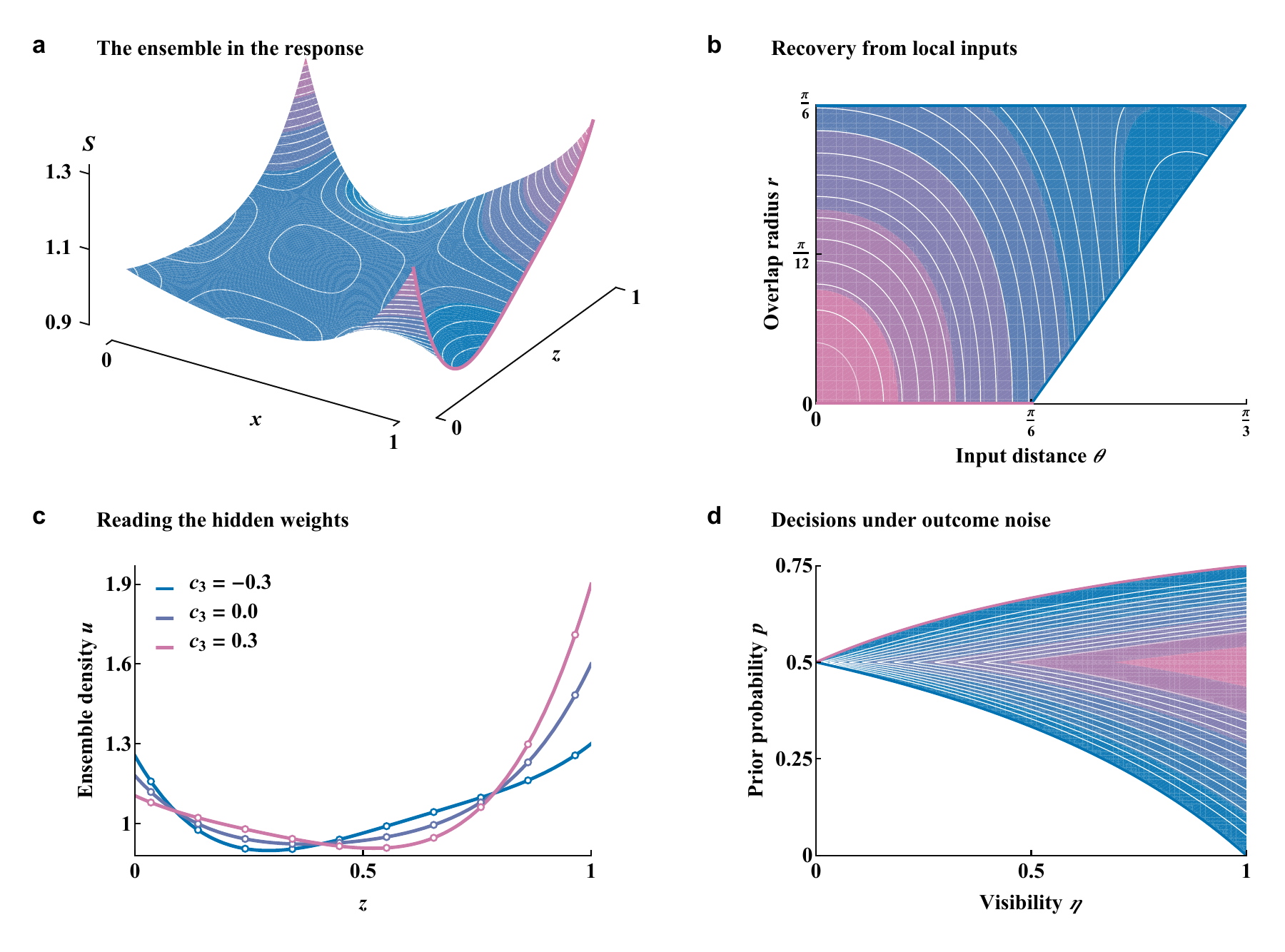} \caption{\textbf{Decision waves and ensemble reconstruction.} All panels use the qutrit density $u(z)=1+0.45\pi_2(z)+c_3\pi_3(z)+0.15\pi_4(z)$, where $z=\Tr(PP_0)$ and $\pi_\ell(z)=P_\ell^{(1,0)}(2z-1)/(\ell+1)$. Unless varied in \textbf{c}, $c_3=-0.3$. \textbf{a}, Normalized response curvature $S(x,z)$ carries the angular weights across overlap thresholds. Its pink edge at $x=1$ is the generating density. \textbf{b}, Local reconstruction at $x_0=3/4$. Response value and slope for inputs within distance $\pi/3$ from $P_0$ (blue upper edge) determine the wave in the coloured domain and the ensemble within distance $\pi/6$ (pink lower edge). Here $r=\arccos\sqrt{x}$ and $\theta=\arccos\sqrt{z}$; the field shown is the exact curvature within this domain of dependence. \textbf{c}, Three positive ensembles give identical success probabilities for every pure input at prior $p=9/16$. Solid curves are the three generating densities; open circles recover their weights from response value and threshold slope by harmonic inversion. The slope resolves the degree-three component hidden at this prior. \textbf{d}, Randomizing the recorded outcome narrows the prior interval in which the measurement can improve on guessing the more probable input. Blue and pink curves bound this interval; the interior colour shows the success advantage over guessing the more probable input, averaged over uniformly distributed pure inputs, with quadratically spaced levels. In \textbf{a}, \textbf{b} and \textbf{d}, blue to pink denotes increasing plotted value and white curves mark equal values.} \label{fig:ensemble-wave}
\end{figure*}

Consider an ensemble measure $\mu$ on the pure states of an $n$-dimensional system, with $n\ge2$. Normalize its operator mean to $\openone_n/n$; an invertible filter on the trusted support gives this form in a steering problem~\cite{Ku2022}. Its weighted projectors define a measurement $G(dP)=nP\,d\mu(P)$. To read the distribution, use this measurement to distinguish a pure input $Q$ from the maximally mixed reference. Assign each recorded outcome to whichever input is more likely. If the reference has prior $p=nx/(1+nx)$, the relevant response is
\begin{equation}
F(x,Q)=\int|x-\Tr(QP)|d\mu(P),\qquad 0<x<1,
\label{eq:main-overlap-response}
\end{equation}
and the success probability is
\begin{equation}
p_{\rm succ}(x,Q;G)=\frac12+\frac{nF(x,Q)}{2(1+nx)}.
\label{eq:main-discrimination}
\end{equation}
Varying the prior moves the threshold $x$ while leaving the measurement fixed. Rotating $Q$ changes which part of the ensemble the decision probes. An outcome associated with $P$ favours the pure input when its overlap exceeds the threshold, and the mixed reference when it falls below. Adding the advantage of the better guess in either case gives the absolute-value response. Its amplitude records a decision score, with the ensemble weights determining where that score bends as the prior varies.

Every state contributes a change of slope when $x$ crosses its overlap with $Q$. Consequently, $\partial_x^2F$ records twice the weight on the surface $\Tr(QP)=x$. These surfaces are spheres in the Fubini--Study geometry of pure states. As their radius changes, their responses satisfy
\begin{equation}
x(1-x)\partial_x^2F-(1-nx)\partial_xF-nF=\frac14\Delta_{\rm FS}F,
\label{eq:main-projective-wave}
\end{equation}
where $\Delta_{\rm FS}$ is the spatial Laplacian on $\mathbb CP^{n-1}$. Here propagation follows the changing decision threshold. Responses to neighbouring inputs are linked because their moving surfaces cross the same distribution. The wavefront through a hidden state consists of the tests at which its outcome switches the preferred input. Replacing $x$ by $\tau=\arcsin(2x-1)$ gives these fronts a constant speed, $1/2$ in Fubini--Study distance. Quantum overlap thus determines how prior variation transports information between input directions.

As $x$ approaches one, the overlap sphere contracts onto $Q$. Dividing the response curvature by the known Haar overlap density gives the mean ensemble weight on that sphere, which tends to the density at $Q$ [Fig.~\ref{fig:ensemble-wave}a]. The corresponding weak limit also recovers ensembles supported on curves or discrete sets. At any interior threshold, the response and its threshold derivative determine all the weights. Figure~\ref{fig:ensemble-wave}c illustrates why the derivative matters: three distinct ensembles give identical decisions at one prior, while their changes with prior recover the missing angular component.

Even a single prior is usually sufficient for pure measurements. At almost every fixed prior in $0<p<n/(n+1)$, uniform dominance over a rank-one measurement for all pure inputs forces the same pure-effect measure, up to splitting and relabelling proportional effects. All rank-one measurements have the same input-averaged success, so inequivalent ones must exchange advantages on open sets of inputs. Wave inversion identifies their maximality under classical processing~\cite{Buscemi2005,Zhu2022} through optimal decisions, complementing tomography from individual outcome probabilities~\cite{Luis1999,Lundeen2009}.

Propagation also limits how far this inference reaches. Value and slope known for inputs in a Fubini--Study ball of radius $R$ at threshold $x_0$ determine the ensemble in the concentric ball of radius $R-\arccos\sqrt{x_0}$, when positive. Thus inputs within radius $\pi/3$ at $x_0=3/4$ recover weights within radius $\pi/6$ [Fig.~\ref{fig:ensemble-wave}b]. Near unit overlap the decision surface is small, so a restricted range of nearby inputs already contains the data needed for local recovery. Choosing the prior sets the angular reach of the readout.

Positivity has a simple expression in the wave. Every encountered weight is nonnegative, so the response is convex in threshold. Together with the endpoint values fixed by the mean, this convexity characterizes positive ensembles. A wave with these properties therefore determines a distribution whose possible divisions can be tested against preparation demands.

\section*{Contact and optimal ensembles}

A shared quantum state imposes those demands: each remote outcome specifies a conditional operator that must be supplied by a portion of the local weights. Attach a Hermitian score to each outcome and assign every hidden state to its highest-scoring outcome. Integrating these maxima gives the ensemble's available response. Convex separation makes a common ensemble possible exactly when its response meets every demanded score (Methods). This formulation applies to arbitrary finite dimensions and outcome numbers.

Allow the total weight to vary while keeping its normalized mean fixed. The least weight satisfying the response constraints, $c_{\mathfrak M}$ for measurement class $\mathfrak M$, measures the cost of the preparations. A cost at or below one gives a normalized local model; above one, their demand exceeds the available probability. Adding white noise to a canonical state multiplies this scale by the visibility. The unit-weight crossing is therefore the steering boundary.

These weights can be computed in arbitrary finite dimensions by evolving a density $u$ relative to uniform pure-state measure. Write $\mathcal Ku$ for its available response and $N$ for the preparation demand. The deficit gives the feedback $G(u)=\mathcal K^*[N-\mathcal Ku]_+$, where $[\,\cdot\,]_+$ denotes the positive part. The spectral update is
\begin{equation}
\begin{aligned}
u_{j+1}&=B\!\left[(1-\alpha_\varepsilon\delta_\varepsilon)u_j\right.\\
&\hspace{12mm}\left.+\alpha_\varepsilon\bigl(\varepsilon^{-1}G(u_j)-1\bigr)\right].
\end{aligned}
\label{eq:dynamics}
\end{equation}
Here $B$ projects onto nonnegative densities with the prescribed normalized mean, $\alpha_\varepsilon$ is the step size, and $\varepsilon,\delta_\varepsilon>0$ control the response penalty and density regularization. Feedback adds weight in directions that supply unmet demands, while the constant cost reduces total weight. With the parameter choices in Methods, iteration converges to a unique density $u_\varepsilon$ for each $\varepsilon$. As $\varepsilon\downarrow0$, its mass approaches $c_{\mathfrak M}$ and every weak limit of the corresponding measures is an optimal ensemble, including those with singular support. Throughout this evolution, the weights generate a decision wave through Eq.~\eqref{eq:main-overlap-response}.

At a saturated bound, preparation fixes the data needed for wave reconstruction. Average a family of tests whose available response depends only on the total weight and mean. Equality of the averaged demand and supply forces their nonnegative gap to vanish along the whole test section. Contact gives the response value; tangency gives its threshold derivative. The wave inverse then determines the ensemble that could attain the bound. Uniform pure-state tests supply such bounds in any finite dimension. A positive reconstructed measure with the prescribed mean that supplies the measurement class attains the boundary. Saturation also forces pure hidden states, so a continuous recovered measure requires infinitely many hidden-state directions.

\begin{figure*}[tbp]
\centering \includegraphics[width=\textwidth]{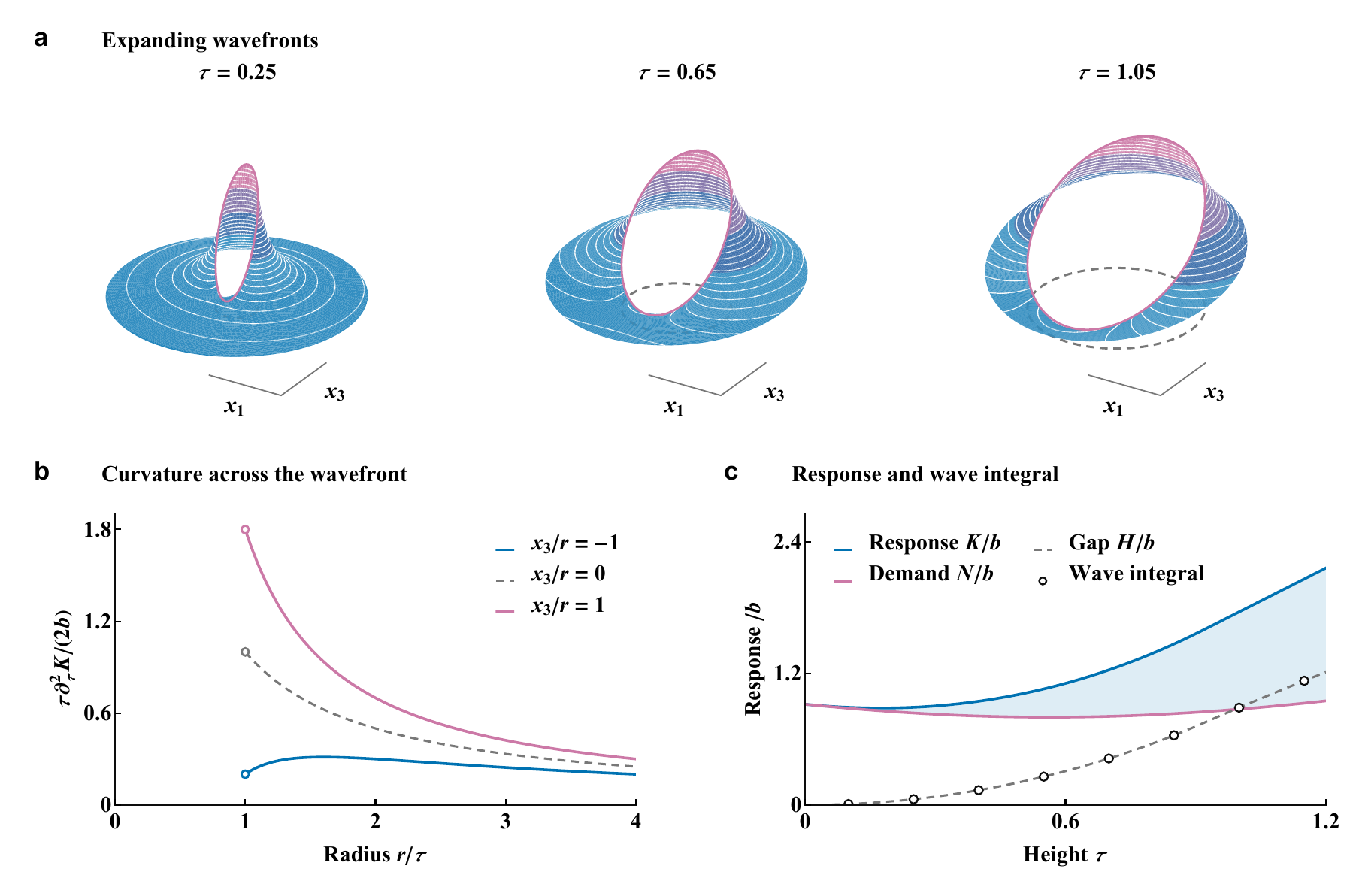} \caption{\textbf{Wave propagation of the qubit response.} The transformed preparation demand is $N/b=\sqrt{x_1^2+x_2^2+(-x_3+h\tau)^2}$, with correlation scale $b>0$ and relative bias $h=0.8$. \textbf{a}, The dimensionless response curvature $\tau\partial_\tau^2K/(2b)=\tau/r\,[1+h\tau x_3/r^2]$ on $x_2=0$ and $r=|\vec x|\ge\tau$, for three heights. The inner boundary is the characteristic circle $r=\tau$, also projected onto the base plane by the gray dashed curve. Blue to pink denotes increasing curvature; white lines mark equal values. All snapshots use the same spatial and colour scales. \textbf{b}, Radial sections of the same curvature along $x_3/r=-1,0,1$. Their common horizontal coordinate $r/\tau$ places the front at one. For $r/\tau<1$ the curvature is zero and coincides with the horizontal axis; open circles mark the outer one-sided values. \textbf{c}, Response $K/b$ (blue), demand $N/b$ (pink) and gap $H/b$ (gray dashed) at $\vec x=(0.8,0,0.45)$. Open circles evaluate the retarded integral of $-\Box N/b$ independently. The positive gap is the excess response available away from contact.} \label{fig:wave}
\end{figure*}

For two qubits, propagation establishes that the recovered ensemble supplies every projective preparation over a broad region. After trusted normalization, let $\vec a$ denote Alice's Bloch vector and $C$ the correlation matrix. The available and demanded binary responses are
\begin{equation}
\begin{aligned}
K(s,\vec y)&=\int|s+\vec y\cdot\vec n|d\mu(\vec n),\\
N(s,\vec y)&=|\vec a s+C\vec y|.
\end{aligned}
\label{eq:pvm}
\end{equation}
Here $\vec n$ is a hidden-state direction, while $s$ and $\vec y$ set the test threshold and direction. Writing the overlap response of Eq.~\eqref{eq:main-overlap-response} in homogeneous Bloch coordinates gives $K$ and turns its propagation into the ordinary wave equation. Bias tilts the contact plane: its value fixes the weights symmetric under $\vec n\mapsto-\vec n$, and its threshold derivative fixes their imbalance.

Starting from zero value and slope at contact, the gap $H=K-N$ obeys $\Box H=-\Box N$. In the effective two or three spatial dimensions, nonnegative forcing produces a nonnegative gap. Contact data consequently extend to $K\ge N$ at every test [Fig.~\ref{fig:wave}]. The trusted mean selects the plane: the first moment of its reconstructed measure is the gradient of the averaged mass bound. Setting that moment to zero locates the unique stationary plane and attains the minimum mass.

In Fig.~\ref{fig:wave}, polarization shifts weight between opposite hemispheres. Moving the threshold sweeps through those unequal weights, producing different curvature amplitudes on the two sides of the wavefront. Once all directions favour the same outcome, the response becomes linear and the curvature vanishes. Meanwhile, the propagated gap records the excess supply away from contact. This excess is what allows the distribution fixed by a tight set of tests to meet all the other projective demands.

These contact solutions cover biased and anisotropic states with a unique optimal distribution~\cite{SpectralV1}. At full correlation rank, the condition $J=CC^{\mathsf T}-\vec a\vec a^{\mathsf T}\succeq0$ permits every bias direction inside the correlation ellipsoid and arbitrary finite anisotropy, subject to state positivity. Planar correlations yield further solutions on an effective circle (Methods).

Once contact has fixed the weights, additional measurement outcomes require further divisions of that distribution, each with the prescribed probability and mean state. On the circle, the binary moment constraints already supply all these divisions: throughout $\rank C\le2$, projective and arbitrary-POVM boundaries coincide. At full rank, an explicit transformed-bias region has the same property. Across the full analytic projective domain, uniqueness gives an exact comparison: the two boundaries agree precisely when the recovered distribution supplies every POVM division.

\section*{Steering boundaries and quantum channels}

\begin{figure*}[tbp]
\centering \includegraphics[width=\textwidth]{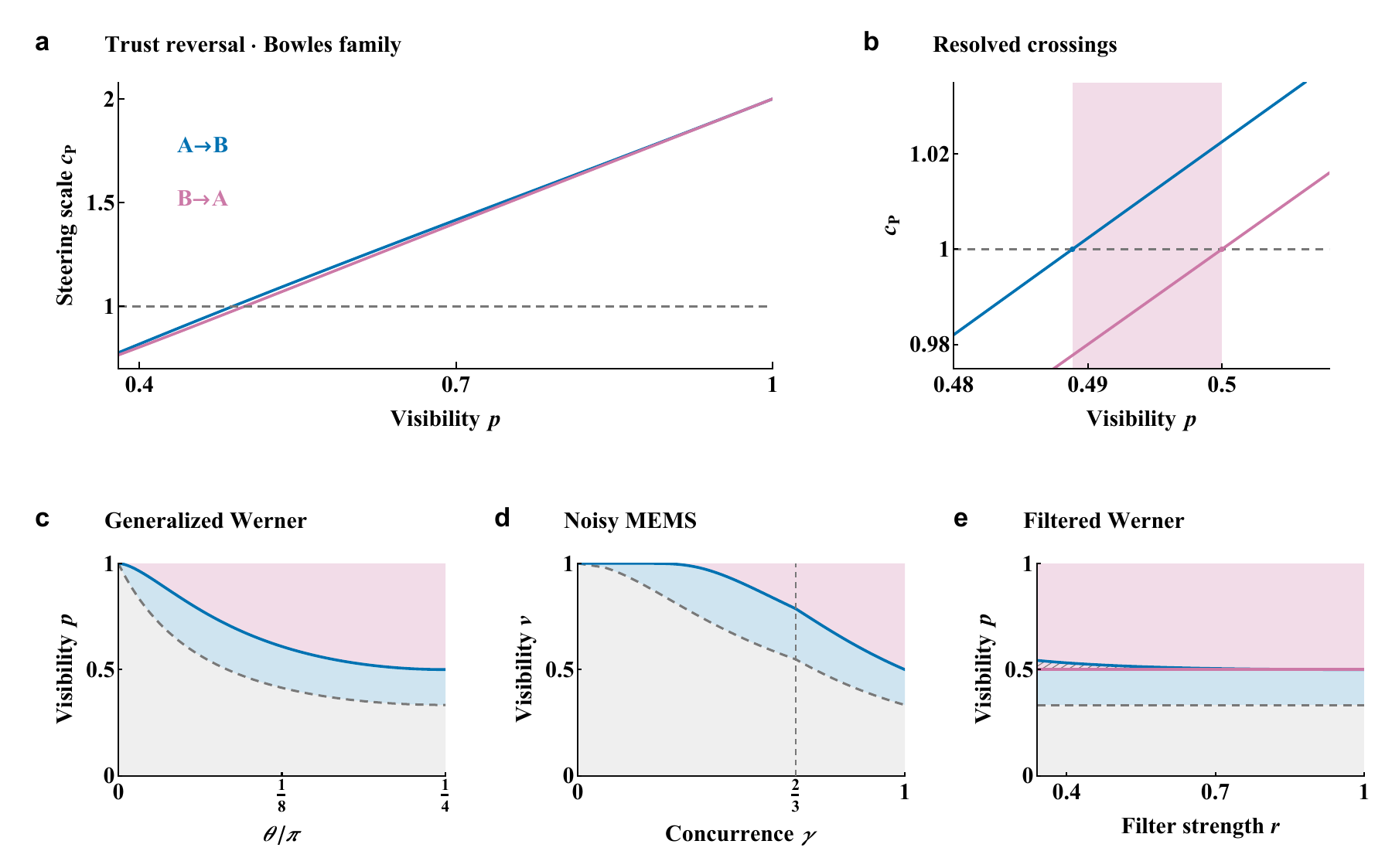} \caption{\textbf{Analytic steering boundaries of mixed-state families.} \textbf{a}, Directional projective scales of the original Bowles state: Alice to Bob (blue) and Bob to Alice (pink). The horizontal dashed line is unit mass. \textbf{b}, The resolved crossings delimit the exact arbitrary-measurement one-way interval $0.488862118569\ldots<p\le1/2$ (pink shading). \textbf{c}, Complete projective boundary for generalized Werner states, with pure component $\cos\theta\ket{00}+\sin\theta\ket{11}$. \textbf{d}, Complete projective boundary for white-noise MEMS as a function of the concurrence $\gamma$ of the initial MEMS; the dotted vertical line marks the branch junction. In \textbf{c} and \textbf{d}, gray is separable, blue shading is entangled and projectively unsteerable, and pink shading is projectively steerable; gray dashed curves are separability boundaries. \textbf{e}, Werner states after an Alice filter $\operatorname{diag}(\sqrt r,1)$. Blue and pink curves are the forward and reverse projective boundaries. The strip between them is one-way from Bob to Alice for projective measurements; the reverse curve also holds for arbitrary POVMs. All panels follow the state definitions in Supplementary Examples~2, 5, 7 and 8.} \label{fig:families}
\end{figure*}

The recovered distributions distinguish preparations with identical global mixedness. Generalized Werner states mix $\cos\theta\ket{00}+\sin\theta\ket{11}$ with white noise~\cite{JirakovaGeneralizedWerner2021}. At fixed visibility their eigenvalues are independent of $\theta$, while their conditional preparations and optimal ensembles change. Contact gives their complete projective boundary: it rises from visibility $1/2$ at maximal entanglement towards one near a product state [Fig.~\ref{fig:families}c]. Equally mixed states can therefore demand sharply different amounts of local weight. Both branches of white-noise maximally entangled mixed states~\cite{MunroMEMS} admit the same scalar contact construction [Fig.~\ref{fig:families}d].

Exchanging the trusted party changes the mean against which these weights are normalized. For the original Bowles family~\cite{BowlesOneWay}, the two reconstructed boundary ensembles supply arbitrary measurements at different crossings, giving the exact one-way interval
\begin{equation}
0.488862118569\ldots<p\le\frac12.
\label{eq:bowles-crossings}
\end{equation}
Within it, Alice steers Bob while Bob cannot steer Alice even with general measurements [Fig.~\ref{fig:families}a,b]. Directional preparation resources underlie experiments~\cite{Bennet2012,Srivastav2022} and applications to cryptography, discrimination and metrology~\cite{Branciard2012,PianiWatrous2015,Yadin2021}. The explicit weights identify where the two directions separate.

Local filtering exposes this dependence on trust in a controlled operation. Apply $\operatorname{diag}(\sqrt r,1)$ to Alice's half of a Werner state. For every $r>0$, the Bob-to-Alice boundary stays at visibility $1/2$ for arbitrary measurements: the trusted filter is absorbed into the normalization. In the opposite direction, the projective boundary moves, reaching $0.518365561644\ldots$ at $r=1/2$ [Fig.~\ref{fig:families}e]. Absorbing the filter into the trusted mean preserves the first boundary; on the untrusted side, it changes the required distribution. This sensitivity complements the use of filtering to reveal hidden nonlocality~\cite{Gisin1996,HirschHiddenNonlocality}.

Pure conditional states impose an especially restrictive demand: every contributing hidden state must lie in their one-dimensional support. Together with correlations in other directions, this can force the required mass above one. For a Bell state undergoing zero-temperature amplitude damping~\cite{KhatriGADC2020}, the constraint preserves steering in both directions throughout partial damping. Complete damping changes the trusted support and yields a product state. Supplementary Section~S5 follows these support changes and derives the remaining state-family boundaries.

The same ensemble constructs a channel's parent measurement. Transposing its pure states converts the preparation divisions into classical processing that reproduces the channel's Heisenberg observables~\cite{Choi1975,Uola2015}. For the biased qubit family defined in Methods, relative bias $h$ displaces the channel's Bloch ball and redistributes the parent weights. In wave coordinates its demanded response becomes isotropic, giving the polarized density explicitly. Bias changes its angular shape, while visibility scales the total weight. Adding output depolarization of visibility $v$ makes all output observables jointly measurable exactly when
\begin{equation}
v\le v_*(h)=\frac{3(1+h)}{2(3+h)},\qquad 0\le h\le\frac45.
\label{eq:channel-threshold}
\end{equation}
Its boundary density gives the common parent explicitly. Entanglement disappears at the smaller visibility $v_{\mathrm E}(h)=(1+h)/(3+h)$~\cite{HorodeckiShorRuskai2003}. Between $v_{\mathrm E}$ and $3v_{\mathrm E}/2$, the channel retains entanglement while every observable family is obtained by dividing one distribution~\cite{Heinosaari2015}. Bias changes that distribution and the noise tolerance together.

\section*{Outcome mixing and information loss}

Once an ensemble supplies a parent measurement, its outcome records carry the distinctions on which the decisions rely. Combining records averages their state directions into a mixed effect. One guess must then serve outcomes that previously favoured different inputs. A nonnegative source in the wave equation records this mixing of directions.

Write each nonzero effect as $G_a=nw_a\tau_a$, where $w_a=\Tr G_a/n$ and $\Tr\tau_a=1$. With $\mathcal L_n$ denoting the wave operator in Eq.~\eqref{eq:main-projective-wave}, the integrated source is
\begin{equation}
\int_0^1\!dx\int d\omega(Q)\,\mathcal L_nF_G
=\frac2n\sum_aw_a(1-\Tr\tau_a^2),
\label{eq:main-integrated-source}
\end{equation}
where $\omega$ is uniform pure-state measure and $F_G=\sum_aw_a|x-\Tr(Q\tau_a)|$. For an initially pure measurement, each merged label contributes the mean squared sine of the Fubini--Study distance between its original directions. The geometry governing propagation thus also measures the distinctions lost when directions share a record.

To quantify its decision cost, average the decrease in success over uniformly distributed pure inputs and threshold density $2(1+nx)/(n+2)$. Under any classical outcome processing, the integrated source increases by $2(n+1)(n+2)/n$ times this average decrease. Both quantities depend on the variance of the directions combined within each new label (Methods). For identical directions, merging records preserves every decision. For distinct directions, a range of thresholds assigns them opposing guesses; combining them removes that advantage. Integrating over the threshold counts these conflicting decisions, and averaging the probe direction turns the loss into their angular variance.

For isotropic outcome noise, the full response separates that loss from the distribution it obscures. Retain an outcome with probability $\eta$; otherwise replace it by an independent label drawn with its trace weight. As $\eta$ decreases, the prior interval in which outcomes can improve on guessing contracts towards equal priors [Fig.~\ref{fig:ensemble-wave}d]. Nevertheless, at every $\eta>0$ the complete success profile determines both the noise strength and the original pure-effect distribution. Its average deficit fixes $\eta$; rescaling threshold and amplitude then recovers the original wave. Individual records have become less reliable, while their full decision response still identifies the measurement that produced them.

\section*{Discussion and conclusions}

Preparation, inference and noise act on the same distribution in different ways. Preparations divide its weights; decisions reveal them; outcome mixing averages their directions. The wave keeps track of these operations because Born overlap supplies both its propagation law and the surface on which a decision changes. Its origin in projective state space makes this description available throughout finite-dimensional quantum theory.

At a saturated preparation bound, a continuum of constraints determines the wave data and can fix the whole ensemble. Uniqueness then has an operational consequence: every additional preparation must be obtained by dividing those same weights. For a continuous boundary measure, finitely many hidden directions cannot provide the required resolution. The qubit solutions make this constraint explicit across biased and anisotropic states, while the corresponding channel parents show how entanglement can survive after incompatible observables have acquired a common realization.

Wave propagation also gives a spatial meaning to incomplete access. Restricting the probe directions restricts the part of the distribution that can be recovered, with the boundary set by the chosen prior. Mixing outcome records instead averages directions within the distribution and reduces their decision contrast. These operations affect the readout in different, calculable ways. The local domain of dependence specifies where the weights are identifiable; the source--loss relation specifies the average decision cost of unresolved directions. Prior control, angular access and outcome resolution therefore enter one experiment through the geometry of its ensemble.

Finding positive contact solutions in higher dimension would extend the exact preparation boundaries. The inverse already identifies the measure selected by contact; the remaining physical task is to determine when its weights supply all required outcomes. For inference, the local propagation law suggests concentrating probes on the part of state space needed by a preparation problem. Quantifying the precision attainable from restricted and noisy decision data would connect this spatially selective reconstruction to experimental resources. In each case the object to be determined is the ensemble distribution, and the wave provides a way to follow how quantum preparations make demands on it and how observations reveal it.

\clearpage
\section*{Methods}

The calculation follows the ensemble from its response to its preparation cost. Convex separation eliminates the outcome assignments, leaving a minimum-mass problem over positive measures with a fixed mean. Pure-state overlap makes the response a wave and supplies its inverse. Contact then determines the candidate measure; positive propagation proves the qubit inequalities, and the mean selects the minimizing contact section. Spectral evolution evaluates the same minimum by adjusting the weights. Finally, conditional averaging of measurement effects gives the wave source and its decision cost.

Let $m$ be Alice's Hilbert-space dimension and $n$ the dimension of Bob's marginal support. Work on that support and apply the trusted filter $\widehat\rho=(\openone_m\otimes\rho_B^{-1/2})\rho(\openone_m\otimes\rho_B^{-1/2})/n$. The inverse is taken on that support. For an Alice effect $M_i=q_i\openone_m+X_i$, with $q_i=\Tr M_i/m$, define $A(X_i)=n\Tr_A[(X_i\otimes\openone_n)\widehat\rho]$. A measure of mass $t$ must be divisible into the conditional operators $[tq_i\openone_n+A(X_i)]/n$. To decide whether such a division exists, score each outcome with a Hermitian operator $Y_i$. Maximizing the score over all divisions assigns each hidden state to a highest-scoring outcome. A common shift of the $Y_i$ cancels, and their overall scale is immaterial. We therefore use $\sum_iY_i=0$ and $\sum_i\Tr Y_i^2=1$. With $z=(M,Y)$ denoting the normalized test, the required and available responses are
\begin{equation}
\begin{aligned}
N_A(z)&=\sum_i\Tr[Y_iA(X_i)],\\
\frac{\kappa(z,P)}{n}&=\max_i\Tr(Y_iP)-\sum_iq_i\Tr(Y_iP).
\end{aligned}
\label{eq:methods-kernel}
\end{equation}
Compactness and convexity of the attainable conditional operators give the equivalent response conditions
\begin{equation}
\begin{gathered}
(\mathcal K\mu)(z)=\int\kappa(z,P)d\mu(P)\ge N_A(z),\\
\int P\,d\mu=\frac{t\openone_n}{n}.
\end{gathered}
\label{eq:source-kernel}
\end{equation}
The inequalities apply to every normalized test in the chosen measurement class. The unknown outcome probabilities have become linear constraints on one measure. Resolving mixed hidden states into pure states preserves them. For an admissible mass $t\le1$, add $(1-t)$ times a pure-state resolution of $\openone_n/n$ and assign it with probabilities $q_i$; this gives a normalized model. Conversely, every normalized model has admissible mass one, and compactness ensures attainment of the least mass. Convex decomposition and classical postprocessing reduce arbitrary Alice POVMs to at most $m^2$ outcomes~\cite{DArianoExtremePOVM}. Supplementary Section~S1 develops these representations.

The binary response permits the ensemble in these constraints to be recovered. We normalize the Fubini--Study metric so that $\cos r\ket\psi+\sin r\ket\xi$ has unit speed for orthogonal unit vectors. Differentiating $q=\Tr(QP)$ along real and imaginary tangent directions gives $|\nabla q|^2=4q(1-q)$ and $\Delta_{\rm FS}q=4(1-nq)$. For $|x-q|$, the second derivative contributes $2\delta(x-q)$. Substituting $\int q\operatorname{sgn}(x-q)d\mu=xF_x-F$ gives Eq.~\eqref{eq:main-projective-wave}. Its principal symbol vanishes on $x=\Tr(QP)$, identifying the characteristic fronts. Its normalized curvature is
\begin{equation}
S(x,Q)=\frac{\partial_x^2F(x,Q)}{2(n-1)(1-x)^{n-2}}.
\label{eq:main-curvature-reconstruction}
\end{equation}
Twice differentiating the wave equation gives
\begin{equation}
x(1-x)S_{xx}+(1-nx)S_x=\frac14\Delta_{\rm FS}S.
\label{eq:methods-shell-wave}
\end{equation}
In the radius $r=\arccos\sqrt x$, this is the spherical-mean equation $S_{rr}+[(2n-3)\cot r-\tan r]S_r=\Delta_{\rm FS}S$. The shrinking support of these averages proves the curvature reconstruction, uniformly for continuous densities and weakly for finite measures.

For the inverse, expand in real orthonormal harmonics with eigenvalues $-4\ell(\ell+n-1)$. Their normalized zonal polynomial is $\pi_\ell(q)=P_\ell^{(n-2,0)}(2q-1)/P_\ell^{(n-2,0)}(1)$~\cite{Koornwinder1973}. The response multiplies the degree-$\ell$ coefficient of the measure by
\begin{equation}
a_\ell(x)=(n-1)\int_0^1|x-q|(1-q)^{n-2}\pi_\ell(q)dq.
\label{eq:methods-response-multiplier}
\end{equation}
Each angular degree obeys $x(1-x)a_\ell''-(1-nx)a_\ell'+[\ell(\ell+n-1)-n]a_\ell=0$, with $a_\ell''=2(n-1)(1-x)^{n-2}\pi_\ell(x)$. Since $a_\ell$ is nonzero, uniqueness for this ordinary differential equation prevents $a_\ell$ and $a_\ell'$ from vanishing together at an interior point. Writing $A_{\ell j}$ and $B_{\ell j}$ for the value and derivative coefficients there gives the recovered moment
\begin{equation}
\mu_{\ell j}=\frac{a_\ell A_{\ell j}+a_\ell'B_{\ell j}}{a_\ell^2+(a_\ell')^2}.
\label{eq:methods-cauchy-inverse}
\end{equation}
Heat summation recovers finite measures from these moments. Each $a_\ell$ is a nonzero polynomial, so outside a countable union of roots a single threshold already gives the inverse $\mu_{\ell j}=A_{\ell j}/a_\ell$.

For the prescribed mass and operator mean, the endpoints are
\begin{equation}
F(0,Q)=t/n,\qquad F(1,Q)=t(n-1)/n.
\label{eq:main-wave-endpoints}
\end{equation}
These values select the physical solution of each angular equation. At degree zero they fix the average $t a_0(x)$; at degree one they force zero traceless moment. For $\ell\ge2$, one independent solution behaves as $x^2$ at zero and the other has a nonzero limit. The zero endpoint therefore selects a multiple of $a_\ell$. Convexity makes $F_{xx}$ positive. Its angular average is $2t(n-1)(1-x)^{n-2}$, so normalized curvature gives positive measures of mass $t$. As $x\to1$, their moments tend to the recovered harmonic coefficients. Compactness supplies a positive limit, and density of the harmonics makes it unique. Its response is $F$, proving the characterization used in the mass minimization.

For a variable threshold $x=f(Q)$, the coordinate $\tau=\arcsin(2x-1)$ gives a wave equation with principal metric $d\tau^2-4g_{\rm FS}$ and smooth lower-order terms on every interior slab. The inequality
\begin{equation}
|\nabla_{\rm FS}f|^2<4f(1-f)
\label{eq:main-spacelike}
\end{equation}
makes the graph a spacelike Cauchy surface. The response of a finite measure is locally Lipschitz, and its transverse derivative exists almost everywhere on that graph. Finite-energy Cauchy uniqueness~\cite{BaerWafo2015} propagates equal data to equal responses, after which Eq.~\eqref{eq:methods-cauchy-inverse} identifies the measure. In the qubit case $F(x,Q)=K(2x-1,-\vec e)/2$ for $Q=(\openone_2+\vec e\cdot\vec\sigma)/2$. A plane $s=\vec q\cdot\vec y$ becomes $f=(1-\vec q\cdot\vec e)/2$, and $4f(1-f)-|\nabla f|^2=1-|\vec q|^2$. The explicit biased-state contact planes are therefore instances of this same reconstruction theorem.

Finite propagation gives a local form of the reconstruction. Let the value and threshold derivative at $x_0$ be known on an open set $U$ of pure test states. In the coordinate $\tau$, the propagation speed is $1/2$. Two finite-measure responses with these data agree wherever
\begin{equation}
d_{\rm FS}(Q,U^c)>\frac12|\tau-\tau_0|,\qquad \tau_0=\arcsin(2x_0-1).
\label{eq:methods-local-wave}
\end{equation}
This is the domain of dependence for the finite-energy wave equation~\cite{BaerWafo2015}. As $x\uparrow1$, the distance travelled tends to $r_0=\arccos\sqrt{x_0}$. For any compact region with distance greater than $r_0$ from $U^c$, the responses and their curvature agree on a neighbourhood throughout the approach to this endpoint. Weak curvature recovery then gives equality of the measures on that region. A ball of input radius $R>r_0$ therefore determines the ensemble in the concentric ball of radius $R-r_0$. Values and prior derivatives of Eq.~\eqref{eq:main-discrimination} supply the required data for a rank-one measurement. Supplementary Section~S2 proves the statement for finite signed measures.

Arbitrary outcome numbers use the linear expectation of a score operator. With orthonormal traceless matrices $T_\alpha$, write $P=\openone_n/n+\sqrt{(n-1)/n}\,\vec v\cdot\vec T$ and $Y_i=s_i\openone_n+\sqrt{n/(n-1)}\,\vec y_i\cdot\vec T$. Purity gives $|\vec v|=1$. For $R_\mu=\int\max_i(s_i+\vec y_i\cdot\vec v)d\mu$, smooth approximation to the maximum yields
\begin{equation}
\left(\partial_{s_i}\partial_{s_j}-\nabla_{\vec y_i}\cdot\nabla_{\vec y_j}\right)R_\mu=0.
\label{eq:methods-general-wave}
\end{equation}
Purity cancels the difference of the two terms through the factor $1-|\vec v|^2$. Centering scores subtracts a linear function and preserves these identities. Matrix identities from $P^2=P$, harmonic multipliers and the reconstruction proofs appear in Supplementary Section~S2.

For a binary effect $0\preceq M\preceq\openone_m$, write $Z=vA^*(H)+(\Tr H)\openone_m/m$. Maximizing $[2\Tr(MZ)-\Tr Z]/n$ gives the demanded binary response
\begin{equation}
N_v(H)=\frac1n\left\|vA^*(H)+\frac{\Tr H}{m}\openone_m\right\|_1.
\label{eq:main-binary-source}
\end{equation}
Haar averaging the available response to $x\openone_n-Q$ gives $a_0(x)$ for every pure direction. Consequently, every normalized ensemble obeys
\begin{equation}
\int N_v(x\openone_n-Q)d\omega(Q)\le x-\frac1n+\frac2n(1-x)^n.
\label{eq:main-general-contact-bound}
\end{equation}
At saturation, its nonnegative gap with the demanded response vanishes everywhere. Convexity of the demand in $x$ fixes the derivative wherever the response is differentiable. At zero visibility the averaged demand is $|x-1/n|<a_0(x)$; convexity in visibility therefore makes every visibility beyond a positive contact violate the same bound. A positive reconstructed spectrum that supplies the measurement class attains the boundary. Mixed hidden states have strictly smaller Haar-averaged response, by Jensen inequality with strictness on an open set of directions. Saturation consequently forces every parent to have rank-one resolution.

For two qubits, write $\widehat\rho=[\openone_4+\vec a\cdot\vec\sigma\otimes\openone_2+\sum_{ij}C_{ij}\sigma_i\otimes\sigma_j]/4$. Maximizing a binary test over Alice's measurement direction gives Eq.~\eqref{eq:pvm}. Only the effective space $E=\operatorname{Ran}C^{\mathsf T}$ is needed: projection onto $E$ followed by radial resolution reduces the hidden support to its unit sphere. At rank at most one, write $C=\vec b\vec v^{\mathsf T}$, $|\vec v|=1$. Equal weights on $\pm\vec v$ give $c_{\mathrm P}=\max\{|\vec a+\vec b|,|\vec a-\vec b|\}$.

At ranks two and three, seek a plane on which the response just reaches the demand. For $\vec q\in E$, $|\vec q|<1$, put $\delta=1-|\vec q|^2$ and $T_{\vec q}=\sqrt\delta\,\openone_3+\vec q\vec q^{\mathsf T}/(1+\sqrt\delta)$. The identity $|\vec q+T_{\vec q}\vec n|=1+\vec q\cdot\vec n$ means that twice the spherical average of $K(\vec q\cdot\vec e,T_{\vec q}\vec e)$ equals the mass of any zero-moment ensemble. Averaging the inequality $K\ge N$ therefore gives $t\ge\Phi(\vec q)$, with
\begin{equation}
\begin{aligned}
\Phi(\vec q)&=2\int_{S^2}\sqrt{\vec e^{\mathsf T}M(\vec q)\vec e}\,d\omega(\vec e),\\
M(\vec q)&=(\vec a+C\vec q)(\vec a+C\vec q)^{\mathsf T}+(1-|\vec q|^2)J.
\end{aligned}
\label{eq:methods-phi}
\end{equation}
This ambient-sphere average applies at both ranks. At equality, a continuous nonnegative gap of zero average vanishes throughout the contact plane, supplying the data for reconstruction.

A Lorentz transformation makes this reconstruction explicit. Set $\gamma=\delta^{-1/2}$ and $S_{\vec q}=\gamma T_{\vec q}$, and use $s=\gamma(\tau+\vec q\cdot\vec x)$, $\vec y=S_{\vec q}\vec x+\gamma\vec q\tau$. The demand becomes $|\vec d\tau+B\vec x|$, where $\vec d=\gamma(\vec a+C\vec q)$ and $B=CS_{\vec q}+\gamma\vec a\vec q^{\mathsf T}$; in particular, $BB^{\mathsf T}-\vec d\vec d^{\mathsf T}=J$. At full rank, $J\succeq0$ ensures positivity. At rank two, it suffices that the two smallest eigenvalues of $J$ sum to a nonnegative value. At full rank, push the measure $2(|B^{\mathsf T}\vec e|+\vec d\cdot\vec e)d\omega(\vec e)$ to $\vec n'=B^{\mathsf T}\vec e/|B^{\mathsf T}\vec e|$. Cosine and hemisphere integration then give the required contact value $|B\vec x|$ and derivative $\vec d\cdot B\vec x/|B\vec x|$. The rank-two construction uses the effective circle. The stated eigenvalue conditions ensure that the reconstructed weights are nonnegative.

Positive propagation extends the reconstructed contact data to all tests. The gap $H=K-N$ obeys
\begin{equation}
\Box H=-\Box N,\qquad H(0,\vec x)=\partial_\tau H(0,\vec x)=0,
\label{eq:wave-gap}
\end{equation}
with $\Box=\partial_\tau^2-\Delta_{\vec x}$. Away from a zero of $N$, set $\vec e=(\vec d\tau+B\vec x)/N$. Differentiation gives
\begin{equation}
-\Box N=\frac{\Tr J-\vec e^{\mathsf T}J\vec e}{N}\ge0. \label{eq:methods-wave-source}
\end{equation}
At rank two, the spatial Laplacian acts on $E$; the same source matrix $J$ occurs in the numerator. Writing $\Gamma=-\Box N$, the three-dimensional solution with zero Cauchy data is $H(\tau,\vec x)=\int_0^\tau t\int_{S^2}\Gamma(\tau-t,\vec x+t\vec e)d\omega(\vec e)dt$ for $\tau\ge0$. The effective two-dimensional solution uses the positive disk propagator. Applying these formulas on both sides of the contact plane proves $K\ge N$ everywhere. Regularization of the source norm treats its zeros.

Transforming back sends $\vec n'$ to $(S_{\vec q}\vec n'-\gamma\vec q)/[\gamma(1-\vec q\cdot\vec n')]$ and multiplies its weight by $\gamma(1-\vec q\cdot\vec n')$. Its first moment becomes $\nabla_E\Phi$ and its mass $\Phi-\vec q\cdot\nabla_E\Phi$. Zero moment therefore selects a measure attaining the bound:
\begin{equation}
\nabla_E\Phi(\vec q_*)=0,\qquad c_{\mathrm P}=\Phi(\vec q_*). \label{eq:contact-scale}
\end{equation}
Strict concavity and inward increase at the boundary ensure a unique interior stationary point. Cauchy inversion fixes the corresponding optimal measure. Supplementary Section~S4 treats both effective ranks and the boundary analysis.

For general qubit measurements, each outcome must match both a probability and a Bloch vector. At most four effects are required~\cite{DArianoExtremePOVM}. Optimizing their required scores gives a weighted Euclidean-distance problem, and the winning regions of their affine response branches determine the parent outcome moments. On an effective circle, those moment conditions imply that every POVM is supplied at the projective mass. For full-rank spectra, spherical cone moments give an explicit sufficient bias ball depending on the transformed anisotropy; at isotropy its radius is $4/5$. Throughout the full-rank analytic domain, the squared positive POVM deficit of the unique projective spectrum integrates to zero exactly when the two measurement classes have the same scale. Section~S3 of the Supplementary Information gives the reduced evaluation of the demanded scores and height variables, and Section~S4 proves the region-moment results.

To compute the measure, choose normalized invariant measure $\omega$ on $\mathbb CP^{n-1}$ and a full-support probability measure $\sigma$ on the compact normalized tests. Let $\vec v(\psi)$ be the unit Bloch coordinates used above. Densities with the correct marginal form the cone $\mathcal C=\{u\ge0:\int\vec v u\,d\omega=0\}$. The mass and response constraints enter one convex functional,
\begin{equation}
\begin{aligned}
J_\varepsilon(u)&=\int u\,d\omega+\frac{\delta_\varepsilon}{2}\|u\|_{L^2(\omega)}^2\\
&\hspace{12mm}+\frac{1}{2\varepsilon}\|[N-\mathcal Ku]_+\|_{L^2(\sigma)}^2.
\end{aligned}
\label{eq:methods-functional}
\end{equation}
Here $N=[N_A]_+$ and $\delta_\varepsilon=\varepsilon^{2n-1}$. Differentiation gives the gradient $1-\varepsilon^{-1}G(u)+\delta_\varepsilon u$, and projecting a step onto $\mathcal C$ yields Eq.~\eqref{eq:dynamics}. To evaluate the projection $B$, minimize $\tfrac12\int[g-\vec\eta\cdot\vec v]_+^2d\omega$ over the finite-dimensional multiplier $\vec\eta$, then put $B(g)=[g-\vec\eta\cdot\vec v]_+$. Stationarity enforces the marginal and the positive part enforces positivity. A minimizing multiplier exists because every nonzero traceless Hermitian combination takes both signs on pure states.

A kernel bound $\kappa(z,\psi)\le H(z)$ gives $L_\varepsilon=\delta_\varepsilon+\|H\|_{L^2(\sigma)}^2/\varepsilon$. With step $1/L_\varepsilon$, the difference of two updates contracts by at most $1-\delta_\varepsilon/L_\varepsilon$. Indeed, the secant slope of the positive-part function lies between zero and one, so the gradient difference contains $\delta_\varepsilon\openone+\varepsilon^{-1}\mathcal K^*\Theta\mathcal K$, with $0\le\Theta\le\openone$; orthogonal projection is nonexpansive. Each regularized problem consequently has a unique attracting density.

Singular optimal measures are recovered by the chosen decay rate of $\delta_\varepsilon$. Heat smoothing at time $\varepsilon^2$ preserves their mass and zero moment, changes the response by $O(\varepsilon)$, and gives squared density norm $O(\varepsilon^{-(2n-2)})$. Substitution into Eq.~\eqref{eq:methods-functional} bounds the regularized minimum by $c_{\mathfrak M}+O(\varepsilon)$. Bounded mass gives weakly convergent subsequences; vanishing response penalty, kernel continuity and full support of $\sigma$ make every limit admissible at every test. Its mass is therefore the minimum:
\begin{equation}
c_{\mathfrak M}=\lim_{\varepsilon\downarrow0}\int u_\varepsilon d\omega.
\label{eq:spectral-equation}
\end{equation}
Circle-supported spectra, finite atomic measures and other singular supports are included. Supplementary Section~S3 derives the evolution and its limits.

Axial symmetry reduces the contact equation to a scalar root. For $\vec a=A\vec e_3$, correlation magnitudes $(T,T,|L|)$ and $L^2\ge A^2$, put $X=T^2(1-q^2)$ and $Y=(L+Aq)^2-X$. Then
\begin{equation}
\begin{aligned}
0&=\int_0^1\frac{-T^2q(1-x^2)+A(L+Aq)x^2}{\sqrt{X+Yx^2}}\,dx,\\
c_{\mathrm P}&=2\int_0^1\sqrt{X+Yx^2}\,dx.
\end{aligned}
\label{eq:methods-axial}
\end{equation}
Elementary inverse-trigonometric or inverse-hyperbolic expressions evaluate the mass, and reduced-rank equations give degenerate endpoints. Both Bowles unit-mass crossings lie within the POVM bias ball: convexity supplies models below them and projective tests detect steering above them. Whitening, parameter ranges and endpoints for all ten families appear in Supplementary Section~S5.

For the biased family, $0\le v,h\le1$, the canonical Bloch vector is $\vec a=2vh\vec e_3/[(3-h)(1+h)]$. The correlation matrix is diagonal, with
\[
\begin{aligned}
C_{11}=C_{22}&=-\frac{v\sqrt{9-h^2}}{(3-h)(1+h)},\\
C_{33}&=-\frac{v(3-h^2)}{(3-h)(1+h)}.
\end{aligned}
\]
Stationarity gives $\vec q_*=-h\vec e_3/3$ and mass $c_{\mathrm P}=2v(3+h)/[3(1+h)]$, with density
\begin{equation}
u_*(n_3)=c_{\mathrm P}\,\frac{(1-h^2/9)(1+h^2/3-4hn_3/3)}{(1-hn_3/3)^4}.
\label{eq:methods-bias-density}
\end{equation}
Its first moment vanishes. The transformed demand is isotropic with relative bias $h$, so POVM equivalence holds for $h\le4/5$.

In the binary decision, the difference of the prior-weighted outcome probabilities is $nw_a[x-\Tr(Q\tau_a)]/(1+nx)$. Summing absolute values proves Eq.~\eqref{eq:main-discrimination}. To compare an arbitrary POVM $H$ with rank-one $G$, fix $x\notin\mathcal E_n$, where $\mathcal E_n$ is the countable union of multiplier zeros. Strict Jensen inequality gives $\int F_Hd\omega\le a_0(x)=\int F_Gd\omega$, with equality exactly when $H$ is rank one. If $F_H\ge F_G$ everywhere, continuity and the equal averages force identical responses; division by the nonzero harmonic multipliers then gives the same pure-state measure. Conversely, any mixed effect admits a pure refinement that increases the average strictly while never lowering a response. Thus rank-one measurements are precisely the maximal measurements for these fixed-prior tasks. Two inequivalent rank-one measurements have different continuous profiles of equal average, so their difference takes both signs on open sets.

The qutrit densities in Fig.~\ref{fig:ensemble-wave}c all have mass one, operator mean $\openone_3/3$ and positive weights. Their responses coincide at $x=3/7$ because $a_3(3/7)=0$, giving identical discrimination probabilities at $p=9/16$. Since $a_3'(3/7)\ne0$, the prior derivatives distinguish them. At every interior threshold, value and derivative together recover all rank-one weights. Supplementary Section~S2 also derives reconstruction from all equal-prior binary input pairs for $n\ge3$ using the cosine transform~\cite{AubrunLancien2016}.

A mixed normalized effect $\tau$ obeys $|\nabla\Tr(Q\tau)|^2=4[\Tr(Q\tau^2)-\Tr(Q\tau)^2]$, while its Laplacian remains $4[1-n\Tr(Q\tau)]$. Substitution gives the nonnegative wave source
\begin{equation}
\begin{aligned}
\mathcal L_nF_G={}&2\sum_aw_a\Tr[Q(\tau_a-\tau_a^2)]\\
&\hspace{8mm}\times\delta\!\left(x-\Tr(Q\tau_a)\right)\ge0.
\end{aligned}
\label{eq:main-measurement-source}
\end{equation}
Integrating in $x$ and $Q$ yields $2/n$ times the weighted linear entropy. Under classical processing $\widetilde G_b=\sum_a C(b|a)G_a$, the weighted linear entropy increases by $\sum_{a,b}w_aC(b|a)\Tr[(\tau_a-\widetilde\tau_b)^2]$. Merging distinct normalized effects therefore strictly increases the integrated source. Supplementary Section~S2 includes the continuous-outcome proofs. For a rank-one initial measurement, the purity loss within a merged outcome is the conditional mean of $1-\Tr(PP')=\sin^2d_{\rm FS}(P,P')$. Thus the source and discrimination loss are averages of the separation between the original pure directions.

The additional source has an exact discrimination cost. Write $\Delta p=p_{\rm succ}(x,Q;G)-p_{\rm succ}(x,Q;\widetilde G)$ for the success lost under the classical processing above, and let $\overline{\Delta p}$ average over Haar-distributed $Q$ and threshold density $2(1+nx)/(n+2)$. Then
\begin{equation}
\overline{\Delta p}=\frac{\sum_{a,b}w_a C(b|a)\Tr[(\tau_a-\widetilde\tau_b)^2]}{(n+1)(n+2)}.
\label{eq:methods-discrimination-loss}
\end{equation}
Indeed, $\int_0^1|x-q|dx=q^2-q+1/2$. At fixed $Q$, preserving the mean of each merged outcome makes the integrated response loss equal to $\sum_{a,b}w_a C(b|a)[\Tr Q(\tau_a-\widetilde\tau_b)]^2$. Haar averaging a traceless Hermitian $D$ gives $\int(\Tr QD)^2d\omega=\Tr D^2/[n(n+1)]$. The chosen threshold density cancels the prefactor in Eq.~\eqref{eq:main-discrimination}, yielding Eq.~\eqref{eq:methods-discrimination-loss}. Comparing with Eq.~\eqref{eq:main-integrated-source} gives the source increase $2(n+1)(n+2)\overline{\Delta p}/n$. Integrating the source only over threshold also gives a nonnegative increase in every input direction, equal to $2\sum_{a,b}w_a C(b|a)\Tr[Q(\tau_a-\widetilde\tau_b)^2]$. Supplementary Section~S2 gives these identities for general outcome spaces.

Isotropic outcome randomization gives $G_\eta(dP)=n[\eta P+(1-\eta)\openone_n/n]d\mu(P)$. Put $b=(1-\eta)/n$. At $\eta>0$, its response is
\begin{equation}
F_\eta(x,Q)=\eta F\!\left(\frac{x-b}{\eta},Q\right).
\label{eq:methods-noise-response}
\end{equation}
For $b<x<b+\eta$ this reads an interior section of the original wave. Outside it, all outcome scores have the same sign and $F_\eta=|x-1/n|$. The corresponding prior interval is
\begin{equation}
p_-(\eta)=\frac{1-\eta}{2-\eta},\qquad p_+(\eta)=\frac{1+(n-1)\eta}{2+(n-1)\eta}.
\label{eq:main-noise-priors}
\end{equation}
Within this interval, Haar averaging gives a strict improvement for some pure inputs, since $a_0(y)>|y-1/n|$ for every $0<y<1$. Since each normalized effect has purity $1/n+(1-1/n)\eta^2$, the integrated source $\Sigma_\eta$ and average loss from the original rank-one measurement are
\begin{equation}
\begin{aligned}
\Sigma_\eta&=\frac{2(n-1)}{n^2}(1-\eta^2),\\
\overline{\Delta p}&=\frac{(n-1)(1-\eta^2)}{n(n+1)(n+2)}.
\end{aligned}
\label{eq:methods-noise-loss}
\end{equation}
The source can be read without differentiating the response. For any normalized measurement, Haar averaging $\int_0^1|x-q|dx=q^2-q+1/2$ gives
\begin{equation}
\Sigma_G=2(n+1)\int_0^1\left[a_0(x)-\int F_G(x,Q)d\omega(Q)\right]dx.
\label{eq:methods-source-area}
\end{equation}
Thus the area deficit of the input-averaged response determines $\Sigma_\eta$ and hence the nonnegative $\eta$. For $\eta>0$, Eq.~\eqref{eq:methods-noise-response} then recovers $F$ on $0<x<1$ and hence its generating measure. Supplementary Section~S2 proves this joint identification for general outcome spaces.

To obtain a parent measurement for a channel, use $\rho_{\mathcal E}=(\mathcal E\otimes\mathrm{id})(\ketbra{\Phi^+})$ with the output on Alice. The identity $2\Tr_A[(M_i\otimes\openone_2)\rho_{\mathcal E}]=\mathcal E^*(M_i)^{\mathsf T}$ converts the unit-mass measure into $G(d\vec n)=2\tau_{\vec n}^{\mathsf T}d\mu$~\cite{Uola2015}. The outcome probabilities of the steering model are its classical postprocessing. For $\rho(1,h)$, the channel drift is $\vec a$ and its Bloch matrix is $C\operatorname{diag}(1,-1,1)$. Output depolarization produces $\rho(v,h)$ and gives Eq.~\eqref{eq:channel-threshold}. Below the threshold $v_*(h)$ in Eq.~\eqref{eq:channel-threshold}, multiply the boundary parent by $\lambda=v/v_*(h)$ and add the effect $(1-\lambda)\openone_2$. Assign this extra outcome probability $\Tr M_i/2$. The resulting effects are exactly $\mathcal E_h^*\circ\mathcal D_v^*(M_i)$.

The wave plots use Jacobi harmonics of the qutrit density specified in Fig.~\ref{fig:ensemble-wave}. Spherical means give its normalized curvature; harmonic inversion of the response and slope reconstructs its weights. The local reconstruction panel displays the domain determined by finite propagation. Supplementary Fig.~S1 also evaluates the response directly by pure-state quadrature, integrating the relative phase analytically. For the axial spectral evolution in Supplementary Fig.~S2, a positive Bernstein basis represents the density, with its first moment imposed as a linear constraint. Projected gradient updates minimize the discretized response-deficit functional.

In Fig.~\ref{fig:wave}, angular integration gives $K/b=r+\tau^2/r-hx_3(\tau-\tau^3/(3r^2))/r$ for $0\le\tau\le r$, and $K/b=2\tau-2hx_3/3$ for $\tau\ge r$, where $r=|\vec x|$. The first two panels show its curvature; panel c evaluates the retarded integral by angular and radial quadrature. Figure~\ref{fig:families} follows the scalar contact equation and its one-dimensional boundary roots.

\backmatterhead{Note added}
Version 1 of this manuscript, submitted on 13 September 2026~\cite{SpectralV1}, characterized the convex set of conditional-state differences in Eq.~(3) and Supplementary Eq.~(S65). For the same pure-state ensemble, the parent-POVM zonoid used in Ref.~\cite{ChitambarZhangV1}, submitted on 5 October 2026, is affinely equivalent to this set. Uniqueness of the optimal pure-state LHS spectrum for the critical Werner state, discussed there, was already included as a special case of Supplementary Theorem S9 of our v1. The present revision develops wave propagation and ensemble reconstruction in arbitrary finite dimension, together with their applications to measurement resolution and noise.

\backmatterhead{Data availability}
The analytical results are provided in this article and its Supplementary Information. The numerical data underlying the figures are available from the corresponding authors upon reasonable request.

\backmatterhead{Code availability}
The scripts used for numerical calculations and figure generation are available from the corresponding authors upon reasonable request.

\backmatterhead{Funding}
This work is supported by the Quantum Science and Technology--National Science and Technology Major Project (Grant No. 2024ZD0301000) and the National Natural Science Foundation of China (Grant No. 12275136). L.-C.K. acknowledges funding from the National Research Foundation, Singapore, and the Ministry of Education, Singapore.

\backmatterhead{Author contributions}
All authors contributed to the conception of the study and the conceptual formulation of the theory. Y.-X.Z. carried out the theoretical derivations and numerical calculations and prepared the initial manuscript. L.-C.K. provided scientific guidance, helped shape the research direction and advised on the notation. J.-L.C. provided scientific guidance and contributed to the theoretical derivations. All authors participated in the interpretation of the results and the revision of the manuscript.

\backmatterhead{Competing interests}
The authors declare no competing interests.

\backmatterhead{Additional information}
Supplementary Information accompanies this article. Correspondence and requests for materials should be addressed to Leong-Chuan Kwek or Jing-Ling Chen.

\end{document}


\raggedbottom

\title{Supplemental Material for ``Quantum ensembles as waves on state space''}

\author{Yu-Xuan Zhang} \affiliation{School of Physics, Nankai University, Tianjin 300071, People's Republic of China} \affiliation{Centre for Quantum Technologies, National University of Singapore, Singapore 117543, Singapore}

\author{Leong-Chuan Kwek} \email{kwekleongchuan@nus.edu.sg} \affiliation{Centre for Quantum Technologies, National University of Singapore, Singapore 117543, Singapore} \affiliation{National Institute of Education, Nanyang Technological University, Singapore 637616, Singapore}

\author{Jing-Ling Chen} \email{chenjl@nankai.edu.cn} \affiliation{Theoretical Physics Division, Chern Institute of Mathematics, Nankai University, Tianjin 300071, People's Republic of China}

\date{\today}

\maketitle

\section{S1. Common ensembles and steering responses} \label{sec:steering-scale}

A local hidden-state description supplies all conditional preparations in a steering experiment from one distribution of local states. We express this requirement through the response of that distribution to outcome scores. Filtering on the trusted support fixes its operator mean; convex separation then makes the existence of every outcome division equivalent to response inequalities. The least total weight satisfying them defines the steering scale. This reduction identifies both the distribution represented by the wave in Section~S2 and the preparation constraints used to select it in Sections~S3 and S4.

\subsection{Common ensembles and the steering scale} \label{sec:lhs-whitening-scale}

Let $m,n\geq1$ be integers, let $\rho$ be a state on $\C^m\otimes\C^n$, and write $B=\Tr_A\rho$ for Bob's reduced state. Let $\mathfrak M$ be a family of finite-outcome POVMs on Alice. For $M=(M_1,\ldots,M_{r_M})\in\mathfrak M$,
\begin{equation}
M_a\succeq0\quad(1\leq a\leq r_M), \qquad \sum_{a=1}^{r_M}M_a=\openone_m. \label{eq:finite-povm}
\end{equation}
Bob's unnormalized conditional states are
\begin{equation}
\sigma_{a|M}=\Tr_A[(M_a\otimes\openone_n)\rho]\succeq0, \qquad \sum_{a=1}^{r_M}\sigma_{a|M}=B. \label{eq:conditional-states}
\end{equation}
An Alice-to-Bob local hidden-state (LHS) model for $\mathfrak M$~\cite{JonesWisemanDoherty} consists of a probability space $(\Lambda,\Sigma,\nu)$, a measurable family of density operators $\tau_\lambda$ on Bob, and measurable response probabilities $p(a|M,\lambda)$ satisfying
\begin{equation}
\sigma_{a|M} =\int_\Lambda p(a|M,\lambda)\tau_\lambda\,d\nu(\lambda) \label{eq:lhs-state-form}
\end{equation}
for every $M\in\mathfrak M$ and every outcome $a$, with
\begin{equation}
p(a|M,\lambda)\geq0, \qquad \sum_{a=1}^{r_M}p(a|M,\lambda)=1. \label{eq:response-probabilities}
\end{equation}
Summing the outcome equations gives
\begin{equation}
\int_\Lambda\tau_\lambda\,d\nu(\lambda) =\sum_{a=1}^{r_M}\sigma_{a|M} =B. \label{eq:lhs-bob-marginal}
\end{equation}
Both $\nu$ and $\tau_\lambda$ are common to all $M\in\mathfrak M$; only the outcome divisions depend on the measurement.

Absorb the weights and states into a positive operator measure $H$:
\[
dH(\lambda)=\tau_\lambda\,d\nu(\lambda).
\]
The model equations become
\begin{equation}
H(\Lambda)=B, \qquad \sigma_{a|M} =\int_\Lambda p(a|M,\lambda)\,dH(\lambda). \label{eq:common-positive-operator-measure}
\end{equation}
Conversely, suppose that a positive operator measure $H$ and response probabilities satisfy Eqs.~\eqref{eq:response-probabilities} and \eqref{eq:common-positive-operator-measure}. Its trace measure
\[
d\nu(\lambda)=\Tr dH(\lambda)
\]
is a probability measure since $\Tr B=1$. Positivity makes $H$ absolutely continuous with respect to $\nu$, so the finite-dimensional Radon--Nikodym theorem gives
\[
dH(\lambda)=\tau_\lambda\,d\nu(\lambda),
\]
where $\tau_\lambda$ is a density operator almost everywhere, and may be chosen so on the null set as well. Substitution into Eq.~\eqref{eq:common-positive-operator-measure} recovers the LHS formulation.

Only the support of $B$ is needed. If $P$ projects onto $\supp B$, positivity gives
\[
\|[\openone_m\otimes(\openone_n-P)]\rho^{1/2}\|_2^2=\Tr[(\openone_n-P)B]=0.
\]
Hence $\rho=(\openone_m\otimes P)\rho(\openone_m\otimes P)$, and every conditional state has this support. Each value of $H$ has the same support because $0\preceq H(E)\preceq H(\Lambda)=B$. Conversely, a parent on the support extends by zero to the original space. We henceforth retain $n$ for the support dimension, so $B\succ0$.

For an invertible operator $K$ on Bob's space, define
\[
\rho_K= \frac{(\openone_m\otimes K)\rho(\openone_m\otimes K^\dagger)} {\Tr(KBK^\dagger)}.
\]
Then
\[
\Tr_A[(M_a\otimes\openone_n)\rho_K] = \frac{K\sigma_{a|M}K^\dagger}{\Tr(KBK^\dagger)}.
\]
Accordingly, if $H$ is a common parent measure for $\rho$, then
\[
dH_K(\lambda) = \frac{K\,dH(\lambda)\,K^\dagger}{\Tr(KBK^\dagger)}
\]
is a parent for $\rho_K$ with the original response probabilities. Applying $K^{-1}$ recovers a parent for $\rho$. Thus invertible filtering on Bob preserves the existence of an LHS model.

Choosing $K=B^{-1/2}$ gives the whitened state
\begin{equation}
\widehat\rho =\frac1n(\openone_m\otimes B^{-1/2}) \rho (\openone_m\otimes B^{-1/2}), \qquad \Tr_A\widehat\rho =\frac1nB^{-1/2}BB^{-1/2} =\frac{\openone_n}{n}. \label{eq:whitened-state}
\end{equation}
Its trusted marginal is maximally mixed.

For each $M\in\mathfrak M$, decompose every effect uniquely into its scalar and traceless parts:
\begin{equation}
q_{a|M}=\frac{\Tr M_a}{m}, \qquad X_{a|M}=M_a-q_{a|M}\openone_m. \label{eq:effect-splitting}
\end{equation}
Let $\mathcal H_m^0$ be the real vector space of traceless Hermitian operators on Alice, and let $\mathcal H_n$ be the real vector space of Hermitian operators on Bob. Equations~\eqref{eq:finite-povm} and \eqref{eq:effect-splitting} give
\begin{equation}
q_{a|M}\geq0, \qquad X_{a|M}\in\mathcal H_m^0, \qquad \sum_{a=1}^{r_M}q_{a|M}=1, \qquad \sum_{a=1}^{r_M}X_{a|M}=0. \label{eq:centered-effect-properties}
\end{equation}
Real-linear maps from $\mathcal H_m^0$ to $\mathcal H_n$ describe this centered dependence on Alice's effects. They form the finite-dimensional real vector space
\[
\mathcal V_{m,n} =\operatorname{Lin}_{\mathbb R}(\mathcal H_m^0,\mathcal H_n)
\]
whose elements will be denoted by $A$.

A whitened state $\widehat\rho$ determines the particular map $A_{\widehat\rho}\in\mathcal V_{m,n}$ defined by
\begin{equation}
A_{\widehat\rho}(X) =n\Tr_A[(X\otimes\openone_n)\widehat\rho], \qquad X\in\mathcal H_m^0. \label{eq:whitened-map-A}
\end{equation}
For Hermitian $X$, cyclicity under the partial trace gives
\[
A_{\widehat\rho}(X)^\dagger =n\Tr_A[\widehat\rho(X\otimes\openone_n)] =A_{\widehat\rho}(X),
\]
so $A_{\widehat\rho}$ takes values in $\mathcal H_n$. For an orthonormal basis $E_1,\ldots,E_{m^2-1}$ of $\mathcal H_m^0$, the condition $\Tr_A\widehat\rho=\openone_n/n$ gives
\begin{equation}
\widehat\rho =\frac{\openone_m\otimes\openone_n}{mn} +\frac1n\sum_{j=1}^{m^2-1} E_j\otimes A_{\widehat\rho}(E_j). \label{eq:whitened-standard-form}
\end{equation}
Thus the map determines the whitened state, and its conditional preparations are
\begin{equation}
\Tr_A[(M_a\otimes\openone_n)\widehat\rho] =\frac1n\left[ q_{a|M}\openone_n +A_{\widehat\rho}(X_{a|M}) \right]. \label{eq:whitened-conditional-states}
\end{equation}
Equation~\eqref{eq:whitened-conditional-states} also defines a parent relation for arbitrary $A\in\mathcal V_{m,n}$. A parent model consists of a hidden space $\Lambda$, a positive operator measure $H$ and response probabilities obeying
\begin{equation}
H(\Lambda)=\openone_n, \qquad \int_\Lambda p(a|M,\lambda)\,dH(\lambda) =q_{a|M}\openone_n+A(X_{a|M}) \label{eq:parent-model}
\end{equation}
for every $M\in\mathfrak M$ and every outcome $a$. By Eq.~\eqref{eq:centered-effect-properties}, the operators on the right-hand side sum to $\openone_n$.

For the measure representation, let
\[
\mathcal D_n=\{\tau\in\mathcal H_n:\tau\succeq0,\ \Tr\tau=1\}.
\]
By the Radon--Nikodym representation above, a parent may be represented on $\mathcal D_n$. Thus $A$ admits a parent model if and only if a positive measure $\nu$ and response probabilities satisfy
\begin{equation}
\int_{\mathcal D_n}\tau\,d\nu(\tau)=\openone_n, \qquad \int_{\mathcal D_n}p(a|M,\tau)\tau\,d\nu(\tau) =q_{a|M}\openone_n+A(X_{a|M}) \label{eq:s2-response-equations}
\end{equation}
for every $M\in\mathfrak M$ and every outcome $a$. Equivalently, use positive outcome measures $\nu_{a|M}$ that sum to $\nu$, replacing $p(a|M,\tau)\,d\nu(\tau)$ by $d\nu_{a|M}(\tau)$ in the outcome integrals. Since $\nu_{a|M}\leq\nu$, the derivatives
\[
p(a|M,\tau)=\frac{d\nu_{a|M}}{d\nu}(\tau)
\]
recover response probabilities after redefinition on a null set, and
\[
dH(\tau)=\tau\,d\nu(\tau)
\]
is the corresponding normalized parent measure.
\begin{lemma}
\label{lem:parent-model-maps} The maps admitting a parent model form a closed convex subset of $\mathcal V_{m,n}$ containing the zero map. For every $A\in\mathcal V_{m,n}$, there exists $s_0>0$ such that $s_0A$ admits a parent model.
\end{lemma}
\begin{proof}
For the zero map, a single hidden state suffices:
\[
H(\Lambda)=\openone_n, \qquad p(a|M)=q_{a|M}.
\]
For $0\leq\theta\leq1$, combine parents for $A_1,A_2$ with weights $\theta,1-\theta$ on the disjoint union of their hidden spaces. The total remains $\openone_n$, and the outcome integrals combine linearly in Eq.~\eqref{eq:parent-model}, giving a parent for $\theta A_1+(1-\theta)A_2$.

Every map has a positive multiple admitting a parent. To construct one, choose positive definite operators $F_1,\ldots,F_{n^2}$ that form a real basis of $\mathcal H_n$ and satisfy
\[
\sum_{j=1}^{n^2}F_j=\openone_n.
\]
Such a basis is obtained by taking sufficiently small traceless perturbations of $\openone_n/n^2$ whose perturbations span the traceless Hermitian operators and sum to zero. Write
\[
A(X)=\sum_{j=1}^{n^2}\ell_j(X)F_j
\]
for real-linear functionals $\ell_j$ on $\mathcal H_m^0$. Since the density operators on $\C^m$ form a compact set, choose $s_0>0$ so small that
\[
s_0\left|\ell_j(m\eta-\openone_m)\right|\leq1
\]
for every density operator $\eta$ and every $j$. On the hidden space $\{1,\ldots,n^2\}$, set
\[
H(\{j\})=F_j, \qquad p(a|M,j)=q_{a|M}+s_0\ell_j(X_{a|M}).
\]
If $q_{a|M}>0$, write $M_a=mq_{a|M}\eta_{a|M}$ for a density operator $\eta_{a|M}$. Then
\[
X_{a|M}=q_{a|M}(m\eta_{a|M}-\openone_m), \qquad p(a|M,j)=q_{a|M}\left[1+s_0\ell_j(m\eta_{a|M}-\openone_m)\right]\geq0.
\]
If $q_{a|M}=0$, positivity gives $M_a=X_{a|M}=0$. The responses sum to one by Eq.~\eqref{eq:centered-effect-properties}, and substitution gives $\sum_jp(a|M,j)F_j=q_{a|M}\openone_n+s_0A(X_{a|M})$. Thus these effects and responses form a parent model for $s_0A$.

For closedness, let parent-model maps $A_k$ converge to $A$. Represent their models by measures $\nu_k$ as in Eq.~\eqref{eq:s2-response-equations}. Each has mass $n$, so compactness of $\mathcal D_n$ gives a weakly convergent subsequence with limit $\nu$ satisfying
\[
\openone_n=\int_{\mathcal D_n}\tau\,d\nu(\tau).
\]
For a fixed $M$, its finitely many outcome measures also have masses bounded by $n$. A further subsequence yields positive limits summing to $\nu$ and having operator moments $q_{a|M}\openone_n+A(X_{a|M})$. These divisions may be chosen separately for each measurement, while $\nu$ remains common to all. Their Radon--Nikodym derivatives therefore give a parent model for $A$.
\end{proof}
Define the steering scale by
\begin{equation}
c_{\mathfrak M}(A) =\inf\left\{ t>0:\frac{A}{t}\text{ admits a parent model for }\mathfrak M \right\}. \label{eq:steering-scale}
\end{equation}
Lemma~\ref{lem:parent-model-maps} shows that $c_{\mathfrak M}(A)$ is finite for every $A\in\mathcal V_{m,n}$. Since the parent-model maps form a convex set containing the zero map, if $A/t_0$ admits a parent model, then so does $A/t$ for every $t\geq t_0$. Thus every $t>c_{\mathfrak M}(A)$ may be used in Eq.~\eqref{eq:steering-scale}.

For every $s\geq0$,
\begin{equation}
c_{\mathfrak M}(sA)=s\,c_{\mathfrak M}(A). \label{eq:steering-scale-homogeneity}
\end{equation}
For $s>0$ this follows directly by changing variables in Eq.~\eqref{eq:steering-scale}, while $c_{\mathfrak M}(0)=0$ because the zero map admits a parent model.

Convexity follows by choosing $t_i>c_{\mathfrak M}(A_i)$ and combining the parents of $A_i/t_i$. With $t=\theta t_1+(1-\theta)t_2$,
\[
\frac{\theta A_1+(1-\theta)A_2}{t} = \frac{\theta t_1}{t}\frac{A_1}{t_1} +\frac{(1-\theta)t_2}{t}\frac{A_2}{t_2}, \qquad \frac{\theta t_1}{t}+\frac{(1-\theta)t_2}{t}=1.
\]
Lemma~\ref{lem:parent-model-maps} makes the left-hand map feasible, so $c_{\mathfrak M}(\theta A_1+(1-\theta)A_2)\le t$. Sending $t_i\downarrow c_{\mathfrak M}(A_i)$ gives
\begin{equation}
c_{\mathfrak M}\!\left(\theta A_1+(1-\theta)A_2\right) \leq \theta c_{\mathfrak M}(A_1) +(1-\theta)c_{\mathfrak M}(A_2) \qquad(0\leq\theta\leq1). \label{eq:steering-scale-convexity}
\end{equation}
Thus $c_{\mathfrak M}$ is a finite convex positively homogeneous function on $\mathcal V_{m,n}$.

Parent-model maps are exactly the unit sublevel set:
\begin{equation}
A\text{ admits a parent model for }\mathfrak M \quad\Longleftrightarrow\quad c_{\mathfrak M}(A)\leq1. \label{eq:parent-model-scale}
\end{equation}
A parent gives this inequality by taking $t=1$. Conversely, if $c_{\mathfrak M}(A)\le1$, every $t>1$ admits a parent for $A/t$. Letting $t\downarrow1$ and using closedness gives a parent for $A$.

Whitening leaves only unitary freedom on Bob. Let $\rho_1,\rho_2$ have marginals $B_1,B_2\succ0$ and be related by an invertible filter $K$ on Bob. Taking the marginal gives
\[
B_2=\frac{KB_1K^\dagger}{\Tr(KB_1K^\dagger)}.
\]
Hence
\[
V=\frac{B_2^{-1/2}KB_1^{1/2}}{\sqrt{\Tr(KB_1K^\dagger)}}, \qquad VV^\dagger =\frac{B_2^{-1/2}KB_1K^\dagger B_2^{-1/2}} {\Tr(KB_1K^\dagger)} =\openone_n,
\]
so $V$ is unitary. Substituting this relation into the two whitening formulas gives
\begin{equation}
\widehat\rho_2 =(\openone_m\otimes V)\widehat\rho_1 (\openone_m\otimes V^\dagger). \label{eq:whitened-states-unitary}
\end{equation}
Unitary freedom leaves the scale unchanged. Under Eq.~\eqref{eq:whitened-states-unitary},
\[
A_{\widehat\rho_2}(X)=VA_{\widehat\rho_1}(X)V^\dagger.
\]
Conjugating a parent measure by $V$ preserves its total $\openone_n$ and gives a parent model for the conjugated map. Applying the same argument with $V^\dagger$ gives
\[
c_{\mathfrak M}(A_{\widehat\rho_2})=c_{\mathfrak M}(A_{\widehat\rho_1}).
\]
\begin{theorem}
\label{thm:lhs-scale-condition} Let $\widehat\rho$ be the whitened state associated with $\rho$. Then $\rho$ admits an LHS model for the measurements in $\mathfrak M$ if and only if
\[
c_{\mathfrak M}(A_{\widehat\rho})\leq1.
\]
\end{theorem}
\begin{proof}
Invertible filtering preserves the existence of an LHS model, so it is enough to consider $\widehat\rho$. By Eq.~\eqref{eq:whitened-conditional-states}, multiplying a common parent measure for $\widehat\rho$ by $n$ gives Eq.~\eqref{eq:parent-model} for $A_{\widehat\rho}$. Conversely, dividing such a parent by $n$ gives a common parent measure for $\widehat\rho$. The result follows from Eq.~\eqref{eq:parent-model-scale}.
\end{proof}
\subsection{Parent measurements and dual bounds} \label{sec:parent-multiplier}

Assume now that $\mathfrak M$ is compact and its POVMs have a common finite number $r$ of labeled outcomes, allowing zero effects. Scalar measures in this subsection are finite regular Borel measures. Outcome scores give both a parent formula and a dual multiplier formula for the scale.

Test the model equations with Hermitian operators $Y_1,\ldots,Y_r$. A parent must supply every score; positive combinations of these demands give lower bounds through an operator multiplier $\Omega$. By Eq.~\eqref{eq:centered-effect-properties}, adding one Hermitian operator to all $Y_a$ changes neither centered target nor kernel. Subtracting their common mean and fixing the overall scale imposes
\begin{equation}
\sum_{a=1}^rY_a=0, \qquad \sum_{a=1}^r\Tr(Y_a^2)=1. \label{eq:s2-test-normalization}
\end{equation}
Let $\mathcal Z_{\mathfrak M}$ be the compact space of pairs $z=(M,Y)$, where $M\in\mathfrak M$ and $Y=(Y_1,\ldots,Y_r)$ satisfies Eq.~\eqref{eq:s2-test-normalization}. For $z=(M,Y)$ and $\tau\in\mathcal D_n$, define
\begin{align}
N_A(z) &=\sum_{a=1}^r\Tr\!\left[Y_aA(X_{a|M})\right], \label{eq:s2-target}\\
k(z;\tau) &=\max_{1\leq a\leq r}\Tr(Y_a\tau) -\sum_{a=1}^rq_{a|M}\Tr(Y_a\tau). \label{eq:s2-kernel}
\end{align}
Here $N_A$ is the centered test value; $k$ replaces the response average by its maximum and subtracts the scalar contribution. Both functions are continuous, and $k\geq0$ because $(q_{a|M})_{a=1}^r$ is a probability vector. For positive measures $\nu$ on $\mathcal D_n$ and $\lambda$ on $\mathcal Z_{\mathfrak M}$, define
\begin{equation}
(\mathcal K\nu)(z) =\int_{\mathcal D_n}k(z;\tau)\,d\nu(\tau), \qquad (\mathcal K^*\lambda)(\tau) =\int_{\mathcal Z_{\mathfrak M}}k(z;\tau)\,d\lambda(z), \label{eq:s2-K-pair}
\end{equation}
and, for $\Omega\succeq0$, define
\begin{equation}
w_\Omega(\tau)=\Tr(\Omega\tau). \label{eq:s2-weight}
\end{equation}
\begin{theorem}
\label{thm:s2-three-way} For every $A\in\mathcal V_{m,n}$, the following three conditions are equivalent:
\begin{align*}
\text{\rm(i)}\quad& A\text{ admits a parent model for }\mathfrak M;\\
\text{\rm(ii)}\quad& \exists\,\nu\geq0:\quad \int_{\mathcal D_n}\tau\,d\nu(\tau)=\openone_n,\qquad \mathcal K\nu\geq N_A;\\
\text{\rm(iii)}\quad& \forall\,\Omega\succeq0,\ \lambda\geq0\text{ satisfying }\mathcal K^*\lambda\leq w_\Omega:\quad \int_{\mathcal Z_{\mathfrak M}}N_A\,d\lambda\leq\Tr\Omega.
\end{align*}
\end{theorem}
\begin{proof}
Fix a positive measure $\nu$ with operator moment $\openone_n$ and a measurement $M\in\mathfrak M$. Let $\mathcal C_{M,\nu}$ be the set of operator families $(\int\tau\,d\nu_a)_{a=1}^r$ obtained from positive measures with $\sum_a\nu_a=\nu$. It is convex and compact: every outcome measure has mass at most $n$, so weak compactness preserves the sum and all operator moments. Its support function at $Y$ is $\int_{\mathcal D_n}\max_a\Tr(Y_a\tau)\,d\nu(\tau)$.
Assigning each $\tau$ to a maximizing outcome attains this value; resolving ties by the smallest label makes the assignment measurable. By convex separation, the target family $B_a=q_{a|M}\openone_n+A(X_{a|M})$ belongs to $\mathcal C_{M,\nu}$ exactly when $\sum_a\Tr(Y_aB_a)\le\int\max_a\Tr(Y_a\tau)\,d\nu(\tau)$ for all Hermitian $Y_a$. Both the target and every family in $\mathcal C_{M,\nu}$ sum to $\openone_n$, so a common shift of the tests cancels. After centering, every nonzero test can be normalized, while the zero test gives equality. The inequalities are therefore exactly $N_A\le\mathcal K\nu$. Membership gives the outcome measures, whose Radon--Nikodym derivatives with respect to $\nu$ are the required response probabilities. Applying this argument to every $M$ with the same $\nu$ proves (i)$\Longleftrightarrow$(ii).

To obtain (iii) from (ii), integrate the source inequality against $\lambda\ge0$. Interchanging the nonnegative kernel integrals gives
\[
\int N_A\,d\lambda \le\int \mathcal K\nu\,d\lambda =\int\mathcal K^*\lambda\,d\nu \le\int\Tr(\Omega\tau)\,d\nu(\tau) =\Tr\Omega.
\]
Conversely, suppose that (iii) holds. Consider the convex cone
\begin{equation}
\mathfrak C= \left\{ \left( \mathcal K\nu-g,\, \int_{\mathcal D_n}\tau\,d\nu(\tau) \right): \nu\geq0,\quad g\in C(\mathcal Z_{\mathfrak M})_+ \right\} \subset C(\mathcal Z_{\mathfrak M})\times\mathcal H_n. \label{eq:s2-parent-cone}
\end{equation}
To establish closedness, suppose
\[
\left( \mathcal K\nu_j-g_j,\, \int_{\mathcal D_n}\tau\,d\nu_j(\tau) \right) \longrightarrow(f,R).
\]
Tracing the second coordinate bounds the masses of $\nu_j$. Compactness as in Lemma~\ref{lem:parent-model-maps} yields a weak limit $\nu$ of a subsequence, with operator moment $R$. Joint continuity of $k$ makes $\mathcal K\nu_j\to\mathcal K\nu$ uniform, so
\[
g_j\longrightarrow\mathcal K\nu-f\geq0,
\]
and $(f,R)\in\mathfrak C$.

Condition (ii) is equivalent to
\[
(N_A,\openone_n)\in\mathfrak C.
\]
If this fails, separation from the closed convex cone gives an operator $\Omega\in\mathcal H_n$ and a signed measure $\lambda$ on $\mathcal Z_{\mathfrak M}$ such that
\begin{equation}
\Tr(\Omega R)-\int_{\mathcal Z_{\mathfrak M}}f\,d\lambda\geq0 \qquad((f,R)\in\mathfrak C), \label{eq:s2-cone-separation}
\end{equation}
while
\begin{equation}
\int_{\mathcal Z_{\mathfrak M}}N_A\,d\lambda>\Tr\Omega. \label{eq:s2-strict-separation}
\end{equation}
Taking $\nu=0$ and varying $g\geq0$ in Eq.~\eqref{eq:s2-cone-separation} gives $\lambda\geq0$. Taking $g=0$ and varying $\nu\geq0$ gives
\[
\mathcal K^*\lambda(\tau)\leq\Tr(\Omega\tau)=w_\Omega(\tau) \qquad(\tau\in\mathcal D_n).
\]
Since $\mathcal K^*\lambda\geq0$, this inequality for every $\tau\in\mathcal D_n$ gives $\Omega\succeq0$. Thus $\Omega$ and $\lambda$ satisfy the constraint in (iii), whereas Eq.~\eqref{eq:s2-strict-separation} contradicts its conclusion. Hence (ii) holds, and the first equivalence gives (i).
\end{proof}
Applying Theorem~\ref{thm:s2-three-way} to $A/t$ gives the two representations of the steering scale:
\begin{align}
c_{\mathfrak M}(A) &=\inf\left\{ t>0: \exists\,\nu\geq0:\quad \int_{\mathcal D_n}\tau\,d\nu(\tau)=\openone_n,\qquad \mathcal K\nu\geq N_{A/t} \right\} \label{eq:s2-parent-problem}\\
&=\sup\left\{ \int_{\mathcal Z_{\mathfrak M}}N_A\,d\lambda: \exists\,\Omega\in\mathcal D_n,\ \lambda\geq0:\quad \mathcal K^*\lambda\leq w_\Omega \right\}. \label{eq:s2-multiplier-problem}
\end{align}
The equivalence (i)$\Longleftrightarrow$(ii) gives the parent formula. For multipliers, (iii) reads $\int N_A\,d\lambda\le t\Tr\Omega$. When $\Tr\Omega>0$, divide both multipliers by that trace. A zero-trace pair has nonpositive objective, since (iii) applies to the feasible map $s_0A$ from Lemma~\ref{lem:parent-model-maps}. Infimizing over $t>0$ therefore gives the normalized supremum in Eq.~\eqref{eq:s2-multiplier-problem}.

Let $A\in\mathcal V_{m,n}$, and fix the measurement family. For the normalized test $z=(M,Y)$ of Section~S1, write $N_A(z)$ for its source and $k(z;\tau)$ for the centered kernel. For a trial scale $t$, multiply the parent measure for $A/t$ by $t$. The resulting measure $\nu$ has operator moment $\int\tau\,d\nu=t\openone_n$ and satisfies $\int k(z;\tau)d\nu\ge N_A(z)$. If $\tau=\sum_i p_i\tau_{\psi_i}$, where $\tau_\psi=\ketbra\psi$, convexity gives
\[
k(z;\tau)\le\sum_i p_i k(z;\tau_{\psi_i}).
\]
Resolving every mixed hidden state into pure states therefore preserves the operator moment and all source inequalities. The resulting measure has mass $nt$. Set $\mu=\nu/n$ and $\kappa(z,\psi)=nk(z;\tau_\psi)$ to obtain
\begin{equation}
\mu\ge0,\qquad \mu(\C P^{n-1})=t,\qquad \int\tau_\psi\,d\mu=\frac{t}{n}\openone_n,\qquad (\mathcal K\mu)(z):=\int\kappa(z,\psi)d\mu(\psi)\ge N_A(z). \label{eq:general-support}
\end{equation}
In Eq.~\eqref{eq:general-support}, the state fixes the required score $N_A$, while a candidate measure supplies the maximal score $\mathcal K\mu$ through its outcome divisions. Domination for every test enforces all conditional preparations. The operator moment fixes their shared trusted marginal. Thus one distribution on pure-state space carries the whole parent problem, and minimizing its mass gives $c(A)$.

Choose an orthonormal basis $T_1,\ldots,T_{n^2-1}$ of traceless Hermitian matrices and put $v_j(\psi)=\sqrt{n/(n-1)}\Tr(T_j\tau_\psi)$ for $n\ge2$. These coordinates have unit length for pure states. At $n=1$ the moment vector is empty. The trace is already fixed by the mass, so the remaining moment condition is $\int\vec v\,d\mu=0$.

Write $\mathfrak X_{m,r}$ for the compact family of $r$-outcome POVMs on Alice, with zero effects allowed. The family of all finite-outcome POVMs has scale
\begin{equation}
c_{\mathrm A}(A)=c_{\mathfrak X_{m,m^2}}(A). \label{eq:s4-all-povm-reduction}
\end{equation}
Every finite POVM is a finite convex combination of extreme POVMs, each with at most $m^2$ nonzero effects~\cite{DArianoExtremePOVM}. A common parent for $\mathfrak X_{m,m^2}$ therefore models every finite POVM by relabeling and mixing its responses. The reverse inclusion is measurement restriction.
\section{S2. Wave propagation and ensemble reconstruction}\label{sec:general-wave}

The response retains the distribution whose divisions are constrained in Section~S1. We derive its wave equation from Born overlap, recover the measure from response data, and characterize positivity through threshold convexity. Finite propagation gives the part of the ensemble determined by local data. With this inverse in hand, saturation of an averaged preparation demand fixes a candidate ensemble through contact. We then apply the same response to a measurement: its decision profile determines the pure effects, while mixing outcome records generates a wave source whose integrated strength measures their lost discrimination advantage.

\subsection{Propagation and positive ensembles}

Let $n\ge2$, and probe a pure-state measure with rank-one projectors $Q$. For $0<x<1$, define
\begin{equation}
F_\mu(x,Q)=\int|x-\Tr(QP)|d\mu(P).
\label{eq:projective-response}
\end{equation}
When $\mu$ is a probability measure of operator mean $\openone_n/n$, the effects $G(dP)=nP\,d\mu(P)$ form a measurement. Use it to distinguish $\openone_n/n$, prepared with prior $p=nx/(1+nx)$, from the pure input $Q$, prepared with prior $1-p$. The optimal assignment of each outcome gives
\begin{equation}
p_{\rm succ}(x,Q;G)=\frac12+\frac{n}{2(1+nx)}F_\mu(x,Q).
\label{eq:biased-discrimination-response}
\end{equation}
The difference of the prior-weighted outcome measures is $n[x-\Tr(QP)]d\mu(P)/(1+nx)$; its total variation gives twice the success advantage over one half. Thus changing the prior reads the same measure through different overlap thresholds.

Born overlap $q=\Tr(QP)$ is the probability of obtaining $Q$ from $P$. Changing $x$ moves the division between larger and smaller overlaps. We normalize the Fubini--Study Laplacian $\Delta_{\rm FS}$ so that $\cos t\ket\chi+\sin t\ket\xi$ has unit speed for orthogonal unit vectors.

\begin{theorem}\label{thm:projective-wave-inversion}
For any finite signed measure $\mu$ on $\mathbb CP^{n-1}$, its response satisfies
\begin{equation}
x(1-x)\partial_x^2F_\mu-(1-nx)\partial_xF_\mu-nF_\mu-\frac14\Delta_{\rm FS}F_\mu=0
\label{eq:projective-wave-equation}
\end{equation}
in distributions on $(0,1)\times\mathbb CP^{n-1}$. At every fixed $x_0\in(0,1)$, the value $F_\mu(x_0,Q)$ and derivative $\partial_xF_\mu(x_0,Q)$ for almost every $Q$ determine $\mu$ uniquely.
\end{theorem}
\begin{proof}
At $Q=\ketbra\chi$, differentiate $q=\langle\chi|P|\chi\rangle$ along the real and imaginary directions of an orthonormal basis perpendicular to $\ket\chi$. Summing their squared first derivatives and their second derivatives gives
\begin{equation}
|\nabla q|^2=4(\langle\chi|P^2|\chi\rangle-q^2)=4q(1-q),\qquad \Delta_{\rm FS}q=4(\Tr P-nq)=4(1-nq).
\label{eq:overlap-laplacian}
\end{equation}
Thus $\Delta_{\rm FS}f(q)=4[q(1-q)f''(q)+(1-nq)f'(q)]$. For $f(q)=|x-q|$, the second derivative is $2\delta(x-q)$. Moreover, $\int q\operatorname{sgn}(x-q)d\mu=x\partial_xF_\mu-F_\mu$. Substitution yields Eq.~\eqref{eq:projective-wave-equation}. Smooth approximation justifies the calculation for finite measures. In particular,
\begin{equation}
\partial_x^2F_\mu(x,Q)=2\int\delta(x-\Tr(QP))d\mu(P)
\label{eq:overlap-shell-response}
\end{equation}
measures the weight on a surface of fixed Born overlap when $\mu$ is positive.

For inversion, use real orthonormal harmonics $\varphi_{\ell j}$ on $\mathbb CP^{n-1}$ with eigenvalues $-4\ell(\ell+n-1)$. The invariant addition formula has normalized zonal polynomial~\cite{Koornwinder1973}
\begin{equation}
\pi_\ell(q)=\frac{P_\ell^{(n-2,0)}(2q-1)}{P_\ell^{(n-2,0)}(1)},\qquad \pi_\ell(1)=1.
\label{eq:projective-zonal-polynomial}
\end{equation}
Here $P_\ell^{(\alpha,\beta)}$ denotes a Jacobi polynomial, with $P_\ell^{(n-2,0)}(1)=\binom{\ell+n-2}{\ell}$. Haar-distributed $Q$ gives overlap density $(n-1)(1-q)^{n-2}$ on $[0,1]$, so convolution with the response kernel multiplies degree $\ell$ by
\begin{equation}
a_\ell(x)=(n-1)\int_0^1|x-q|(1-q)^{n-2}\pi_\ell(q)dq.
\label{eq:projective-multiplier-integral}
\end{equation}
Integration gives
\begin{equation}
\begin{aligned}
a_0(x)&=x-\frac1n+\frac2n(1-x)^n,\\
a_1(x)&=\frac{2(1-x)^n(1+nx)-1}{n(n+1)},\\
a_\ell(x)&=\frac{2(n-1)x^2(1-x)^nP_{\ell-2}^{(n,2)}(2x-1)}{\ell(\ell-1)P_\ell^{(n-2,0)}(1)}\qquad(\ell\ge2).
\end{aligned}
\label{eq:projective-multipliers}
\end{equation}
For the last formula, the constant and linear moments of $\pi_\ell$ vanish. Write the absolute-value integral as $2\int_0^x(x-q)(n-1)(1-q)^{n-2}\pi_\ell(q)dq$, then apply the Jacobi Rodrigues formula twice. The boundary factors $q^\ell(1-q)^{\ell+n-2}$ remove the integration terms at zero.

Projecting Eq.~\eqref{eq:projective-wave-equation} onto degree $\ell$ gives the ordinary differential equation
\begin{equation}
x(1-x)a_\ell''-(1-nx)a_\ell'+[\ell(\ell+n-1)-n]a_\ell=0.
\label{eq:projective-multiplier-ode}
\end{equation}
No $a_\ell$ is identically zero, since its second derivative is $2(n-1)(1-x)^{n-2}\pi_\ell(x)$. At an interior point, $a_\ell$ and $a_\ell'$ cannot both vanish: uniqueness for this regular second-order equation would make the solution zero everywhere. Write $A_{\ell j}$ and $B_{\ell j}$ for the harmonic coefficients of the value and derivative at $x_0$. The moments of the unknown measure are therefore
\begin{equation}
\int\varphi_{\ell j}(P)d\mu(P)=\frac{a_\ell(x_0)A_{\ell j}+a_\ell'(x_0)B_{\ell j}}{a_\ell(x_0)^2+a_\ell'(x_0)^2}.
\label{eq:projective-cauchy-inverse}
\end{equation}
For each fixed $P$, the overlap density assigns zero Haar measure to $\Tr(QP)=x_0$. Fubini applied to $|\mu|$ therefore gives the derivative for almost every $Q$. Since harmonic polynomials are dense in continuous functions on $\mathbb CP^{n-1}$, their recovered moments determine $\mu$. Multiplying degree $\ell$ by $e^{-4\epsilon\ell(\ell+n-1)}$ produces a heat-smoothed density converging weakly to the measure as $\epsilon\downarrow0$.
\end{proof}

With $x(1-x)>0$ in the interior, Eq.~\eqref{eq:projective-wave-equation} is hyperbolic and $x$ is its propagation coordinate. Fixed-overlap surfaces carry the fronts. For qubits, $\Tr(QP)=(1+\vec q\cdot\vec p)/2$ makes $x=1/2$ the section of even cosine and odd hemisphere data. Jacobi multipliers give the corresponding inverse on $\mathbb CP^{n-1}$ in higher dimensions.

Reconstruction also follows locally from curvature. Let $\mathcal S_x(P,dQ)$ be the invariant probability measure on the surface $\Tr(QP)=x$, and average it against the spectrum: $\mu_x(dQ)=\int\mathcal S_x(P,dQ)d\mu(P)$. Using the overlap density in Eq.~\eqref{eq:overlap-shell-response} gives
\begin{equation}
\mu_x(dQ)=\frac{\partial_x^2F_\mu(x,Q)}{2(n-1)(1-x)^{n-2}}d\omega(Q),\qquad \mu_x\rightharpoonup\mu\quad(x\uparrow1).
\label{eq:projective-curvature-recovery}
\end{equation}
For singular ensembles the first equality is an identity of measures. A continuous test function averaged over $\mathcal S_x(P,dQ)$ tends uniformly to its value at $P$, because the support radius $\arccos\sqrt x$ tends to zero. Integration against finite $|\mu|$ proves weak convergence. For continuous densities, the normalized curvature converges uniformly to $u(Q)$.

Writing $S(x,Q)=\partial_x^2F_\mu/[2(n-1)(1-x)^{n-2}]$, differentiation of Eq.~\eqref{eq:projective-wave-equation} twice gives the spherical-mean equation
\begin{equation}
x(1-x)\partial_x^2S+(1-nx)\partial_xS=\frac14\Delta_{\rm FS}S.
\label{eq:projective-shell-wave}
\end{equation}
For a density, $S(x,Q)$ is its average over the fixed-overlap surface around $Q$. The radius $r=\arccos\sqrt x$ rewrites this as $\partial_r^2S+[(2n-3)\cot r-\tan r]\partial_rS=\Delta_{\rm FS}S$. Thus the propagation from an interior section can be followed to shrinking surfaces at $x=1$, where their normalized weight recovers the ensemble.

Convexity and endpoint values distinguish the waves generated by positive ensembles.

\begin{theorem}[Positive ensembles and convex wave responses]\label{thm:positive-wave-response}
Let $n\ge2$ and $t\ge0$. A continuous real function $F$ on $[0,1]\times\mathbb CP^{n-1}$ has the representation $F=F_\mu$ for a positive measure with $\int P\,d\mu=t\openone_n/n$ if and only if it satisfies Eq.~\eqref{eq:projective-wave-equation} in distributions, is convex in $x$ for every $Q$, and obeys
\begin{equation}
F(0,Q)=\frac tn,\qquad F(1,Q)=\frac{n-1}{n}t.
\label{eq:wave-endpoint-data}
\end{equation}
The measure is unique and is recovered by Eq.~\eqref{eq:projective-curvature-recovery}.
\end{theorem}
\begin{proof}
A positive measure gives a convex absolute-value response. Its mass and operator mean fix the endpoints, and the preceding theorem supplies the wave equation.

For the converse, project $F$ onto the harmonics $\varphi_{\ell j}$. Each coefficient is continuous at the endpoints and solves Eq.~\eqref{eq:projective-multiplier-ode} in the interior; distributional solutions of this ordinary differential equation are smooth there. At degree zero, the independent solutions $a_0(x)$ and $x-1/n$ show that the angular average is $t a_0(x)$. At degree one, the independent solutions $a_1(x)$ and $1$ and the two zero endpoint values force every coefficient to vanish.

For $\ell\ge2$, Eq.~\eqref{eq:projective-multipliers} gives $a_\ell(x)=c x^2+O(x^3)$ at zero with $c\ne0$. The Wronskian of two solutions is proportional to $x(1-x)^{n-1}$. Reduction of order therefore gives a second solution $a_\ell(x)\int^x y(1-y)^{n-1}/a_\ell(y)^2\,dy$, which tends to a nonzero constant at zero. The zero boundary value consequently makes the coefficient of $F$ equal to $c_{\ell j}a_\ell(x)$. Put $c_{00}=t$ and $c_{1j}=0$.

Convexity makes $\partial_x^2F$ a positive measure on the product space. Its angular average is $t a_0''(x)=2t(n-1)(1-x)^{n-2}$, so disintegration defines positive measures $\mu_x$ of mass $t$ for almost every $x$, with
\begin{equation}
\partial_x^2F\,dx\,d\omega(Q)=2(n-1)(1-x)^{n-2}\,dx\,d\mu_x(Q),\qquad \int\varphi_{\ell j}\,d\mu_x=c_{\ell j}\pi_\ell(x).
\label{eq:positive-wave-disintegration}
\end{equation}
Here the distributional second derivative is a measure on the product space. Along any admissible sequence $x\to1$, compactness supplies weak limits of $\mu_x$. Each limit has harmonic moments $c_{\ell j}$ because $\pi_\ell(1)=1$, and density of harmonics identifies a unique measure $\mu$. Its mass is $t$; zero degree-one moments give $\int P\,d\mu=t\openone_n/n$. Matching all coefficients $c_{\ell j}a_\ell(x)$ and using continuity yields $F_\mu=F$ everywhere.
\end{proof}

Threshold convexity thus represents spectral positivity, and endpoint data represent the operator mean. Weak curvature recovery also evaluates every continuous outcome kernel in Eq.~\eqref{eq:general-support}:
\begin{equation}
(\mathcal K\mu)(z)=\lim_{x\uparrow1}\int\kappa(z,Q)\,d\mu_x(Q)=\lim_{x\uparrow1}\int\kappa(z,Q)\frac{\partial_x^2F(x,Q)}{2(n-1)(1-x)^{n-2}}d\omega(Q).
\label{eq:wave-outcome-recovery}
\end{equation}
For singular spectra, use the curvature measures of Eq.~\eqref{eq:projective-curvature-recovery}. The scale is the least $t$ admitting a convex wave solution with endpoints \eqref{eq:wave-endpoint-data} whose limit in Eq.~\eqref{eq:wave-outcome-recovery} dominates $N_A(z)$ at every test. This follows directly from Theorem~\ref{thm:positive-wave-response} and the response criterion. The nonnegative, zero-moment density iterates in Section~\hyperref[sec:fourier-spectra]{S3} generate convex waves with angular average $t a_0(x)$ and zero degree-one part. Their response deficits measure which preparation inequalities remain unmet.

At most thresholds, the response value alone contains every angular component. Define $\mathcal E_n$ as the union of the zeros in $(0,1)$ of the nonzero polynomials $a_\ell$; it is countable. For $x_0\notin\mathcal E_n$, dividing the value coefficients by $a_\ell(x_0)$ recovers any finite signed measure. At exceptional thresholds, Eq.~\eqref{eq:projective-cauchy-inverse} uses both data. Independently prescribed value and derivative coefficients must obey $a_\ell B_{\ell j}=a_\ell' A_{\ell j}$; the propagated field represents a positive ensemble precisely when it also has the convexity and endpoints of Theorem~\ref{thm:positive-wave-response}.

The reconstruction also extends to thresholds that vary with the test direction.
\begin{theorem}[Reconstruction from a spacelike section]\label{thm:curved-cauchy}
Let $f:\mathbb CP^{n-1}\to(0,1)$ be smooth and satisfy
\begin{equation}
|\nabla_{\rm FS}f|^2<4f(1-f)
\label{eq:spacelike-contact-condition}
\end{equation}
everywhere. The values $F_\mu(f(Q),Q)$ and $\partial_xF_\mu(f(Q),Q)$ for almost every $Q$ determine any finite signed measure $\mu$ uniquely.
\end{theorem}
\begin{proof}
Set $\tau=\arcsin(2x-1)$. Equation~\eqref{eq:projective-wave-equation} becomes
\begin{equation}
\partial_\tau^2F-\frac14\Delta_{\rm FS}F+b_n(\tau)\partial_\tau F-nF=0,\qquad b_n(\tau)=\frac{(n+1)x-3/2}{\sqrt{x(1-x)}}.
\label{eq:projective-wave-time}
\end{equation}
The principal metric is $d\tau^2-4g_{\rm FS}$. On the graph $\tau=h(Q)=\arcsin(2f(Q)-1)$, the bound $|\nabla h|<2$ is uniform by compactness. Future causal curves move at most at speed $1/2$ in $g_{\rm FS}$, so $\tau-h(Q)$ increases strictly. Across an interior time slab containing the graph it changes sign exactly once, making the graph a spacelike Cauchy surface. All coefficients in Eq.~\eqref{eq:projective-wave-time} are smooth on this slab.

Responses of finite measures are locally Lipschitz. Their threshold derivatives exist on the graph for almost every $Q$: at a zero of $f(Q)-\Tr(QP)$, Eq.~\eqref{eq:overlap-laplacian} and the strict spacelike inequality make the gradient nonzero. Each such zero set has Haar measure zero, and Fubini applies to $|\mu|$. For two responses sharing the data, their difference and transverse derivative vanish on the graph. Its tangential derivative also vanishes, giving zero Cauchy data. Finite-energy uniqueness for the smooth wave equation~\cite{BaerWafo2015} makes the responses identical on the slab. Enlarging the slab covers $0<x<1$, after which Theorem~\ref{thm:projective-wave-inversion} identifies their measures.
\end{proof}

For qubits, write $Q=(\openone_2+\vec e\cdot\vec\sigma)/2$. The two responses obey $F(x,Q)=K(2x-1,-\vec e)/2$. The contact plane $s=\vec q\cdot\vec y$ therefore becomes $f(Q)=(1-\vec q\cdot\vec e)/2$. Since the qubit Fubini--Study sphere has radius $1/2$,
\begin{equation}
4f(1-f)-|\nabla_{\rm FS}f|^2=1-|\vec q|^2.
\label{eq:qubit-contact-spacelike}
\end{equation}
Thus $|\vec q|<1$ in the qubit construction is precisely the spacelike condition. Fixed thresholds and the direction-dependent sections of biased qubits belong to the same reconstruction theorem.

Propagation also determines the part of an ensemble that can be recovered from a restricted set of test directions. Write $d_{\rm FS}$ for Fubini--Study distance, and for an open set $U$ define $U_r=\{Q:d_{\rm FS}(Q,U^c)>r\}$, with distance to the empty set taken as infinity.
\begin{corollary}[Local reconstruction]\label{cor:local-wave-reconstruction}
Let $n\ge2$, $x_0\in(0,1)$ and $U\subset\mathbb CP^{n-1}$ be open. Suppose two finite signed measures $\mu,\nu$ have equal response values and threshold derivatives at $x_0$ for almost every $Q\in U$. Put $\tau=\arcsin(2x-1)$ and $\tau_0=\arcsin(2x_0-1)$. Their responses agree on
\begin{equation}
\left\{(x,Q):0<x<1,\ Q\in U_{|\tau-\tau_0|/2}\right\}.
\label{eq:local-wave-domain}
\end{equation}
The measures themselves agree on $U_{r_0}$, where $r_0=\arccos\sqrt{x_0}$. In particular, data on a ball $B(Q_0,R)$ with $R>r_0$ determine the measure on $B(Q_0,R-r_0)$.
\end{corollary}
\begin{proof}
The difference $F_\mu-F_\nu$ is a locally Lipschitz finite-energy solution of Eq.~\eqref{eq:projective-wave-time}. Its Cauchy data vanish on $U$. On every closed interior time slab, the coefficients are smooth and the principal metric is $d\tau^2-4g_{\rm FS}$. Finite propagation~\cite{BaerWafo2015} confines the support arising from the complementary data to points at Fubini--Study distance at most $|\tau-\tau_0|/2$ from $U^c$. This proves Eq.~\eqref{eq:local-wave-domain} by exhausting the open time interval with such slabs.

Let $\varphi$ be a continuous test function with compact support in $U_{r_0}$. Its support has distance strictly greater than $r_0$ from $U^c$, and
\begin{equation}
\lim_{x\uparrow1}\frac{\tau-\tau_0}{2}=\frac{\pi/2-\tau_0}{2}=\arccos\sqrt{x_0}=r_0.
\label{eq:local-reconstruction-radius}
\end{equation}
For all $x$ sufficiently close to one, the response difference vanishes on a neighbourhood of this support. Its second threshold derivative therefore vanishes there as a distribution. Equation~\eqref{eq:projective-curvature-recovery} makes the corresponding curvature measures equal on this neighbourhood for almost every such $x$. Along a sequence approaching one, weak convergence gives $\int\varphi\,d(\mu-\nu)=0$. Continuous compactly supported tests determine the restriction of a finite measure, proving equality on $U_{r_0}$. Finally, the triangle inequality gives $B(Q_0,R-r_0)\subset U_{r_0}$ when $U=B(Q_0,R)$.
\end{proof}

For a normalized rank-one measurement, Eq.~\eqref{eq:biased-discrimination-response} converts success probabilities and their prior derivatives to these local Cauchy data. At $x_0=3/4$, the prior is $3n/(4+3n)$ and $r_0=\pi/6$. Input directions in a Fubini--Study ball of radius $\pi/3$ then determine all pure-effect weight within its concentric ball of radius $\pi/6$.

A qutrit calculation illustrates the two data with a nonuniform angular density. Fix $P_0$ and write $z=\Tr(P_0P)$. Relative to Haar measure, take
\begin{equation}
u(P)=1+0.45\pi_2(z)-0.30\pi_3(z)+0.15\pi_4(z).
\label{eq:qutrit-reconstruction-density}
\end{equation}
By the addition formula and Cauchy--Schwarz, $|\pi_\ell(z)|\le1$, so $u\ge0.1$. Orthogonality gives mass one and operator mean $\openone_3/3$. At $x_0=3/7$, $a_3=x^2(1-x)^3(7x-3)/6$ vanishes but $a_3'=96/2401$. Only the transverse derivative retains the degree-three component on this section. More generally, replacing $-0.30$ by any $c_3\in[-0.30,0.30]$ preserves the bound $u\ge0.1$, mass one and operator mean $\openone_3/3$. All these ensembles have the same response at $x_0$, while their derivatives distinguish $c_3$ (main-text Fig.~1c). Main-text Fig.~1b displays the local domain for the same ensemble: data at $r_0=\pi/6$ on the ball $\theta<\pi/3$ determine the response for $\theta<\pi/6+r$ as $r$ decreases to zero.

\begin{figure}[H]
\centering\includegraphics[width=.92\textwidth]{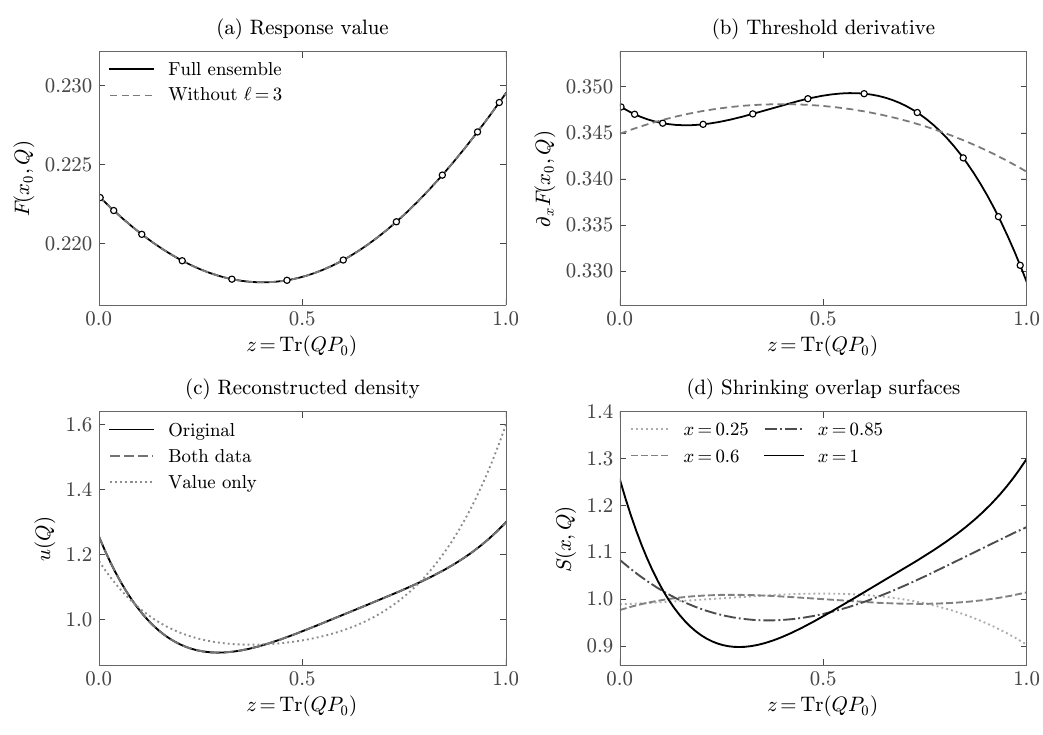}
\caption{Response propagation and reconstruction of the qutrit ensemble in Eq.~\eqref{eq:qutrit-reconstruction-density}, with $z=\Tr(QP_0)$ and $x_0=3/7$. Panels (a,b) compare the full ensemble (solid) with the ensemble whose degree-three coefficient is set to zero (dashed). Their response values coincide because $a_3(x_0)=0$, while their threshold derivatives resolve the missing component. Open circles are independent pure-state quadrature. Panel (c) shows the original density, its reconstruction from both quadrature data, and the alternative density obtained by setting the invisible coefficient to zero (dotted). Panel (d) follows the normalized curvature on shrinking overlap surfaces; its $x=1$ limit is the original density.}\label{fig:sm-qutrit-reconstruction}
\end{figure}

To obtain the response data in Fig.~\ref{fig:sm-qutrit-reconstruction}, choose $Q=(\sqrt z,\sqrt{1-z},0)(\sqrt z,\sqrt{1-z},0)^{\mathsf T}$. Write the first two hidden-state probabilities as $r$ and $s=(1-r)v$. Their Haar density is $2(1-r)dr\,dv$, with a uniform relative phase. The overlap has the form $a+b\cos\phi$, where $a=zr+(1-z)s$ and $b=2\sqrt{z(1-z)rs}$. For $c=x-a$, the phase average of $|c-b\cos\phi|$ is $|c|$ when $|c|\ge b$, and $2[\sqrt{b^2-c^2}+c\arcsin(c/b)]/\pi$ otherwise. Its derivative is $\operatorname{sgn}c$ outside and $2\arcsin(c/b)/\pi$ inside. Gauss--Legendre quadrature evaluates the remaining amplitude integrals, and Jacobi projection extracts their angular coefficients. Combining the response and derivative coefficients through the Cauchy inverse above recovers the density shown in panel (c).

\subsection{Contact and preparation bounds}

Let $n\ge2$, let $P=\ketbra\psi$ denote a rank-one projector on Bob, and let $\mu$ be a finite measure on $\mathbb CP^{n-1}$. For the reconstruction statements, $\mu$ may be signed. Choose traceless Hermitian matrices $T_\alpha$, $1\le\alpha\le d=n^2-1$, with $\Tr(T_\alpha T_\beta)=\delta_{\alpha\beta}$, and put $r_n=\sqrt{(n-1)/n}$. The expansions
\begin{equation}
P=\frac{\openone_n}{n}+r_n\vec v(P)\cdot\vec T,\qquad Y(s,\vec y)=s\openone_n+r_n^{-1}\vec y\cdot\vec T
\label{eq:general-bloch-coordinates}
\end{equation}
give $|\vec v(P)|=1$ and $\Tr(YP)=s+\vec y\cdot\vec v(P)$. The fixed length follows immediately from $\Tr P^2=\Tr P=1$; for $n>2$, the allowed directions form a proper subset of the sphere~\cite{Kimura2003}.

\begin{proposition}\label{prop:general-response-wave}
The binary response and the $r$-outcome response,
\begin{equation}
K_\mu(s,\vec y)=\int|s+\vec y\cdot\vec v(P)|d\mu(P),\qquad R_\mu(\{s_i,\vec y_i\})=\int\max_{1\le i\le r}(s_i+\vec y_i\cdot\vec v(P))d\mu(P),
\label{eq:general-responses}
\end{equation}
satisfy, as distributions,
\begin{equation}
(\partial_s^2-\Delta_{\vec y})K_\mu=0,\qquad (\partial_{s_i}\partial_{s_j}-\nabla_{\vec y_i}\cdot\nabla_{\vec y_j})R_\mu=0\quad(1\le i,j\le r).
\label{eq:general-wave-system}
\end{equation}
Centering the kernel as in Section~S1 preserves these second-order score identities.
\end{proposition}
\begin{proof}
For any smooth function $f$ of the branch values $b_i=s_i+\vec y_i\cdot\vec v$, the difference of the two derivatives in Eq.~\eqref{eq:general-wave-system} is $(1-|\vec v|^2)\partial_i\partial_jf=0$. Smooth approximations to the absolute value and maximum converge locally uniformly, so the identities pass to their distributional limits and then to the integral against a finite measure. Centering subtracts a linear function of the scores, whose second derivatives vanish.
\end{proof}

Purity also imposes matrix identities. Define the matrix gradient by $\nabla_Y=(\openone_n/n)\partial_s+r_n\sum_\alpha T_\alpha\partial_{y_\alpha}$, so that $\nabla_Y\Tr(YP)=P$ and $\Tr\nabla_Y=\partial_s$. The same smooth approximation gives
\begin{equation}
\left[\nabla_{Y_i}\nabla_{Y_j}-(\Tr\nabla_{Y_i})\nabla_{Y_j}\right]R_\mu=0.
\label{eq:matrix-purity-wave}
\end{equation}
Applying the matrix operator to a smooth branch function multiplies its mixed second derivative by $P^2-P$. Taking the trace recovers Eq.~\eqref{eq:general-wave-system}. For a single score, its symbol is $Z^2-(\Tr Z)Z$. A nonzero Hermitian $Z$ makes this vanish precisely at rank one: every nonzero eigenvalue must equal $\Tr Z$, so only one can occur. These identities therefore retain the pure directions supporting the ensemble.

Binary wave data admit an explicit inverse on any sphere, including the effective circle of rank-two qubit correlations.
\begin{proposition}\label{prop:spherical-cauchy-inversion}
Let $d\ge2$ and let $\mu$ be a finite signed measure on $S^{d-1}$. The values and transverse derivatives of $K(s,\vec y)=\int|s+\vec y\cdot\vec v|d\mu(\vec v)$ on any plane $s=\vec q\cdot\vec y$ with $|\vec q|<1$ determine $\mu$ uniquely. Derivatives are required only at almost every direction on that plane.
\end{proposition}
\begin{proof}
Write $\gamma=(1-|\vec q|^2)^{-1/2}$ and $S=\openone_d+(\gamma-1)\vec q\vec q^{\mathsf T}/|\vec q|^2$, with $S=\openone_d$ at $\vec q=0$. The change
\begin{equation}
s=\gamma(\tau+\vec q\cdot\vec x),\qquad \vec y=S\vec x+\gamma\vec q\tau
\label{eq:general-lorentz-change}
\end{equation}
sends the plane to $\tau=0$. Each direction and its weight become
\begin{equation}
\vec v'=\frac{S\vec v+\gamma\vec q}{w(\vec v)},\qquad w(\vec v)=\gamma(1+\vec q\cdot\vec v)>0,\qquad d\nu=w\,d\mu.
\label{eq:general-lorentz-measure}
\end{equation}
Here $\nu$ is the weighted pushforward and $|\vec v'|=1$. In transformed coordinates the response is $\int|\tau+\vec x\cdot\vec v'|d\nu$. Its data at $|\vec e|=1$ are
\begin{equation}
A(\vec e)=\int|\vec e\cdot\vec v'|d\nu(\vec v'),\qquad B(\vec e)=\int\operatorname{sgn}(\vec e\cdot\vec v')d\nu(\vec v').
\label{eq:spherical-cauchy-data}
\end{equation}
For each fixed $\vec v'$, the exceptional directions $\vec e\cdot\vec v'=0$ have zero sphere measure. Fubini's theorem and dominated convergence therefore justify the derivative for almost every $\vec e$, including singular $\nu$.

Cosine inversion recovers even measures~\cite{AubrunLancien2016}; the hemisphere transform recovers odd measures. Their multipliers make both facts explicit. Expanding in orthonormal spherical harmonics with normalized area gives, for $d\ge3$,
\begin{equation}
\begin{aligned}
\lambda_0&=\frac{\Gamma(d/2)}{\sqrt\pi\,\Gamma((d+1)/2)},& \lambda_{2k}&=\frac{(-1)^{k-1}\Gamma(d/2)\Gamma(k-1/2)}{2\pi\,\Gamma(k+(d+1)/2)}\quad(k\ge1),\\
\eta_{2k+1}&=\frac{(-1)^k\Gamma(d/2)\Gamma(k+1/2)}{\pi\,\Gamma(k+(d+1)/2)}\quad(k\ge0).
\end{aligned}
\label{eq:general-cosine-multipliers}
\end{equation}
The other parities vanish. These formulas follow by integrating the normalized zonal Gegenbauer polynomial against $|t|(1-t^2)^{(d-3)/2}$ or $\operatorname{sgn}(t)(1-t^2)^{(d-3)/2}$ and using Rodrigues' formula. On $S^1$, direct Fourier integration gives $\lambda_{2k}=2(-1)^{k+1}/[\pi(4k^2-1)]$ and $\eta_{2k+1}=2(-1)^k/[\pi(2k+1)]$, including $\lambda_0=2/\pi$. Hence $A$ determines every even harmonic moment of $\nu$ and $B$ every odd moment. Polynomial density determines the measure. An explicit weak reconstruction is obtained by dividing these moments by their multipliers, multiplying degree $\ell$ by $e^{-\epsilon\ell(\ell+d-2)}$, and sending $\epsilon\downarrow0$. Finally, the inverse of Eq.~\eqref{eq:general-lorentz-measure} recovers $\mu$.
\end{proof}

The canonical map $A$ supplies the required response of the bipartite state. For $\widehat\rho_v=v\widehat\rho+(1-v)\openone_{mn}/(mn)$, $0\le v\le1$, this map is $vA$. Take its adjoint between real Hilbert--Schmidt spaces, with traceless output on Alice, and define the binary source
\begin{equation}
N_v(H)=\frac1n\left\|vA^*(H)+\frac{\Tr H}{m}\openone_m\right\|_1,\qquad H=H^\dagger,
\label{eq:general-binary-source}
\end{equation}
where $\|Z\|_1=\Tr|Z|$. A probability measure with $\int P\,d\mu=\openone_n/n$ supplies every binary POVM exactly when
\begin{equation}
\int|\Tr(HP)|d\mu(P)\ge N_v(H)\qquad\text{for every Hermitian }H.
\label{eq:general-binary-domination}
\end{equation}
To derive this equivalence, fix a binary effect $0\preceq M\preceq\openone_m$ and score its two outcomes by $H$ and $-H$. Its required score is $[2\Tr(MZ)-\Tr H]/n$, with $Z=vA^*(H)+(\Tr H)\openone_m/m$. Maximization over $M$ gives $[2\Tr Z_+-\Tr Z]/n=\|Z\|_1/n$. The available score is the absolute-value integral. Convex separation for each binary measurement proves both directions. Several-outcome measurements are supplied by the maximum-response condition of Section~S1.

An angular average of rank-one tests bounds the visibility; equality can determine the entire ensemble.
\begin{theorem}\label{thm:projective-contact-boundary}
Let $\mathfrak M$ be a measurement class containing every binary POVM, and let $d\omega(Q)$ be normalized Haar measure on pure states. Every LHS model for $\widehat\rho_v$ obeys, for $0<x<1$,
\begin{equation}
\int N_v(x\openone_n-Q)d\omega(Q)\le a_0(x)=x-\frac1n+\frac2n(1-x)^n.
\label{eq:projective-contact-bound}
\end{equation}
Suppose equality holds at $(v_*,x_*)$ with $v_*>0$ and $0<x_*<1$. Any pure-state ensemble realizing an LHS model there is uniquely determined by
\begin{equation}
F_\mu(x_*,Q)=N_{v_*}(x_*\openone_n-Q),\qquad \partial_xF_\mu(x_*,Q)=\left.\partial_xN_{v_*}(x\openone_n-Q)\right|_{x=x_*}
\label{eq:projective-contact-data}
\end{equation}
and Eq.~\eqref{eq:projective-cauchy-inverse}. If the reconstructed measure is positive, has operator moment $\openone_n/n$, and satisfies the response inequalities for $\mathfrak M$ at $v_*$, then $v_*$ is its exact critical visibility. Every $v>v_*$ in $[0,1]$ violates Eq.~\eqref{eq:projective-contact-bound}.
\end{theorem}
\begin{proof}
For each pure $P$, Haar overlap averaging gives $\int|x-\Tr(QP)|d\omega(Q)=a_0(x)$. Averaging Eq.~\eqref{eq:general-binary-domination} for $H=x\openone_n-Q$ proves the bound. Equality forces the continuous nonnegative gap to vanish for all $Q$. Almost everywhere, $F_\mu$ is differentiable by Theorem~\ref{thm:projective-wave-inversion}. Since the demand is convex and touches it from below, all demand subgradients equal that derivative. These are the two contact data in Eq.~\eqref{eq:projective-contact-data}; inversion fixes the measure.

At fixed $x_*$, the averaged source is convex in visibility. At zero visibility it is $|x_*-1/n|<a_0(x_*)$, since $x_*-\Tr(QP)$ takes both signs with positive probability. Beyond a positive contact $v_*$, convexity therefore makes the average exceed its bound. The reconstructed measure supplies a model at $v_*$ under the stated conditions, and mixing with the maximally mixed model supplies every smaller visibility.
\end{proof}

It suffices that the class contain all binary projective measurements: convex combinations of projections give every binary effect. Varying $x$ moves the contact section, and saturation on any interior section fixes the ensemble.

Saturation also forces pure hidden states. Write a mixed $\tau$ in its spectral decomposition $\tau=\sum_i\lambda_iP_i$. Convexity gives $|x-\Tr(Q\tau)|\le\sum_i\lambda_i|x-\Tr(QP_i)|$. For $0<x<1$, choose $Q$ near an eigenvector of positive weight. Its overlap with that eigenvector exceeds $x$, while overlap with a second positively weighted eigenvector is below $x$. The summands have opposite signs on an open set, making Jensen inequality strict. Hence
\begin{equation}
\int|x-\Tr(Q\tau)|d\omega(Q)<a_0(x)\quad\text{if $\tau$ is mixed}.
\label{eq:mixed-contact-strict}
\end{equation}
Every realizing model at saturation of Eq.~\eqref{eq:projective-contact-bound} is therefore pure almost everywhere, with the unique measure fixed by Theorem~\ref{thm:projective-contact-boundary}. Parents differ only by splitting and relabeling proportional effects. Merging distinct directions loses a required binary response, and a continuous unique measure requires infinitely many parent outcomes.

For an explicit three-level boundary, take the isotropic states $\rho_v=v\ketbra{\Phi_3}+(1-v)\openone_9/9$, where $\ket{\Phi_3}=\sum_{j=1}^3\ket{jj}/\sqrt3$. Their map is $A(X)=X^{\mathsf T}$. On $H=x\openone_3-Q$, the three source eigenvalues before division by three are $x-(1+2v)/3$ and $x-(1-v)/3$, the latter twice. Between their zeros,
\begin{equation}
N_v(x\openone_3-Q)=\frac{3x+4v-1}{9}.
\label{eq:qutrit-isotropic-source}
\end{equation}
Its tangency to $a_0(x)=x-1/3+2(1-x)^3/3$ gives
\begin{equation}
x_*=v_*=1-\frac1{\sqrt3}.
\label{eq:qutrit-isotropic-binary-threshold}
\end{equation}
Direction-independent contact values and derivatives make all nonconstant moments vanish, so the recovered measure is Haar. For Alice's rank-one projection $Q^{\mathsf T}$, assign hidden state $P$ to that outcome when $q=\Tr(QP)\ge x_*$. This cap has mass $(1-x_*)^2=1/3$ and operator moment
\begin{equation}
\int_{q\ge x_*}P\,d\omega(P)=\frac13\left[v_*Q+\frac{1-v_*}{3}\openone_3\right].
\label{eq:qutrit-isotropic-cap-model}
\end{equation}
Indeed, its component along $Q$ is $\int_{x_*}^1 2q(1-q)dq=1/3-2/(9\sqrt3)$; unitary invariance distributes the remaining trace equally among the two perpendicular directions. Complementation gives all rank-two projections, and convex combinations give all binary effects. Thus Eq.~\eqref{eq:qutrit-isotropic-binary-threshold} recovers the known exact dichotomic threshold~\cite{NguyenGuhne2020}. For the class of all projective measurements, including three outcomes, the isotropic threshold is $5/12$~\cite{Wiseman2007}. Both measurement classes use the general response equations, with their respective outcome constraints.

\subsection{State discrimination and outcome mixing}

A normalized pure-state measure gives the parent POVM $G(dP)=nP\,d\mu(P)$. More generally, write a finite POVM as $G_a=nw_a\tau_a$, where $w_a=\Tr G_a/n>0$, $\Tr\tau_a=1$, $\sum_aw_a=1$, and $\sum_aw_a\tau_a=\openone_n/n$. Its response to the same overlap tests is
\begin{equation}
F_G(x,Q)=\sum_aw_a|x-\Tr(Q\tau_a)|.
\label{eq:measurement-response}
\end{equation}
For a POVM on a standard Borel outcome space, use its trace measure and measurable operator density; sums below become integrals. This includes finite and countable outcomes and continuous parents on pure-state space.

Equation~\eqref{eq:biased-discrimination-response} extends to these mixed effects with $F_\mu$ replaced by $F_G$. The difference of the two prior-weighted outcome probabilities is $nw_a[x-\Tr(Q\tau_a)]/(1+nx)$. Summing its absolute value gives the optimal binary decision score for finite or continuous outcomes~\cite{AubrunLancien2016}.

Reconstruction turns a discrimination profile into the pure directions and weights of a measurement. To compare it with arbitrary POVMs, average the profile over pure inputs: every rank-one measurement attains the same maximal average, and a mixed effect lowers it strictly. Combining this average with the inverse gives the following characterization using a fixed reference state and prior. It also gives an operational test of maximality under classical postprocessing~\cite{Buscemi2005,Zhu2022}.

\begin{theorem}[Measurement comparison at a fixed prior]\label{thm:fixed-prior-comparison}
Fix $x\in(0,1)\setminus\mathcal E_n$ and the prior $p=nx/(1+nx)$. Let $G$ be a rank-one POVM and $H$ any POVM. If
\begin{equation}
p_{\rm succ}(x,Q;H)\ge p_{\rm succ}(x,Q;G)\qquad\text{for every pure }Q,
\label{eq:fixed-prior-dominance}
\end{equation}
then $H$ is rank one almost everywhere and has the same pure-effect measure as $G$. Thus $H$ and $G$ are mutually obtainable by classical postprocessing. Rank-one POVMs are precisely the maximal measurements under the comparison \eqref{eq:fixed-prior-dominance}. For two inequivalent rank-one POVMs, each succeeds strictly more often on some nonempty open set of pure inputs.
\end{theorem}
\begin{proof}
For a rank-one $G$, Haar averaging gives $\int F_G(x,Q)d\omega(Q)=a_0(x)$. Resolving each normalized effect of $H$ into pure states and applying Jensen inequality gives
\begin{equation}
\int F_H(x,Q)d\omega(Q)\le a_0(x),
\label{eq:measurement-average-bound}
\end{equation}
with equality exactly when $H$ is rank one almost everywhere. Strictness for every mixed effect is Eq.~\eqref{eq:mixed-contact-strict}; integration preserves strictness whenever mixed effects have positive trace measure. Dominance in Eq.~\eqref{eq:fixed-prior-dominance} is equivalent to $F_H(x,Q)\ge F_G(x,Q)$. Its nonnegative continuous difference has nonpositive average, so it vanishes everywhere. Equality in Eq.~\eqref{eq:measurement-average-bound} makes $H$ rank one, and division by the nonzero multipliers $a_\ell(x)$ identifies its pure-effect measure with that of $G$.

To relate this measure to outcome labels, write a rank-one POVM as $G(d\lambda)=nP_\lambda\,dw(\lambda)$, with $w=\Tr G/n$. Its pushforward under $\lambda\mapsto P_\lambda$ is $\mu_G$, so forgetting the label within each pure direction gives $nP\,d\mu_G(P)$. Conversely, the conditional distribution of $\lambda$ given $P_\lambda$ splits that effect back into its original outcomes. Standard Borel outcome spaces admit these conditional distributions. Equality of $\mu_G$ and $\mu_H$ therefore gives classical postprocessing in both directions, including continuous parents.

A measurement with mixed effects admits a pure refinement. Convexity increases every response under refinement, and strictness in Eq.~\eqref{eq:measurement-average-bound} increases its average. Such a measurement is not maximal. Rank-one measurements are maximal by the first part. Finally, inequivalent rank-one POVMs have different response profiles by inversion, but the same angular average. Their continuous difference takes both signs, and each strict sign persists on an open set.
\end{proof}

The qutrit ensembles in Eq.~\eqref{eq:qutrit-reconstruction-density} illustrate the exceptional priors. Changing their degree-three weight leaves every success probability unchanged at $p=9/16$, while the nonzero threshold derivative distinguishes the resulting rank-one parents. More generally, success probabilities and their prior derivatives recover both Cauchy data at every interior threshold. If an arbitrary measurement has the same profile as a rank-one measurement there, Eq.~\eqref{eq:measurement-average-bound} first forces its effects to be rank one; Cauchy inversion then identifies the pure-effect measure.

Mixed effects also change the propagation law. Let $\mathcal L_n$ be the operator on the left of Eq.~\eqref{eq:projective-wave-equation}.
\begin{proposition}[Mixed effects as wave sources]\label{prop:measurement-wave-source}
For every POVM $G$, the distribution $\mathcal L_nF_G$ is the nonnegative measure
\begin{equation}
\mathcal L_nF_G=2\sum_aw_a\Tr[Q(\tau_a-\tau_a^2)]\delta\bigl(x-\Tr(Q\tau_a)\bigr).
\label{eq:measurement-wave-source}
\end{equation}
Its total weight is
\begin{equation}
\int_0^1\!dx\int d\omega(Q)\,\mathcal L_nF_G=\frac2n\sum_aw_a\bigl(1-\Tr\tau_a^2\bigr).
\label{eq:measurement-source-purity}
\end{equation}
It vanishes exactly when all nonzero effects are rank one.
\end{proposition}
\begin{proof}
For $q=\Tr(Q\tau)$, tangent differentiation gives $|\nabla q|^2=4[\Tr(Q\tau^2)-q^2]$ and $\Delta_{\rm FS}q=4(1-nq)$. Applying $\mathcal L_n$ to $|x-q|$ leaves $2\Tr[Q(\tau-\tau^2)]\delta(x-q)$, which is nonnegative because $0\preceq\tau\preceq\openone_n$. Integration over $x$ and $\int Q\,d\omega=\openone_n/n$ give Eq.~\eqref{eq:measurement-source-purity}. Endpoint level sets have zero Haar measure unless their coefficient already vanishes. Finally, $\Tr\tau^2=1$ for a positive unit-trace matrix exactly at rank one.
\end{proof}

The integrated source can also be obtained from the mean response. In an eigenbasis of $\tau$, the Haar overlaps $q_i=\Tr(Q\ketbra i)$ satisfy $\int q_i^2d\omega=2/[n(n+1)]$ by the overlap density. Symmetry and $\sum_iq_i=1$ then give $\int q_iq_jd\omega=1/[n(n+1)]$ for $i\ne j$. Expanding in this basis yields $\int[\Tr(Q\tau)]^2d\omega=(1+\Tr\tau^2)/[n(n+1)]$. Together with $\int_0^1|x-q|dx=q^2-q+1/2$, comparison with a pure effect gives
\begin{equation}
\Sigma_G:=\int_0^1\!dx\int d\omega(Q)\,\mathcal L_nF_G=2(n+1)\int_0^1\left[a_0(x)-\int F_G(x,Q)d\omega(Q)\right]dx.
\label{eq:source-response-area}
\end{equation}
Thus the area between the input-averaged response and the rank-one average determines the total source.

The integrated source is determined by POVM purity~\cite{Zhu2022}. Its change under classical processing follows by expanding the mixture within each new outcome. For $\widetilde G_b=\sum_aC(b|a)G_a$, $\sum_bC(b|a)=1$, put $\widetilde w_b=\sum_aw_aC(b|a)$ and $\widetilde\tau_b=\sum_aw_aC(b|a)\tau_a/\widetilde w_b$, omitting zero weights. Expanding the squared differences gives
\begin{equation}
\sum_b\widetilde w_b(1-\Tr\widetilde\tau_b^2)-\sum_aw_a(1-\Tr\tau_a^2)=\sum_{a,b}w_aC(b|a)\Tr[(\tau_a-\widetilde\tau_b)^2]\ge0.
\label{eq:coarse-graining-source}
\end{equation}
Thus merging distinct normalized effects strictly increases the integrated source. Equality holds precisely when every retained outcome combines only proportional effects. The weights $w_aC(b|a)/\widetilde w_b$ are the conditional probabilities of the old records given the new label $b$, for a maximally mixed input. Thus $\widetilde\tau_b$ is their conditional mean and the right side is their conditional Hilbert--Schmidt variance, averaged over the new labels. For an initially rank-one measurement it is exactly the angular spread of the pure directions that a coarser record no longer resolves.

This variance can be expressed in the same Fubini--Study distance that governs propagation. For a rank-one initial measurement, let $\mu_b$ be the conditional probability distribution of its pure directions given a retained label $b$, so $\widetilde\tau_b=\int P\,d\mu_b(P)$. Then
\begin{equation}
\begin{aligned}
1-\Tr\widetilde\tau_b^2&=\iint\bigl[1-\Tr(PP')\bigr]d\mu_b(P)d\mu_b(P'),\\
&=\iint\sin^2d_{\rm FS}(P,P')d\mu_b(P)d\mu_b(P').
\end{aligned}
\label{eq:source-angular-separation}
\end{equation}
The first equality expands the square of the conditional mean; the second uses $\Tr(PP')=\cos^2d_{\rm FS}(P,P')$. Multiplying by $2\widetilde w_b/n$ and summing gives the total wave source. It is therefore fixed by the pairwise angular separation of the pure directions behind the same recorded label.

The same variance has a direct interpretation as lost discrimination performance. Let $\Delta p(x,Q)$ denote the decrease in success probability when $G$ is replaced by its classical processing $\widetilde G$. It is nonnegative because decisions made after processing can be made from the original outcomes.
\begin{proposition}[Discrimination loss and wave source]\label{prop:discrimination-source-loss}
For each pure input $Q$, put $D_{ab}=\tau_a-\widetilde\tau_b$. The threshold-integrated response loss and source increase obey
\begin{equation}
\int_0^1(F_G-F_{\widetilde G})(x,Q)dx=\sum_{a,b}w_a C(b|a)(\Tr QD_{ab})^2,
\label{eq:directional-response-loss}
\end{equation}
\begin{equation}
\int_0^1\mathcal L_n(F_{\widetilde G}-F_G)(x,Q)dx=2\sum_{a,b}w_a C(b|a)\Tr(QD_{ab}^2)\ge0.
\label{eq:directional-source-increase}
\end{equation}
Average $\Delta p$ over Haar-distributed $Q$ and independently over $x\in[0,1]$ with probability density $2(1+nx)/(n+2)$, and denote this average by $\overline{\Delta p}$. Then
\begin{equation}
\overline{\Delta p}=\frac{\sum_{a,b}w_a C(b|a)\Tr D_{ab}^2}{(n+1)(n+2)},
\label{eq:averaged-discrimination-loss}
\end{equation}
and the total source increase is
\begin{equation}
\int_0^1\!dx\int d\omega(Q)\,\mathcal L_n(F_{\widetilde G}-F_G)=\frac{2(n+1)(n+2)}n\,\overline{\Delta p}.
\label{eq:source-discrimination-identity}
\end{equation}
The average loss vanishes exactly when every retained outcome combines only proportional effects. The identities hold for finite POVMs and for standard Borel outcome spaces, with sums replaced by their trace-measure and conditional-distribution integrals.
\end{proposition}
\begin{proof}
For $0\le q\le1$, direct integration gives $\int_0^1|x-q|dx=q^2-q+1/2$. At fixed $Q$, put $q_a=\Tr(Q\tau_a)$ and $\widetilde q_b=\Tr(Q\widetilde\tau_b)$. Since $\widetilde w_b\widetilde q_b=\sum_a w_a C(b|a)q_a$, cancellation of the linear and constant terms gives
\begin{equation}
\int_0^1(F_G-F_{\widetilde G})(x,Q)dx=\sum_a w_a q_a^2-\sum_b\widetilde w_b\widetilde q_b^2=\sum_{a,b}w_a C(b|a)(q_a-\widetilde q_b)^2.
\label{eq:decision-threshold-variance}
\end{equation}
This proves Eq.~\eqref{eq:directional-response-loss}. Integrating Eq.~\eqref{eq:measurement-wave-source} over threshold and using the preserved operator mean gives twice the expectation in $Q$ of $\sum_a w_a\tau_a^2-\sum_b\widetilde w_b\widetilde\tau_b^2$. Expanding the matrix squares expresses this operator as $\sum_{a,b}w_a C(b|a)D_{ab}^2$, proving Eq.~\eqref{eq:directional-source-increase}.

Multiplying Eq.~\eqref{eq:biased-discrimination-response} by the threshold density cancels $1+nx$, giving
\[
\overline{\Delta p}=\frac n{n+2}\int_0^1\!dx\int d\omega(Q)\,[F_G-F_{\widetilde G}](x,Q).
\]
Equation~\eqref{eq:source-response-area} identifies the last integral with $(\Sigma_{\widetilde G}-\Sigma_G)/[2(n+1)]$, proving Eq.~\eqref{eq:source-discrimination-identity}. Equations~\eqref{eq:measurement-source-purity} and \eqref{eq:coarse-graining-source} express the same source difference as $(2/n)\sum_{a,b}w_aC(b|a)\Tr D_{ab}^2$, which gives Eq.~\eqref{eq:averaged-discrimination-loss}.

Every term $\Tr D_{ab}^2$ is nonnegative. Vanishing average loss therefore requires $\tau_a=\widetilde\tau_b$ for every pair with $w_a C(b|a)>0$, precisely the condition that each retained outcome combines proportional effects. Conversely, this condition makes every term zero. For general outcome spaces, disintegrate the joint trace measure before and after processing. All normalized effects are bounded, so Fubini's theorem and the same conditional-variance identities prove the integral statements.
\end{proof}

Consider a rank-one POVM $G(dP)=nP\,d\mu(P)$, where $\mu$ is a probability measure with operator mean $\openone_n/n$. Retain an outcome with probability $\eta$ and otherwise draw a new label independently from $\mu$. This classical processing gives
\begin{equation}
G_\eta(dP)=n\tau_P(\eta)d\mu(P),\qquad \tau_P(\eta)=\eta P+(1-\eta)\openone_n/n,\qquad 0\le\eta\le1.
\label{eq:isotropic-outcome-noise}
\end{equation}
The processing keeps the trace weights. For a basis measurement it gives the familiar blurred effects $\eta\ketbra a+(1-\eta)\openone_n/n$.
\begin{corollary}[Identifying a measurement under outcome noise]\label{cor:noise-identification}
Let $n\ge2$ and let $G_\eta$ be as in Eq.~\eqref{eq:isotropic-outcome-noise}. Its binary success probabilities for all pure inputs and all priors determine the integrated source
\begin{equation}
\Sigma_\eta=\int_0^1\!dx\int d\omega(Q)\,\mathcal L_nF_\eta(x,Q)=\frac{2(n-1)}{n^2}(1-\eta^2)
\label{eq:noise-source-calibration}
\end{equation}
and hence the noise visibility $\eta$. If $\eta>0$, they also determine the original pure-effect measure $\mu$. Outcome-dependent decisions can improve on guessing the more probable input only for
\begin{equation}
\frac{1-\eta}{2-\eta}<p<\frac{1+(n-1)\eta}{2+(n-1)\eta}.
\label{eq:noise-prior-window}
\end{equation}
At every prior strictly inside this interval, some pure inputs have a strict improvement. The average loss relative to $G$, with the averaging of Proposition~\ref{prop:discrimination-source-loss}, is
\begin{equation}
\overline{\Delta p}=\frac{(n-1)(1-\eta^2)}{n(n+1)(n+2)}.
\label{eq:noise-universal-loss}
\end{equation}
At $\eta=0$, the response is $|x-1/n|$ and is independent of $\mu$.
\end{corollary}
\begin{proof}
Put $b=(1-\eta)/n$. Since $\Tr[Q\tau_P(\eta)]=b+\eta\Tr(QP)$, at $\eta>0$ the response satisfies
\begin{equation}
F_\eta(x,Q)=\eta F\left((x-b)/\eta,Q\right),
\label{eq:noise-response-rescaling}
\end{equation}
where the defining integral for $F$ extends to every real threshold. For $x\le b$ or $x\ge b+\eta$, all integrands have the same sign before taking the absolute value. The prescribed mean then gives $F_\eta=|x-1/n|$, and Eq.~\eqref{eq:biased-discrimination-response} reduces to $p_{\rm succ}=\max\{p,1-p\}$. Converting $b<x<b+\eta$ to $p=nx/(1+nx)$ gives Eq.~\eqref{eq:noise-prior-window}.

For any interior rescaled threshold $y\in(0,1)$, Haar averaging the original response gives $a_0(y)=\int|y-\Tr(QP)|d\omega(Q)$, independently of $P$ and of the ensemble. The Haar overlap has positive density on $(0,1)$, so strict Jensen inequality gives $a_0(y)>|y-1/n|$. Some pure inputs therefore have a response strictly above the guessing value. Continuity gives an open set of such inputs, proving the assertion throughout the interval.

Directly, $\Tr\tau_P(\eta)^2=1/n+(1-1/n)\eta^2$. Equations~\eqref{eq:measurement-source-purity} and \eqref{eq:source-response-area} give the integrated source from the effect purity or directly from the response. Equation~\eqref{eq:noise-source-calibration} therefore gives $\eta=[1-n^2\Sigma_\eta/(2(n-1))]^{1/2}$. At positive $\eta$, the interval $b<x<b+\eta$ lies inside $0<x<1$ and Eq.~\eqref{eq:noise-response-rescaling} recovers the full original response. Curvature recovery then determines $\mu$, including singular measures. The original rank-one measurement has zero source, so Eq.~\eqref{eq:source-discrimination-identity} gives Eq.~\eqref{eq:noise-universal-loss}. These computations use bounded integrands and hold directly for standard Borel outcome spaces. At zero visibility every normalized effect is $\openone_n/n$, which gives the stated response.
\end{proof}

The mean advantage over guessing also follows directly. For $b<x<b+\eta$, put $y=(x-b)/\eta$ and use $\int F(y,Q)d\omega(Q)=a_0(y)$. At prior $p=nx/(1+nx)$,
\begin{equation}
\int p_{\rm succ}(x,Q;G_\eta)d\omega(Q)-\max\{p,1-p\}=\frac12+\frac{n\eta}{2(1+nx)}a_0(y)-\max\{p,1-p\}.
\label{eq:noise-mean-decision-advantage}
\end{equation}
Outside the interval it vanishes. The average is independent of the original pure-effect distribution. Main-text Fig.~1d evaluates it for $n=3$, for which $a_0(y)=y-1/3+2(1-y)^3/3$. Each randomized record has this reduced decision advantage, while the full response identifies the original distribution at positive visibility.

With at least three levels, equal-prior binary tasks with arbitrary input pairs provide another reconstruction.
\begin{proposition}\label{prop:traceless-response-unique}
For $n\ge3$, the values $\int|\Tr(YP)|d\mu(P)$ for all traceless Hermitian $Y$ determine any finite signed pure-state measure $\mu$. More generally, if a density matrix $\sigma$ has rank at least three, the values $\int|\Tr[Y(P-\sigma)]|d\mu(P)$ for traceless $Y$ determine $\mu$.
\end{proposition}
\begin{proof}
For the central case, the cosine transform gives the even part of the measure on the Bloch sphere. Its physical support contains no antipodal pair: $\vec v(P)=-\vec v(Q)$ would give $P+Q=2\openone_n/n$, whose two sides have ranks at most two and $n$, respectively. Restricting the recovered even measure to the physical pure-state support and multiplying by two therefore recovers the original measure.

For the general center, use the unit direction $(P-\sigma)/\|P-\sigma\|_2$ and weight $\|P-\sigma\|_2$, where $\|\cdot\|_2$ is the Hilbert--Schmidt norm. No two distinct projectors give collinear differences: $P-\sigma=c(Q-\sigma)$ with $c\ne1$ would express $\sigma$ as a linear combination of two rank-one matrices. Thus this direction map is injective and its image contains no antipodal pair. The same even-part restriction, followed by division by the strictly positive weight, gives $\mu$.
\end{proof}

For equally likely inputs $\rho_0,\rho_1$, the rank-one parent succeeds with probability $1/2+(n/4)\int|\Tr[(\rho_0-\rho_1)P]|d\mu$. Every nonzero traceless Hermitian matrix is a positive multiple of a difference of states. Hence, for $n\ge3$, all equal-prior binary discrimination probabilities determine the same pure-state weights. This uses arbitrary pairs of inputs; Eq.~\eqref{eq:biased-discrimination-response} instead uses one fixed maximally mixed input and a variable pure input.

\section{S3. Spectral dynamics and minimum ensemble mass}\label{sec:fourier-spectra}

To select the wave required by a bipartite state, we minimize the mass of its generating ensemble subject to the response inequalities of Section~S1. An unmet score supplies a variation of the weights: states that can provide the missing response acquire weight, while a mass cost removes excess. Projection maintains positivity and the operator mean. We derive the resulting evolution and prove that its regularized solutions approach the minimum, including singular optimal ensembles. Finite measurement restrictions give bounds on the same quantity. Explicit qubit response formulas then reduce the variables needed for its evaluation and prepare the contact construction of Section~S4.

\subsection{Spectral dynamics in arbitrary finite dimensions} \label{subsec:spectral-bipartite}

To derive the density equation once for these different spaces, denote the hidden-state space by $\mathcal X$ and its invariant probability measure by $\omega$. Thus $\mathcal X=\C P^{n-1}$ in the general problem and is an effective sphere after qubit rank reduction. The moment coordinates are the unit Bloch coordinates of Section~\hyperref[sec:steering-scale]{S1}; on a sphere they are simply $\vec v(x)=x$. Write $k=\kappa$ and $N=[N_A]_+$ in the general problem. The positive part preserves the inequalities because the centered kernel is nonnegative. With a fixed full-support probability measure $\sigma$ on normalized tests, define
\[
(\mathcal Ku)(z)=\int k(z;x)u(x)d\omega(x),\qquad (\mathcal K^*f)(x)=\int k(z;x)f(z)d\sigma(z).
\]
We assume continuous $N\in L^2(\sigma)$ and continuous $k$, with $k(z;\cdot)\le H(z)$ and hidden-state Lipschitz constant at most $L(z)$, where $H,L\in L^2(\sigma)$. The parent construction in Section~S1 supplies a feasible measure of finite mass. These conditions hold for the general bipartite kernels and for both qubit kernels below.

Integration measures can be fixed before the state is supplied. Haar measure on $U(m)$ gives Alice's PVMs $M_a=U\ketbra aU^\dagger$. For POVMs, $r=m^2$ outcomes suffice by Eq.~\eqref{eq:s4-all-povm-reduction}: uniformly distributed isometries $V:\C^m\to\C^{rm}$ give effects $M_a=V_a^\dagger V_a$ from their $m$-row blocks. Combine either distribution with normalized area on the centered Hermitian-test sphere to define $\sigma$. Continuity includes zero effects and limiting tests.

A trial density may supply some tests and fall short on others. Its deficit is the positive difference between the required and available responses. Weighting each deficit by the response of a hidden state and integrating over tests defines
\begin{equation}
r(u;z)=[N(z)-(\mathcal Ku)(z)]_+,\qquad G(u)=\mathcal K^*r(u), \label{eq:general-linear-residual}
\end{equation}
Nonnegative densities of the required moment form $\mathcal C=\{u\in L^2(\omega):u\ge0,\ \int\vec v u\,d\omega=0\}$, with mass $t(u)=\int u\,d\omega$. A squared-deficit integral enforces response constraints: full support of $\sigma$ detects every open set of violated tests, and positive deficits cannot cancel. Balance this penalty against mass with
\[
J_\varepsilon(u)=t(u)+\frac{\|r(u)\|_2^2}{2\varepsilon} +\frac{\delta_\varepsilon}{2}\|u\|_2^2, \qquad u\in\mathcal C,
\]
Mass enters linearly. Squared deficits penalize unmet responses, while the quadratic density term makes the functional strictly convex. Taking $\varepsilon\downarrow0$ forces deficits to vanish; the decay rate of $\delta_\varepsilon$ allows the attracting densities to approach either continuous or singular optimal measures.

Differentiating gives gradient $1-\varepsilon^{-1}G(u)+\delta_\varepsilon u$. Deficit feedback assigns weight according to how much a hidden state can supply, leading to stationarity on $\mathcal C$:
\[
\delta_\varepsilon u_\varepsilon(x) =\left[\varepsilon^{-1}G(u_\varepsilon)(x)-1 -\vec\eta_\varepsilon\cdot\vec v(x)\right]_+, \qquad \int\vec v u_\varepsilon\,d\omega=0.
\]
The multiplier $\vec\eta_\varepsilon$ fixes the marginal: weights are positive where feedback exceeds mass and moment costs. Choose $\delta_\varepsilon=\varepsilon^{D+1}$ with $D\ge\dim\mathcal X$. Projection onto $\mathcal C$ is finite-dimensional after positivity is imposed. At a fixed multiplier the closest nonnegative function is $[g-\vec\eta\cdot\vec v]_+$; zero moment selects the multiplier. The following integral computes that projection.

For $g\in L^2(\omega)$ and a moment multiplier $\vec\eta\in\R^{d_v}$, where $d_v$ is the number of moment coordinates, define
\begin{equation}
\begin{aligned}
E_g(\vec\eta)&=\frac12\int[g(x)-\vec\eta\cdot\vec v(x)]_+^2d\omega(x),\\
B_\beta(g)(x)&= \frac{\displaystyle\int_{\R^{d_v}}[g(x)-\vec\eta\cdot\vec v(x)]_+ e^{-\beta E_g(\vec\eta)}d\vec\eta} {\displaystyle\int_{\R^{d_v}}e^{-\beta E_g(\vec\eta)}d\vec\eta}, \qquad B(g)=\lim_{\beta\to\infty}B_\beta(g).
\end{aligned}
\label{eq:general-moment-root}
\end{equation}
For $d_v=0$, these maps are $[g]_+$. For $d_v>0$, differentiating $E_g$ shows that a minimizing multiplier imposes zero moment on the positive part. The map $B$ is the orthogonal projection onto $\mathcal C$. Every finite-$\beta$ integral $B_\beta$ also lies in $\mathcal C$, and
\begin{equation}
\|B_\beta(g)-B(g)\|_2^2\le\frac{d_v}{\beta}. \label{eq:general-projection-error}
\end{equation}
Each supplied response is linear in $u$, while deficits and moment projection make the density equation nonlinear. Its regularized solution can be evaluated by the contraction below.
\begin{theorem}
\label{thm:linear-spectral-equation} Let $c$ be the minimum mass of a positive zero-moment measure satisfying $\mathcal K\mu\ge N$. Choose $D\ge\dim\mathcal X$, $0<\varepsilon\le1$, and $\delta_\varepsilon=\varepsilon^{D+1}$. There is a unique density $u_\varepsilon\in\mathcal C$ satisfying
\begin{equation}
\delta_\varepsilon u_\varepsilon =B\!\left(\varepsilon^{-1}G(u_\varepsilon)-1\right), \qquad c=\lim_{\varepsilon\downarrow0}\int u_\varepsilon\,d\omega. \label{eq:linear-spectral-equation}
\end{equation}
Set $M=\|H\|_{L^2(\sigma)}$ and
\[
L_\varepsilon=\delta_\varepsilon+M^2/\varepsilon, \qquad R_\varepsilon=\frac{1+M\|N\|_2/\varepsilon}{\delta_\varepsilon}.
\]
The map
\begin{equation}
T_\varepsilon(u)=B\!\left[ \left(1-\frac{\delta_\varepsilon}{L_\varepsilon}\right)u -\frac1{L_\varepsilon}+\frac{G(u)}{\varepsilon L_\varepsilon}\right] \label{eq:linear-contraction}
\end{equation}
contracts $L^2$ distances by at most $1-\delta_\varepsilon/L_\varepsilon$. To evaluate its fixed point, put
\[
N_\varepsilon=\left\lceil\frac{L_\varepsilon}{\delta_\varepsilon} \log\frac{2(1+R_\varepsilon)}\varepsilon\right\rceil, \qquad \beta_\varepsilon=\max\left\{1, \frac{4d_vL_\varepsilon^2}{\delta_\varepsilon^2\varepsilon^2}\right\}.
\]
Starting from $u_0=0$, replace $B$ in Eq.~\eqref{eq:linear-contraction} by $B_{\beta_\varepsilon}$ and iterate $N_\varepsilon$ times. Every iterate is nonnegative with zero first moment, and
\begin{equation}
\|u_{N_\varepsilon}-u_\varepsilon\|_2\le\varepsilon, \qquad \left|\int u_{N_\varepsilon}d\omega-\int u_\varepsilon d\omega\right|\le\varepsilon. \label{eq:linear-finite-error}
\end{equation}
Consequently $c=\lim_{\varepsilon\downarrow0}\int u_{N_\varepsilon}d\omega$.
\end{theorem}
Take $D=2n-2$ for general bipartite systems; $D=2$ gives $\delta_\varepsilon=\varepsilon^3$ for qubits. Input data enter through $N$, with kernel, measures and moment projection fixed. Every weak limit of the resulting measures is optimal.
\begin{proof}
We first establish the integral projection and its error bound. Every nonzero linear combination of the moment coordinates takes both signs. Thus
\[
\chi_v^2=\min_{|\vec e|=1}\int[-\vec e\cdot\vec v]_+^2d\omega>0, \qquad E_g(\vec\eta)\ge\frac{\chi_v^2}{4}|\vec\eta|^2-\frac12\|g\|_2^2.
\]
For the sphere $S^{r-1}$, $\chi_v^2=1/(2r)$. The bound ensures both a minimizing multiplier $\vec\eta_*$ and convergence of the integrals. The density $b_*=[g-\vec\eta_*\cdot\vec v]_+$ has zero moment by $\nabla E_g(\vec\eta_*)=0$. For $w\in\mathcal C$,
\[
\langle g-b_*,w-b_*\rangle =-\int[g-\vec\eta_*\cdot\vec v]_-w\,d\omega\le0, \qquad [s]_-=[-s]_+.
\]
This variational inequality identifies $b_*$ as the orthogonal projection.

Write $\mathbb E_\beta$ for expectation with density proportional to $e^{-\beta E_g}$, and put $b_{\vec\eta}=[g-\vec\eta\cdot\vec v]_+$, $q_{\vec\eta}=[g-\vec\eta\cdot\vec v]_-$, and $b=\mathbb E_\beta b_{\vec\eta}$. Integration by parts gives
\[
\mathbb E_\beta\nabla E_g=0,\qquad \mathbb E_\beta[\vec\eta\cdot\nabla E_g]=d_v/\beta.
\]
The first identity proves the zero moment of $b$. The second, together with $q_{\vec\eta}b_{\vec\eta}=0$, gives for every $w\in\mathcal C$
\[
\langle g-b,w-b\rangle =\langle\mathbb E_\beta q_{\vec\eta},b-w\rangle \le\frac{d_v}{\beta}-\mathbb E_\beta\|b_{\vec\eta}-b\|_2^2.
\]
In particular,
\begin{equation}
\langle g-B_\beta(g),w-B_\beta(g)\rangle\le d_v/\beta. \label{eq:general-integral-projection}
\end{equation}
Apply this with $w=b_*$ and add the projection inequality with $w=B_\beta(g)$. The result is Eq.~\eqref{eq:general-projection-error}, which also proves the limit in Eq.~\eqref{eq:general-moment-root}. If $\vec v(x)\ne0$, then $B_\beta(g)(x)>0$, since a half-space of multipliers gives a positive integrand at $x$.

From the kernel bound, $\|\mathcal K\|\le M$. Coercivity and strong convexity with constant $\delta_\varepsilon$ give a unique minimizer of $J_\varepsilon$ on the closed cone $\mathcal C$. Its variational inequality is equivalent to
\[
u_\varepsilon=B[u_\varepsilon-\alpha\nabla J_\varepsilon(u_\varepsilon)] \qquad(\alpha>0).
\]
Taking $\alpha=1/\delta_\varepsilon$ and using positive homogeneity of $B$ gives the spectral equation; taking $\alpha=1/L_\varepsilon$ gives the fixed point of Eq.~\eqref{eq:linear-contraction}.

For the contraction, secant slopes of the scalar positive-part function give measurable $0\le\theta(z)\le1$ with
\[
G(u)-G(w)=-\mathcal K^*\Theta\mathcal K(u-w), \qquad(\Theta f)(z)=\theta(z)f(z).
\]
Since $0\preceq\mathcal K^*\Theta\mathcal K\preceq M^2\openone$, the two arguments of $B$ in Eq.~\eqref{eq:linear-contraction} differ by
\[
\left[\openone- \frac{\delta_\varepsilon\openone+\varepsilon^{-1}\mathcal K^*\Theta\mathcal K} {L_\varepsilon}\right](u-w).
\]
All eigenvalues of the bracket lie in $[0,1-\delta_\varepsilon/L_\varepsilon]$. Orthogonal projection cannot increase distances, giving the contraction factor.

To bound the minimizer, test its variational inequality at zero. Then $\langle\nabla J_\varepsilon(u_\varepsilon),u_\varepsilon\rangle\le0$, and strong monotonicity gives
\[
\delta_\varepsilon\|u_\varepsilon\|_2^2 \le-\langle\nabla J_\varepsilon(0),u_\varepsilon\rangle \le(1+M\|N\|_2/\varepsilon)\|u_\varepsilon\|_2.
\]
To obtain the mass limit, let $\mu_*$ be a minimizing measure. It exists because bounded mass gives weak compactness on $\mathcal X$, while the moment and continuous-kernel conditions are closed. Heat-kernel smoothing produces a density $u_\tau$ of the same mass $c$. The moment remains zero because its coordinate functions are Laplacian eigenfunctions. There are constants $A_0,A_1$ independent of $\tau$ such that
\[
\|u_\tau\|_2^2\le A_0c^2\tau^{-\dim\mathcal X/2},\qquad \|\mathcal Ku_\tau-\mathcal K\mu_*\|_2 \le A_1c\sqrt\tau\,\|L\|_2.
\]
At $\tau=\varepsilon^2$ and $D\ge\dim\mathcal X$, both penalty terms are $O(\varepsilon)$. Thus
\begin{equation}
J_\varepsilon(u_\varepsilon)\le c+C\varepsilon, \qquad \|[N-\mathcal Ku_\varepsilon]_+\|_2 \le\sqrt{2\varepsilon(c+C\varepsilon)}, \label{eq:linear-smoothing-bound}
\end{equation}
where $C=c^2(A_0+A_1^2\|L\|_2^2)/2$. Along a weakly convergent subsequence, every fixed test integral converges. Fatou's lemma makes the limit feasible almost everywhere in the test measure; continuity and full support extend feasibility to every test. Its mass is at least $c$, while $t(u_\varepsilon)\le J_\varepsilon(u_\varepsilon)\le c+C\varepsilon$. Thus all limiting masses equal $c$, and every weak limit is an optimal ensemble.

Finally, put $q_\varepsilon=1-\delta_\varepsilon/L_\varepsilon$ and $e_j=\|u_j-u_\varepsilon\|_2$. The integral projection error and the contraction give
\[
e_{j+1}\le q_\varepsilon e_j+\sqrt{d_v/\beta_\varepsilon},\qquad e_N\le q_\varepsilon^NR_\varepsilon+ \frac{L_\varepsilon}{\delta_\varepsilon}\sqrt{d_v/\beta_\varepsilon}.
\]
Each term is at most $\varepsilon/2$ for the displayed choices of $N_\varepsilon$ and $\beta_\varepsilon$. Since $\omega$ has unit mass, the density error bounds the mass error as well. The same argument gives
\[
\|[N-\mathcal Ku_{N_\varepsilon}]_+\|_2 \le\sqrt{2\varepsilon(c+C\varepsilon)}+M\varepsilon, \qquad t(u_{N_\varepsilon})\le c+(C+1)\varepsilon.
\]
\end{proof}
Some scalar qubit test coordinates alter the available response without changing the source. Write these heights as $\alpha\in\mathcal A_z$, with compact domain varying continuously in the retained test $z$. Since $N(z)$ is independent of $\alpha$, all height constraints are equivalent to the minimum-response constraint
\begin{equation}
K(z;\mu)=\min_{\alpha\in\mathcal A_z}\int k(z,\alpha;x)d\mu(x), \qquad c=\inf_{\substack{\mu\ge0,\ \int\vec v\,d\mu=0\\
K(z;\mu)\ge N(z)\ \text{for all }z}} \mu(\mathcal X). \label{eq:general-concave-kernel}
\end{equation}
Each fixed-height response is linear in the measure, so its minimum $K(z;\mu)$ is concave. At the current density $u$, a height $\alpha$ attaining the minimum defines the linear kernel $k_{z,u}(x)=k(z,\alpha;x)$. Its integral $K(z;u)=\int k_{z,u}(x)u(x)d\omega(x)$ is the minimum response. In the qubit formulas below, region-moment equations determine these heights. The selected kernel, or a convex combination when several heights attain the minimum, is a supergradient of $K(z;\cdot)$. It therefore replaces the fixed kernel in the deficit integral. Assuming the preceding bounds uniformly in $\alpha$, set
\begin{equation}
r(u;z)=[N(z)-K(z;u)]_+, \qquad G(u)(x)=\int k_{z,u}(x)r(u;z)d\sigma(z), \label{eq:general-eliminated-residual}
\end{equation}
Here $-G(u)$ is a subgradient of $\|r(u)\|_2^2/2$. Balancing it against mass and the marginal gives the reduced density equation. The next recurrence incorporates the active kernel's dependence on $u$.
\begin{theorem}
\label{thm:general-integral-root} Under the preceding assumptions, choose $D\ge\dim\mathcal X$ and set
\[
\delta_\varepsilon=\varepsilon^{D+1},\qquad \beta_\varepsilon=\varepsilon^{-(2D+12)},\qquad N_\varepsilon=\lceil\varepsilon^{-(2D+6)}\rceil, \qquad\gamma_j=\frac1{\delta_\varepsilon(j+2)}.
\]
The functional $J_\varepsilon$ formed with the reduced residual in Eq.~\eqref{eq:general-eliminated-residual} has a unique minimizer $u_\varepsilon^*\in\mathcal C$, whose mass tends to $c$. Starting from $u_0=0$, define
\begin{equation}
u_{j+1}=B_{\beta_\varepsilon}\!\left[ (1-\gamma_j\delta_\varepsilon)u_j-\gamma_j+ \frac{\gamma_j}{\varepsilon}G(u_j)\right], \qquad j=0,\ldots,N_\varepsilon-1. \label{eq:general-root-iteration}
\end{equation}
These are nonnegative zero-moment densities, and
\begin{equation}
c=\lim_{\varepsilon\downarrow0}\int u_{N_\varepsilon}\,d\omega. \label{eq:general-finite-integral-value}
\end{equation}
\end{theorem}
\begin{proof}
Use the same functional $J_\varepsilon$ with the reduced residual. Convexity of its squared deficit and the quadratic term give a unique minimizer $u_\varepsilon^*$. The source and kernel bounds imply
\begin{equation}
\|1-\varepsilon^{-1}G(u)\|_2\le L_\varepsilon, \qquad L_\varepsilon=1+\varepsilon^{-1}\|H\|_2\|N\|_2. \label{eq:general-subgradient-bound}
\end{equation}
At the minimizer an appropriate choice of the active-kernel supergradient gives $h_\varepsilon^*=1-\varepsilon^{-1}G(u_\varepsilon^*)$ satisfying
\[
\langle h_\varepsilon^*+\delta_\varepsilon u_\varepsilon^*, w-u_\varepsilon^*\rangle\ge0\quad(w\in\mathcal C).
\]
Taking $w=0$ and $w=2u_\varepsilon^*$ bounds the norm by $\delta_\varepsilon\|u_\varepsilon^*\|_2\le L_\varepsilon$. Uniform Lipschitz control of active kernels gives the same heat-smoothing bound for their minimum. The preceding mass-limit argument therefore yields $t(u_\varepsilon^*)\to c$.

For the finite recurrence, put $h_j=1-\varepsilon^{-1}G(u_j)$ and $D_j=\|u_j-u_\varepsilon^*\|_2^2$. Equation~\eqref{eq:general-integral-projection}, the variational inequality at the minimizer, and monotonicity of subgradients give
\[
\begin{aligned}
D_{j+1} &\le\|(1-\gamma_j\delta_\varepsilon)(u_j-u_\varepsilon^*) -\gamma_j(h_j-h_\varepsilon^*)\|_2^2 +\frac{2d_v}{\beta_\varepsilon}\\
&\le\left(\frac{j+1}{j+2}\right)^2D_j +\frac{4L_\varepsilon^2}{\delta_\varepsilon^2(j+2)^2} +\frac{2d_v}{\beta_\varepsilon}.
\end{aligned}
\]
Multiplication by $(j+2)^2$ and summation yield
\[
D_N\le\frac{D_0+4NL_\varepsilon^2/\delta_\varepsilon^2}{(N+1)^2} +\frac{2d_v}{\beta_\varepsilon(N+1)^2}\sum_{k=2}^{N+1}k^2, \qquad D_0\le L_\varepsilon^2/\delta_\varepsilon^2.
\]
With the stated exponents, $D_{N_\varepsilon}=O(\varepsilon^2)$; hence finite iteration and the regularized minimum have the same mass limit.
\end{proof}
The attaining branches in the height equations supply the kernels used in $G(u_j)$. Recalculating the heights after each density update gives Eq.~\eqref{eq:general-root-iteration}. Height elimination preserves the feasible measures, so the recurrence converges to the same minimum mass.

A compact unitary symmetry group $\mathsf U$ can reduce the integrations further. Suppose it preserves the source and test measure and satisfies $k(Uz,U\alpha;Ux)=k(z,\alpha;x)$. Group averaging preserves spectral mass and moment, and concavity preserves feasibility. It also decreases $J_\varepsilon$, so its unique minimizer is invariant. The projection $B_\beta$ commutes with this action, and group-averaged subgradients keep every iterate invariant. For a linear kernel the resulting expression is
\begin{equation}
\overline\kappa([z],[x])=\int_{\mathsf U}\kappa(z,Ux)dU, \qquad \overline Ku([z])=\int_{\mathcal X/\mathsf U} \overline\kappa([z],[x])u([x])d\overline\omega([x]), \label{eq:general-symmetry-kernel}
\end{equation}
where $dU$ is normalized Haar measure and $\overline\omega$ is the measure on orbits. For eliminated coordinates the averaging applies to each linear kernel before their minimum is taken. Heat smoothing commutes with the symmetry, so the same mass convergence holds on the reduced domain. The qubit reductions below specify the source and all height integrals needed in the density recurrence.

\begin{figure}[H]
\centering\includegraphics[width=.92\textwidth]{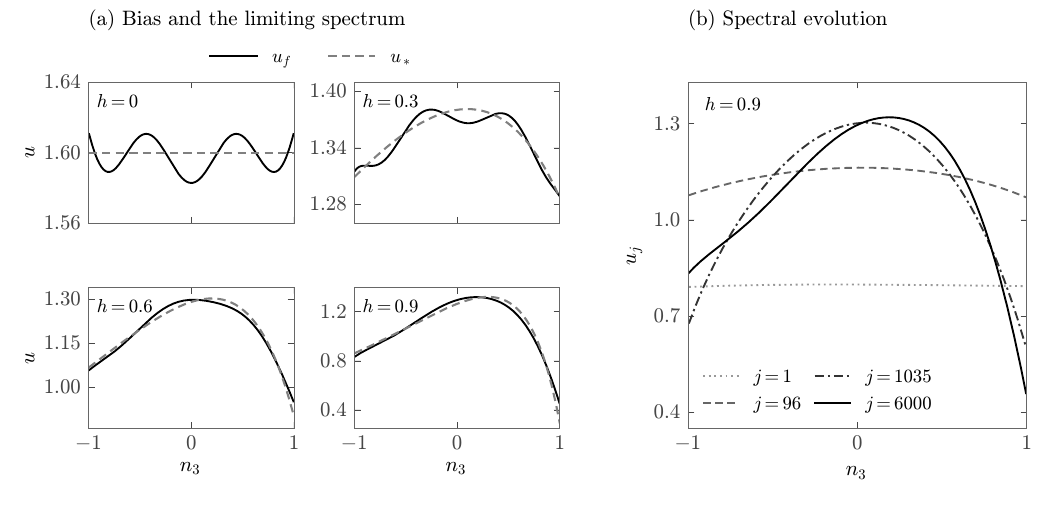}
\caption{Spectral evolution for the biased family at visibility $v=0.8$. (a) Finite iterates $u_f$ (solid) and analytic densities $u_*$ (dashed) at four relative biases $h$. Each vertical range resolves the variation of that spectrum. (b) Successive densities for $h=0.9$, showing how the iteration forms unequal angular tails while keeping the first moment zero.}\label{fig:sm-spectral-evolution}
\end{figure}

For axial qubit spectra, azimuthal integration reduces the response to an integral over $z=n_3$. A positive finite basis gives a direct discretization of the same functional:
\begin{equation}
u(z)=\sum_{k=0}^d b_k B_k(z),\qquad B_k(z)=(d+1)\binom dk\left(\frac{1+z}{2}\right)^k\left(\frac{1-z}{2}\right)^{d-k}.
\label{eq:bernstein-spectral-representation}
\end{equation}
Relative to $dz/2$, each $B_k$ has mass one and first moment $(2k-d)/(d+2)$. Thus $b_k\ge0$ ensures positivity, while $\sum_k b_k(2k-d)=0$ fixes the first moment. Quadrature forms the response matrix and density Gram matrix. An accelerated projected-gradient iteration, with restart and a step set by a Hessian bound, minimizes the resulting finite convex functional. Projection enforces the nonnegative coefficients and their single moment constraint. Figure~\ref{fig:sm-spectral-evolution} follows these densities and compares them with the analytic solution in Section~S5.

\subsection{Bounds from finite measurement families} \label{sec:finite-global-bounds}

Finite measurement families give parent models and dual lower bounds for approximating the continuous class.

For a nonempty finite family $D=\{M^1,\ldots,M^N\}$, a parent can be resolved into simultaneous outcome strings $\alpha=(\alpha_1,\ldots,\alpha_N)$. Its positive operators $H_\alpha$ obey
\begin{equation}
\sum_\alpha H_\alpha=t\openone_n, \qquad \sum_{\alpha:\,\alpha_x=a}H_\alpha =q_{a|M^x}t\openone_n+A(X_{a|M^x}). \label{eq:s4-finite-primal}
\end{equation}
Minimizing $t\ge0$ subject to these constraints gives $c_D(A)$. A parent for $A/t$ gives
\[
H_\alpha=t\int\prod_xp(\alpha_x|M^x,\lambda)dH(\lambda).
\]
Conversely, dividing the operators in Eq.~\eqref{eq:s4-finite-primal} by $t>0$ gives a parent with deterministic responses. At $t=0$ they all vanish, and feasibility means that $A$ vanishes on the centered effects of $D$.

The dual maximizes the linear functional
\begin{equation}
\Phi(A')=\sum_{x,a}\Tr[Y_a^xA'(X_{a|M^x})] \label{eq:s4-dual-functional}
\end{equation}
over Hermitian $Y_a^x$ and $\Omega\in\mathcal D_n$ satisfying
\begin{equation}
\sum_aq_{a|M^x}Y_a^x=0, \qquad \Omega-\sum_xY_{\alpha_x}^x\succeq0 \quad\text{for every }\alpha. \label{eq:s4-finite-dual}
\end{equation}
This is Eq.~\eqref{eq:s2-multiplier-problem} with a finite measurement family. To see the equivalence, group its multiplier measure by measurement and integrate $Y_a-\sum_bq_{b|M^x}Y_b$ within each group. The objective is preserved, while convexity of the maximum preserves feasibility. Conversely, centering and normalizing each finite family $(Y_a^x)_a$ gives an atomic multiplier measure. Its kernel constraint is precisely Eq.~\eqref{eq:s4-finite-dual}, since
\[
\sum_x\max_a\Tr(Y_a^x\tau) =\max_\alpha\Tr\!\left[\left(\sum_xY_{\alpha_x}^x\right)\tau\right].
\]
Compactness gives both extrema. Bounded $t$ bounds every positive $H_\alpha$. For dual compactness, omit zero effects and average the matrix inequalities with $q_{a|M^x}$ to obtain
\[
-\frac{1-q_{a|M^x}}{q_{a|M^x}}\Omega \preceq Y_a^x\preceq\Omega.
\]
Both optimizing sequences therefore lie in compact sets. Restore omitted effects with zero operators; averaging over nonzero outcomes supplies their additional dual inequalities.

Write
\begin{equation}
\mathcal B_{\mathfrak M} =\{A\in\mathcal V_{m,n}:c_{\mathfrak M}(A)\le1\}. \label{eq:s4-unit-body}
\end{equation}
Measurement restriction enlarges this closed convex set. Moreover, the finite restrictions determine it completely:
\begin{align}
\mathcal B_{\mathfrak M} &=\bigcap_{D\subseteq\mathfrak M,\ D\text{ finite}}\mathcal B_D, \label{eq:s4-body-finite-completeness}\\
c_{\mathfrak M}(A) &=\sup_{D\subseteq\mathfrak M,\ D\text{ finite}}c_D(A). \label{eq:s4-finite-completeness}
\end{align}
For the converse inclusion, the positive measures with operator moment $\openone_n$ form a weakly compact set. Admitting responses for a fixed measurement is a closed condition, by the outcome-measure argument in Lemma~\ref{lem:parent-model-maps}. Feasibility for every finite $D$ therefore gives a common measure by the finite-intersection property. Applying this set equality to $A/t$ gives the scale equality.

Let $m\ge2$, and take a finite $D$ whose centered effects span $\mathcal H_m^0$. The bounds
\[
-q_{a|M}\openone_n\preceq A(X_{a|M}) \preceq(1-q_{a|M})\openone_n\qquad(A\in\mathcal B_D)
\]
then make $\mathcal B_D$ compact and $c_D(A)>0$ for $A\ne0$. Choose finitely many maps $V_i$ on its boundary and corresponding optimal dual functionals $\Phi_i$, so that
\begin{equation}
c_D(V_i)=\Phi_i(V_i)=1. \label{eq:s4-finite-contact}
\end{equation}
Their parent models give convex-combination upper bounds, and their functionals give lower bounds:
\begin{align}
L(A)&=\max\{0,\max_i\Phi_i(A)\}, \label{eq:s4-polyhedral-lower}\\
U(A)&=\inf\left\{\sum_i\theta_i: A=\sum_i\theta_iV_i,\quad\theta_i\ge0\right\}. \label{eq:s4-polyhedral-upper}
\end{align}
Here $U=+\infty$ when the decomposition is impossible. Convexity and homogeneity give $L\le c_D\le U$; evaluating these bounds requires a finite maximum and a linear program.
\begin{theorem}
\label{thm:s4-global-carrier} Suppose that the centered effects of $D$ span $\mathcal H_m^0$. If $\{A:L(A)\le1\}$ is bounded, let $W_j$ be its vertices. For $\kappa\ge1$, the finite conditions
\begin{equation}
W_j=\sum_i\theta_{ij}V_i, \qquad \theta_{ij}\ge0, \qquad \sum_i\theta_{ij}\le\kappa \quad\text{for every }j \label{eq:s4-carrier-factor}
\end{equation}
are equivalent to
\begin{equation}
L(A)\le c_D(A)\le U(A)\le\kappa L(A) \qquad(A\in\mathcal V_{m,n}). \label{eq:s4-global-body-bracket}
\end{equation}
For every $\kappa>1$, a finite choice of maps and optimal functionals satisfying these conditions exists.
\end{theorem}
\begin{proof}
Every point with $L\le1$ is a convex combination of the $W_j$, so Eq.~\eqref{eq:s4-carrier-factor} gives $U\le\kappa$ there. Boundedness makes $L(A)>0$ for $A\ne0$, and rescaling by $L(A)$ proves the bound on every ray. Conversely, $L(W_j)=1$ at each vertex, so the bound implies Eq.~\eqref{eq:s4-carrier-factor}. Thus $\max_jU(W_j)$ is the least factor.

For existence, $c_D$ is continuous and its unit level set is compact. The contracted body $\kappa^{-1/2}\mathcal B_D$ lies in the interior of $\mathcal B_D$. Cover it by finitely many small cubes contained in that interior, and normalize the nonzero vertices of these cubes onto $c_D=1$. The resulting $V_i$ satisfy
\[
\kappa^{-1/2}\mathcal B_D \subseteq\operatorname{conv}(\{0\}\cup\{V_i\}_i) \subseteq\mathcal B_D,
\]
which gives $c_D\le U\le\sqrt\kappa\,c_D$. At each point of $c_D=1$, an optimal functional equals one and remains greater than $\kappa^{-1/2}$ in a neighborhood. A finite subcover supplies functionals with $\kappa^{-1/2}c_D\le L\le c_D$. Take the union of the two finite collections, retaining optimal parents and functionals at every selected map. Adding data only improves both bounds. Thus $U\le\kappa L$, and $\{L\le1\}\subseteq\sqrt\kappa\,\mathcal B_D$ is bounded.
\end{proof}
Keeping earlier data as $\kappa\downarrow1$ makes $L$ increase and $U$ decrease uniformly on $c_D=1$, with the same relative bound on every nonzero ray. The construction and vertex criterion also admit feasible parents and dual functionals at different input maps.

To pass from finite measurements to all $r$-outcome POVMs, let
\[
\mathfrak X_{m,r} =\{(M_1,\ldots,M_r):M_a\succeq0,\ \sum_aM_a=\openone_m\}
\]
and contract the traceless parts by
\begin{equation}
(\mathsf S_\eta M)_a =q_{a|M}\openone_m+\eta X_{a|M} \qquad(0<\eta\le1). \label{eq:s4-measurement-shrink}
\end{equation}
\begin{proposition}
\label{thm:s4-shrinking-lift} If a finite $D\subseteq\mathfrak X_{m,r}$ satisfies
\begin{equation}
\mathsf S_\eta(\mathfrak X_{m,r})\subseteq\operatorname{conv}D, \label{eq:s4-shrink-containment}
\end{equation}
then
\begin{equation}
\eta\mathcal B_D\subseteq\mathcal B_{\mathfrak X_{m,r}} \subseteq\mathcal B_D, \qquad c_D\le c_{\mathfrak X_{m,r}}\le c_D/\eta. \label{eq:s4-shrinking-body}
\end{equation}
Such a finite $D$ exists for every $0<\eta<1$.
\end{proposition}
\begin{proof}
A parent for $D$ also models $\operatorname{conv}D$ by mixing response probabilities. Under Eq.~\eqref{eq:s4-shrink-containment}, it therefore models every $\mathsf S_\eta M$. Its outcome operators are $q_{a|M}\openone_n+(\eta A)(X_{a|M})$, proving the first inclusion; restriction gives the second.

For existence, apply the finite-cube construction above in the trace-one affine space to choose density operators $\tau_j$ with
\begin{equation}
\{\eta\tau+(1-\eta)\openone_m/m:\tau\in\mathcal D_m\} \subseteq\operatorname{conv}\{\tau_j\}_j\subseteq\mathcal D_m. \label{eq:s4-density-polytope}
\end{equation}
The nonnegative arrays $w_{aj}$ satisfying $\sum_{a,j}w_{aj}\tau_j=\openone_m$ form a compact polytope, since their total weight is $m$. Each vertex defines a POVM $M_a^x=\sum_jw_{aj}^{(x)}\tau_j$. For an arbitrary $M$, apply Eq.~\eqref{eq:s4-density-polytope} to each normalized nonzero effect $M_a/(mq_{a|M})$. The resulting coefficients express $\mathsf S_\eta M$ as a point of this polytope, hence as a convex combination of its vertex POVMs.
\end{proof}
For a given $D$, its finite facets determine the admissible contraction. Write them as $\sum_a\Tr[F_a^{(j)}M_a]\le b_j$, including affine equalities as pairs of inequalities. Maximize each facet over contracted POVMs and take its SDP dual:
\begin{equation}
\exists Z_j\in\mathcal H_m:\quad \Tr Z_j\le b_j, \qquad Z_j\succeq\eta F_a^{(j)} +(1-\eta)\frac{\Tr F_a^{(j)}}m\openone_m \quad\text{for every }a,j. \label{eq:s4-measurement-factor-test}
\end{equation}
Strict feasibility at $M_a=\openone_m/r$ makes this equivalent to Eq.~\eqref{eq:s4-shrink-containment}. Maximizing $0\le\eta\le1$ gives the largest contraction factor supported by $D$.

Combining the two approximations yields
\begin{equation}
L(A)\le c_{\mathfrak X_{m,r}}(A) \le U(A)/\eta\le(\kappa/\eta)L(A). \label{eq:s4-shrinking-bracket}
\end{equation}
Thus finite directions cost a factor $\kappa$, and finite measurement coverage a factor $\eta^{-1}$.
\begin{corollary}
\label{cor:s4-arbitrary-relative-accuracy} For $r\ge2$ and every $\varepsilon>0$, finite measurements and finite parent and dual data can be chosen so that
\begin{equation}
L(A)\le c_{\mathfrak X_{m,r}}(A) \le U(A)/\eta\le(1+\varepsilon)L(A) \qquad(A\in\mathcal V_{m,n}). \label{eq:s4-arbitrary-relative-accuracy}
\end{equation}
\end{corollary}
\begin{proof}
Choose $(1+\varepsilon)^{-1}<\eta<1$ and a finite $D$ as in Proposition~\ref{thm:s4-shrinking-lift}. Add sufficiently small binary perturbations of $\openone_m/2$ in a basis of $\mathcal H_m^0$, padded with zeros, so that its centered effects span this space. Theorem~\ref{thm:s4-global-carrier} with $\kappa=\eta(1+\varepsilon)>1$ then gives the result.
\end{proof}
Taking $r=m^2$ and using Eq.~\eqref{eq:s4-all-povm-reduction} gives the same finite relative bounds for $c_{\mathrm A}$.

\subsection{Qubit response kernels} \label{subsec:spectral-pvm}

After whitening Bob, a two-qubit state is described by a Bloch vector $\vec a$ and a real correlation matrix $C$:
\begin{equation}
\begin{aligned}
\widehat\rho&=\frac14\left[\openone_2\otimes\openone_2+ \vec a\cdot\vec\sigma\otimes\openone_2+ \sum_{i,j=1}^3C_{ij}\sigma_i\otimes\sigma_j\right],\\
A_{\vec a,C}\!\left(\frac{\vec x\cdot\vec\sigma}{2}\right) &=\frac12\left[(\vec a\cdot\vec x)\openone_2+ (C^{\mathsf T}\vec x)\cdot\vec\sigma\right].
\end{aligned}
\label{eq:qubit-data}
\end{equation}
Write $c_{\mathrm P}$ and $c_{\mathrm A}$ for the PVM and POVM scales. Pure hidden states are $\tau_{\vec n}=(\openone_2+\vec n\cdot\vec\sigma)/2$, $\vec n\in S^2$, with normalized area $d\omega=dS/(4\pi)$.

For the projectors $P_\pm(\vec e)=(\openone_2\pm\vec e\cdot\vec\sigma)/2$, the difference between the two conditional operators is determined by $(\vec a\cdot\vec e,C^{\mathsf T}\vec e)$. The scalar component gives the imbalance of outcome probabilities; the vector component gives the Bloch response. In an LHS model these quantities come from the response difference $f_{\vec e}=p(+|\vec e,\vec n)-p(-|\vec e,\vec n)$, with $|f_{\vec e}|\le1$:
\[
\int f_{\vec e}(\vec n)(1,\vec n)\,d\mu =(\vec a\cdot\vec e,C^{\mathsf T}\vec e).
\]
A test $(s,\vec y)$ assigns the hidden direction $\vec n$ the score $f_{\vec e}(\vec n)(s+\vec y\cdot\vec n)$. Its largest value is $|s+\vec y\cdot\vec n|$, attained by choosing the sign of the parenthesis. Integrated over the spectrum, this is the support function of all conditional differences it supplies. Domination of the target score for every test is therefore equivalent to the existence of a response function. Maximizing that target score, $\vec e\cdot(\vec a s+C\vec y)$, over the projective direction gives $|\vec a s+C\vec y|$ and removes the measurement direction as well:
\begin{equation}
\begin{gathered}
N_{\mathrm P}(s,\vec y)=|\vec a s+C\vec y|,\qquad k_{\mathrm P}(s,\vec y;\vec n)=|s+\vec y\cdot\vec n|,\\
(\mathcal K\mu)(s,\vec y)=\int k_{\mathrm P}(s,\vec y;\vec n)d\mu(\vec n),\\
c_{\mathrm P}=\min\{\mu(S^2):\mu\ge0,\ \int\vec n\,d\mu=0,\ \mathcal K\mu\ge N_{\mathrm P}\}.
\end{gathered}
\label{eq:pvm-support}
\end{equation}
Homogeneity restricts the tests to $(s,\vec y)\in S^3$. Use normalized area on this sphere as the test measure. With $N=N_{\mathrm P}$ and $k=k_{\mathrm P}$, Eq.~\eqref{eq:linear-spectral-equation} gives the PVM scale at every rank of $C$. The kernel is at most $\sqrt2$ and its hidden-state Lipschitz constant is at most one, so the hypotheses of that theorem hold uniformly.

For $r=\rank C\ge2$, only the subspace $E=\operatorname{Ran}C^{\mathsf T}$ is needed. Choose an orthogonal coordinate map $V:\R^r\to E$, taking $V=\openone_3$ at full rank, and write
\begin{equation}
\vec n=V\vec m,\qquad \vec m\in S^{r-1},\qquad C_r=CV. \label{eq:qubit-rank-reduction}
\end{equation}
Indeed, restricting tests to $E$ projects a hidden-state spectrum onto the unit ball of $E$. Every point $\vec x$ of that ball is the mean of the antipodal points $\pm\vec x/|\vec x|$ with weights $(1\pm|\vec x|)/2$; at zero any equal antipodal pair suffices. This replacement preserves mass and first moment and increases a convex kernel. The resulting spectrum lies on the sphere in $E$, where an arbitrary test depends only on its projection onto $E$, as does the source. The reduced and original problems therefore have the same minimum. This argument also applies to the maximum of affine functions in a POVM kernel.

Absorb the bias component in $\operatorname{Ran}C$ into the vector test by defining
\begin{equation}
\vec p_0=C_r^+\vec a,\qquad \vec a_\perp=\vec a-C_r\vec p_0,\qquad \vec y=V(\vec x-s\vec p_0),\qquad g(\vec m)=1-\vec p_0\cdot\vec m, \label{eq:pvm-inactive-test-change}
\end{equation}
where $+$ denotes the Moore--Penrose inverse. The source and kernel become
\begin{equation}
\widetilde N_{\mathrm P}(s,\vec x) =\sqrt{|C_r\vec x|^2+|\vec a_\perp|^2s^2},\qquad \widetilde k_{\mathrm P}(s,\vec x;\vec m) =|\vec x\cdot\vec m+s g(\vec m)|. \label{eq:pvm-singular-sheared-source}
\end{equation}
If $\vec a_\perp\ne0$, the tests remain $(s,\vec x)\in S^r$. If $\vec a_\perp=0$, the source depends only on $\vec x$, and the scalar height $s$ can be eliminated. In particular, this elimination applies to every invertible $C$.

Let $u\ge0$ be a zero-moment density on $S^{r-1}$, of mass $t$. All sphere measures below have total mass one. For a fixed $\vec x\in S^{r-1}$, set
\begin{equation}
\begin{aligned}
F_u(s;\vec x)&=\int|\vec x\cdot\vec m+s g(\vec m)|u(\vec m)d\omega_{r-1},\\
D_u(s;\vec x)&=\int g(\vec m)\operatorname{sgn}[\vec x\cdot\vec m+s g(\vec m)] u(\vec m)d\omega_{r-1}.
\end{aligned}
\label{eq:pvm-height-derivative}
\end{equation}
Fix $u$ and $\vec x$. Changing $s$ moves the plane separating the two outcome regions, whose signed $g$-weighted mass difference is $D_u$. Convexity of $F_u$ makes $D_u$ nondecreasing and places the smallest response at its zero. The moment identities $\int gu=t$ and $\int\vec m u=0$ give $F_u(s;\vec x)\ge t|s|$ by Jensen inequality, while $F_u(0;\vec x)\le t$. Hence a minimum occurs in $[-1,1]$.

Before the minimizing heights the derivative is negative; after them it is positive. Integrating its sign on $[-1,1]$ selects their midpoint in that interval:
\begin{equation}
\zeta_u(\vec x)=-\frac12\int_{-1}^1\operatorname{sgn}D_u(s;\vec x)\,ds. \label{eq:pvm-median-integral}
\end{equation}
Integrating $|D_u|$ from the two endpoints to this height gives the decrease of $F_u$ on both sides. The minimum response is consequently
\begin{equation}
F_u(\zeta_u(\vec x);\vec x)= \frac{F_u(-1;\vec x)+F_u(1;\vec x)}2 -\frac12\int_{-1}^1|D_u(s;\vec x)|\,ds. \label{eq:pvm-median-integrated-kernel}
\end{equation}
A sphere integral evaluates $D_u$; its sign integral selects the height; Eq.~\eqref{eq:pvm-median-integrated-kernel} then gives the minimum response entering the residual.

For $u>0$ almost everywhere, this height is unique. At a zero of $D_u$, the affine argument takes both signs on the sphere. Its zero set cannot coincide with $g=0$, because the constant terms would then force $\vec x=0$. A crossing therefore occurs where $g\ne0$. Changing $s$ moves this crossing through a set of positive weight, so $D_u$ cannot vanish on an interval. For a nonnegative $u$, Eq.~\eqref{eq:pvm-median-integral} still selects an attaining height.

The retained test is $\xi=\vec x$ when $\vec a_\perp=0$ and $\xi=(s,\vec x)$ otherwise. In either case $h_u(\xi;\vec m)$ is the linear kernel attained at that test. The source, kernel, and test space are
\begin{equation}
\begin{array}
{c|c|c} &\vec a_\perp=0&\vec a_\perp\ne0\\
\hline \mathcal Q_{\mathrm P}&S^{r-1}&S^r\\
\mathcal S_{\mathrm P}(\xi)&|C_r\vec x|&\widetilde N_{\mathrm P}(s,\vec x)\\
(\mathcal K_{\mathrm P}u)(\xi)&F_u(\zeta_u(\vec x);\vec x)&F_u(s;\vec x)\\
h_u(\xi;\vec m)&|\vec x\cdot\vec m+\zeta_u(\vec x)g(\vec m)| &\widetilde k_{\mathrm P}(s,\vec x;\vec m)
\end{array}
\label{eq:pvm-median-kernel}
\end{equation}
In either column, a zero-moment spectrum is feasible exactly when $\mathcal K_{\mathrm P}\mu\ge\mathcal S_{\mathrm P}$ for all retained tests. With normalized area $d\sigma_{\mathrm P}$ on $\mathcal Q_{\mathrm P}$, its attaining kernel gives
\begin{equation}
r_u(\xi)=[\mathcal S_{\mathrm P}(\xi)-(\mathcal K_{\mathrm P}u)(\xi)]_+, \qquad (\mathcal G_{\mathrm P}u)(\vec m)= \int_{\mathcal Q_{\mathrm P}}h_u(\xi;\vec m)r_u(\xi)d\sigma_{\mathrm P}(\xi). \label{eq:pvm-reduced-source-integral}
\end{equation}
With $G=\mathcal G_{\mathrm P}$ and $D=2$, Eq.~\eqref{eq:general-root-iteration} has mass limit $c_{\mathrm P}$ as in Eq.~\eqref{eq:general-finite-integral-value}. Both retained test spaces use the source and kernel in Eq.~\eqref{eq:pvm-median-kernel}. On the analytic domain of Section~S4, the minimum response reaches the source; equality and height stationarity determine the spectrum and contact tests together.

The source is bounded, and $2+|\vec p_0|$ bounds both kernels and their hidden-state Lipschitz constants. Restricting heights to $[-1,1]$ gives continuity. Zero heights initialize $u_0=0$, and subsequent recurrence densities are positive. The mass theorem therefore applies. For ranks zero and one, use Eq.~\eqref{eq:pvm-support}; Section~S4 evaluates their elementary scale.

\label{subsec:spectral-povm}

For general qubit POVMs, source maximization and outcome divisions involve at most four branches.

Four outcomes suffice for all qubit POVMs. Write their effects and Hermitian tests as
\[
M_i=q_i\openone_2+\vec x_i\cdot\vec\sigma,\qquad Y_i=(\alpha_i\openone_2+\vec\beta_i\cdot\vec\sigma)/2, \qquad i=1,\ldots,4.
\]
The measurement conditions are $|\vec x_i|\le q_i$, $\sum_iq_i=1$, and $\sum_i\vec x_i=0$. At fixed weights $q_i$, the vectors $\vec x_i$ occur only in the source. With $\vec z_i=\alpha_i\vec a+C\vec\beta_i$, their elimination gives
\begin{equation}
\max_{\substack{|\vec x_i|\le q_i\\
\sum_i\vec x_i=0}} \sum_i\vec x_i\cdot\vec z_i =\min_{\vec w}\sum_iq_i|\vec z_i-\vec w|. \label{eq:povm-measurement-elimination}
\end{equation}
Indeed, subtracting any $\vec w$ from all source points leaves the left-hand sum unchanged and bounds it by the distance sum on the right. At a minimizing center, the subgradient condition gives $\vec s_i\in\partial|\cdot|(\vec z_i-\vec w)$ with $\sum_iq_i\vec s_i=0$. The measurement vectors $\vec x_i=q_i\vec s_i$ attain that bound.

A common test shift cancels from the centered inequality. Homogeneity fixes $\sum_i\xi_i=0$ and $\sum_i|\xi_i|^2=1$, where $\xi_i=(\alpha_i,\vec\beta_i)$, so $\xi\in S^{11}$. For $q_i>0$, write the source as
\begin{equation}
\mathcal S(q,\xi)=\min_{\vec w}\sum_iq_i|\vec z_i-\vec w|, \qquad \vec z_i=\alpha_i\vec a+C\vec\beta_i, \label{eq:povm-fixed-source}
\end{equation}
At each hidden direction the response probabilities divide its weight among the four outcomes. Their largest test score is the maximum of the affine scores. Subtracting the $q$-weighted mean centers this response:
\begin{equation}
\mathcal K(q,\xi;\mu)=\int_{S^2} \left[\max_i(\alpha_i+\vec\beta_i\cdot\vec n) -\sum_iq_i(\alpha_i+\vec\beta_i\cdot\vec n)\right]d\mu(\vec n). \label{eq:povm-singular-kernel}
\end{equation}
Minimizing the mass of a positive zero-moment measure subject to $\mathcal K(q,\xi;\mu)\ge\mathcal S(q,\xi)$ gives the POVM scale. Source maximization accounts for all allowed effect vectors; the kernel accounts for assigning hidden weight to four outcomes. Uniform probability on the closed weight simplex and normalized area on $S^{11}$ give the test measure. A kernel bound of two, uniform hidden-state Lipschitz continuity and a continuous source, including zero weights, permit Eq.~\eqref{eq:linear-spectral-equation} at every rank. We next evaluate the distance source and integrate out its unused heights.

To evaluate the source, first combine coincident $\vec z_i$ and add their weights, leaving $m\le4$ distinct points with positive weights $(\vec z_i,q_i)$. This grouping is used only for the source; the spectral kernel retains its original outcome branches. Put $\delta_{ij}=|\vec z_i-\vec z_j|^2$. If $m=1$, the source is zero.

A minimizing center may coincide with a source point $\vec z_k$. The remaining terms then contribute subgradient $\sum_{j\ne k}q_j(\vec z_k-\vec z_j)/|\vec z_k-\vec z_j|$, whose squared norm is
\begin{equation}
\Gamma_k=\sum_{i,j\ne k}q_iq_j \frac{\delta_{ik}+\delta_{jk}-\delta_{ij}}{2\sqrt{\delta_{ik}\delta_{jk}}}. \label{eq:povm-source-contact}
\end{equation}
Adding the radius-$q_k$ subgradient ball of the coincident term includes zero precisely when $\Gamma_k\le q_k^2$. Then $\vec z_k$ is a minimizing center and
\begin{equation}
\mathcal S=\sum_{j\ne k}q_j\sqrt{\delta_{jk}}. \label{eq:povm-source-contact-value}
\end{equation}
Collinear points always have a weighted median at a source point, so this condition also resolves all collinear configurations.

Otherwise all centers avoid the source points, which must be noncollinear. Equality in convexity of the distance sum between two centers would force every source point onto their common line. Strict convexity therefore gives a unique center. With $r_i=|\vec z_i-\vec w|$, zero gradient gives
\[
\vec w=\frac{\sum_i(q_i/r_i)\vec z_i}{\sum_iq_i/r_i}.
\]
Set $\lambda_i=q_ir_m/(q_mr_i)$, so that $\lambda_m=1$, and write
\begin{equation}
T=\sum_i\lambda_i,\qquad R_i=\left|\sum_j\lambda_j(\vec z_i-\vec z_j)\right|^2. \label{eq:povm-source-quartic-data}
\end{equation}
Because $\vec w=\sum_i\lambda_i\vec z_i/T$, the quadratic expression $R_i$ equals $T^2|\vec z_i-\vec w|^2=T^2r_i^2$. Substituting $\lambda_i$ eliminates distances and leaves $m-1$ quartic equations:
\begin{equation}
q_m^2\lambda_i^2R_i-q_i^2R_m=0, \qquad \lambda_i>0,\quad i=1,\ldots,m-1. \label{eq:povm-source-quartic-root}
\end{equation}
Conversely, a positive root defines a barycenter and obeys $R_i=T^2|\vec z_i-\vec w|^2$. If $R_m=0$, the equations force all distances to vanish, contradicting distinctness. Every positive root therefore recovers the same stationary center and the same distance ratios, so the positive root is unique.

Recovering $r_i=\sqrt{R_i}/T$ gives the source value
\begin{equation}
\mathcal S=\frac1T\sum_iq_i\sqrt{R_i}. \label{eq:povm-source-algebraic-value}
\end{equation}
Thus source-point tests or the unique positive quartic root evaluate the source. Projecting a center onto the convex hull decreases every distance, so minimization takes place in that compact hull and depends continuously on points and weights.

For $r=\rank C\ge2$, apply the rank reduction \eqref{eq:qubit-rank-reduction}, with the same $C_r$, $\vec p_0$, $\vec a_\perp$, and $g$ used for PVMs. If $\vec a_\perp\ne0$, retain $\xi_i=(\alpha_i,\vec\beta_i)\in\R^{r+1}$ with zero sum and unit total squared norm. Then $\xi\in S^{3r+2}$, the source points are $\vec z_i=\alpha_i\vec a+C_r\vec\beta_i$, and the spectral kernel is Eq.~\eqref{eq:povm-singular-kernel} on $S^{r-1}$. The algebraic source formulas and this linear kernel give the reduced characterization directly.

If $\vec a_\perp=0$, absorb the bias by writing
\begin{equation}
\vec\beta_i=\vec y_i-\alpha_i\vec p_0,\qquad \vec z_i=C_r\vec y_i. \label{eq:povm-source-shear}
\end{equation}
Only $q$ and $\vec y$ remain in the source. Normalize retained vectors by $\sum_i\vec y_i=0$ and $\sum_i|\vec y_i|^2=1$, giving $\vec y\in S^{3r-1}$. Evaluate $\mathcal S(q,\vec y)$ from the points $C_r\vec y_i$. At fixed $(q,\vec y)$, the minimum available response enforces every height constraint.

For a zero-moment density $u$ of mass $t$, let $\bar\alpha_q=\sum_iq_i\alpha_i$ and $\bar{\vec y}_q=\sum_iq_i\vec y_i$. Define
\begin{equation}
\begin{aligned}
h_\alpha(\vec m) &=\max_i\{(\vec y_i-\bar{\vec y}_q)\cdot\vec m +(\alpha_i-\bar\alpha_q)g(\vec m)\},\\
H_u(\alpha)&=\int h_\alpha(\vec m)u(\vec m)d\omega_{r-1},\qquad \mathcal K(q,\vec y;u)=\min_{\sum_i\alpha_i=0}H_u(\alpha).
\end{aligned}
\label{eq:povm-height-elimination}
\end{equation}
Each centered branch has weighted mean zero, so $h_\alpha\ge0$. A common height shift preserves branch differences and the centered maximum; zero sum fixes this freedom.

Moment constraints bound the minimizing heights. Put $q_{\min}=\min_iq_i$ and $\operatorname{osc}\alpha=\max_i\alpha_i-\min_i\alpha_i$. Integrating one branch gives
\begin{equation}
H_u(\alpha)\ge t(\max_i\alpha_i-\bar\alpha_q) \ge tq_{\min}\operatorname{osc}\alpha,\qquad H_u(0)\le t. \label{eq:povm-height-radius}
\end{equation}
For $t>0$, comparison with zero heights restricts the minimum to
\begin{equation}
\mathcal A_q=\{\alpha:\sum_i\alpha_i=0, \ \operatorname{osc}\alpha\le1/q_{\min}\}. \label{eq:povm-height-domain}
\end{equation}
At $t=0$ every kernel integral vanishes, and zero heights suffice. For positive mass, this compact domain also makes the reduced response continuous in its test and density.

We evaluate the minimum for $u>0$ almost everywhere. Equal retained vectors $\vec y_i$ can be assigned equal heights: replacing their heights by the weighted average lowers the maximum and preserves the subtracted mean. Combine these groups and their weights for the height calculation, keeping their vector coordinates fixed. With $m\le4$ distinct vectors remaining, choose $\alpha_m=0$ and write $\ell=m-1$. The region where branch $i$ attains the maximum is
\begin{align}
D_i(\alpha)&=\{\vec m:\vec y_i\cdot\vec m+\alpha_i g(\vec m) \ge\vec y_k\cdot\vec m+\alpha_k g(\vec m) \ \text{for all }k\}, \label{eq:povm-height-cells}\\
\partial_iH_u(\alpha)&=\int_{D_i(\alpha)}g(\vec m)u(\vec m)d\omega_{r-1}-q_it. \label{eq:povm-height-stationarity}
\end{align}
Moving $\alpha_i$ changes the boundary of its winning region in Eq.~\eqref{eq:povm-height-cells}. Integrating $gu$ over that region and subtracting $q_it$ gives the derivative. Here $g$ may have either sign, because it arises from a source-preserving test change. Since $\int gu=t$, stationarity assigns signed mass $q_it$ to each region. Distinct branches tie on sphere-null sets, justifying differentiation under the integral and Eq.~\eqref{eq:povm-height-stationarity}.

Nested sign integrals eliminate successive heights. With two free heights, first fix the second and minimize in the first by its sign integral. Repeating that minimization for each second height gives a convex one-variable profile. Iterating this procedure removes all free heights. Define
\[
H_u^{(0)}(\alpha_1,\ldots,\alpha_\ell) =H_u(\alpha_1,\ldots,\alpha_\ell,0),\qquad H_u^{(j)}(\beta) =\min_{\alpha_1,\ldots,\alpha_j} H_u(\alpha_1,\ldots,\alpha_j,\beta,0),
\]
where $j=1,\ldots,\ell$ and $\beta=(\alpha_{j+1},\ldots,\alpha_\ell)$ is the fixed tail. Thus $H_u^{(j-1)}(s,\beta)$ already contains the minimizing first $j-1$ heights, while $s=\alpha_j$ is the next free height. The final value $H_u^{(\ell)}$ is the minimum response.

For a fixed tail, the minimizing free heights exist by Eq.~\eqref{eq:povm-height-radius}. They are unique as well. Suppose two choices attain the same partial minimum. Equality in convexity implies that their maxima have a common winning branch almost everywhere. On the connected cap $g>0$, the difference of these maxima divided by $g$ is continuous and takes values in the finite set of height differences. It is therefore constant. Each free branch meets this cap, since its stationary signed mass in Eq.~\eqref{eq:povm-height-stationarity} is $q_it>0$. Summing the free stationarity equations leaves total signed mass $t\sum_{i>j}q_i>0$ in the fixed branches, so at least one fixed branch also meets the cap. Its height difference is zero, which fixes the constant to zero. Every free height difference is then zero.

A finite integration interval follows from the same mass bound. The branch formula gives $|h_\alpha-h_0|\le\operatorname{osc}\alpha\,|g|$. Setting the free heights to zero therefore gives a trial value at most $t[1+(1+|\vec p_0|)\operatorname{osc}(\beta,0)]$, because $\int|g|u\le t(1+|\vec p_0|)$. Hence all free minimizing heights lie in $[-R_\beta,R_\beta]$, with
\[
R_\beta=\frac{1+(1+|\vec p_0|)\operatorname{osc}(\beta,0)}{q_{\min}}.
\]
Local uniformity of this bound and uniqueness make partial minimizers continuous in the tail. Bounding a profile difference by the minimizers at its two endpoints then identifies its derivative: Eq.~\eqref{eq:povm-height-stationarity} for branch $j$, evaluated at the earlier minimizing heights and fixed $(s,\beta,0)$.

Each one-variable profile is convex, so its nondecreasing derivative determines its unique minimum by the same sign integral as in the PVM calculation:
\begin{equation}
\begin{aligned}
s_j(\beta)&=-\frac12\int_{-R_\beta}^{R_\beta} \operatorname{sgn}\!\left[\frac{\partial}{\partial s} H_u^{(j-1)}(s,\beta)\right]ds,\\
H_u^{(j)}(\beta)&=H_u^{(j-1)}(s_j(\beta),\beta), \qquad j=1,\ldots,\ell.
\end{aligned}
\label{eq:povm-height-integral}
\end{equation}
At step $j$, every trial tail $(s,\beta)$ requires the earlier heights. Recover them backwards: $\alpha_{j-1}=s_{j-1}(s,\beta)$, then $\alpha_{j-2}=s_{j-2}(\alpha_{j-1},s,\beta)$, continuing to $\alpha_1$. The $j$th region moment at these heights is the derivative in the sign integrand. Starting from the original $H_u^{(0)}$ and ending with the empty tail gives the full minimizing vector
\[
\alpha_m^*=0,\qquad \alpha_\ell^*=s_\ell,\qquad \alpha_j^*=s_j(\alpha_{j+1}^*,\ldots,\alpha_\ell^*) \quad(j=\ell-1,\ldots,1).
\]
Assign each group height to its original outcomes and subtract the ordinary mean to restore $\sum_{i=1}^4\alpha_i^*=0$. The resulting attaining kernel and its value are
\begin{equation}
h_*=h_{\alpha_*},\qquad \mathcal K(q,\vec y;u)=H_u^{(\ell)} =\int h_*u\,d\omega_{r-1}. \label{eq:povm-exact-height-kernel}
\end{equation}
If there is only one group, equal heights give $\mathcal K=0$. Thus the reduced POVM characterization is
\begin{equation}
u\ge0,\qquad\int\vec m u\,d\omega_{r-1}=0,\qquad \mathcal K(q,\vec y;u)\ge\mathcal S(q,\vec y) \quad\text{for all retained tests}. \label{eq:new-all-povm}
\end{equation}
As in the original spectral problem, replacing $u\,d\omega_{r-1}$ by a positive measure gives the measure formulation of these inequalities.

Eliminating heights allows the kernel bound to grow as $1/q_{\min}$ near the boundary of the weight simplex. For the mass integral, choose a full-support test measure that makes this bound square-integrable. On $\mathcal Q_r=\Delta_4^\circ\times S^d$, with $\Delta_4^\circ=\{q_i>0:\sum_iq_i=1\}$, use
\begin{equation}
d\nu_r=\frac{\Gamma(12)}{\Gamma(3)^4} q_1^2q_2^2q_3^2q_4^2\,dq_1dq_2dq_3\,d\sigma_d, \qquad d=
\begin{cases}
3r-1,&\vec a_\perp=0,\\
3r+2,&\vec a_\perp\ne0.
\end{cases}
\label{eq:povm-reduced-test-measure}
\end{equation}
Here $q_4=1-q_1-q_2-q_3$ and $\sigma_d$ is normalized sphere area; at full rank write $\mathcal Q=\mathcal Q_3$ and $\nu=\nu_3$. For $\vec a_\perp=0$, use the attaining kernel in Eq.~\eqref{eq:povm-exact-height-kernel}; for $\vec a_\perp\ne0$, use the integrand in Eq.~\eqref{eq:povm-singular-kernel}. In both cases set
\begin{equation}
r_u=[\mathcal S-\mathcal K(u)]_+,\qquad (\mathcal G_{\mathrm A}u)(\vec m)= \int_{\mathcal Q_r}h_*(\vec m)r_u\,d\nu_r. \label{eq:povm-source-integral}
\end{equation}
Using $G=\mathcal G_{\mathrm A}$ and $D=2$ in Eq.~\eqref{eq:general-root-iteration} gives the POVM density and the mass limit \eqref{eq:general-finite-integral-value}. Evaluate each source by Eq.~\eqref{eq:povm-source-contact-value} or \eqref{eq:povm-source-algebraic-value}. When heights leave the source unchanged, use region moments and nested sign integrals to obtain the attaining kernel; otherwise retain the original affine kernel. Zero heights initialize $u_0=0$, with positive densities at subsequent steps.

The source bounds are $\|C_r\|_{\mathrm{op}}$ when $\vec a_\perp=0$ and $\|[\vec a\ C_r]\|_{\mathrm{op}}$ otherwise. Equation~\eqref{eq:povm-height-domain} bounds the eliminated kernels and their Lipschitz constants by $2+(1+|\vec p_0|)/q_{\min}$. This bound is square-integrable because
\[
\int q_{\min}^{-2}d\nu_r \le\sum_{i=1}^4\int q_i^{-2}d\nu_r=220.
\]
Retained-height kernels are bounded by $2\sqrt2$ and have Lipschitz constant at most two. These estimates supply continuity and integrability for the mass theorem. Zero-weight POVMs follow as limits of $q_i(s)=(1-s)q_i+s/4$, $\vec x_i(s)=(1-s)\vec x_i$. Below rank two, Eqs.~\eqref{eq:povm-fixed-source} and \eqref{eq:povm-singular-kernel}, with uniform simplex measure and sphere area, give the unreduced formula.

\section{S4. Contact solutions for two-qubit steering} \label{sec:explicit-regions}

For qubits, contact gives an analytic solution of the minimum-mass problem in Section~S1. A suitable average of preparation scores bounds its mass. Saturation fixes the value and normal derivative on a plane, so the inverse from Section~S2 determines the candidate weights. Positive propagation proves their response inequality away from contact, and the trusted first moment selects the plane attaining the bound. This gives the analytic spectrum and its uniqueness. Dividing that spectrum into outcome regions then determines when arbitrary POVMs have the same boundary as projective measurements.

\subsection{Contact planes and projective steering} \label{subsec:pvm-contact-spectra}

At correlation rank at most one, write $C=\vec b\vec v^{\mathsf T}$ with $|\vec v|=1$. The effective hidden-state sphere consists of $\pm\vec v$, and zero first moment gives equal weights. A spectrum of mass $t$ has kernel $t\max\{|s|,|\vec y\cdot\vec v|\}$. The maximum of the convex source on the unit square in $(s,\vec y\cdot\vec v)$ occurs at a vertex. Consequently
\begin{equation}
c_{\mathrm P}(\vec a,\vec b\vec v^{\mathsf T}) =\max\{|\vec a+\vec b|,|\vec a-\vec b|\}. \label{eq:qubit-rank-one-value}
\end{equation}
This includes $c_{\mathrm P}=|\vec a|$ when $C=0$.

For $r=\rank C\ge2$, retain $E=\operatorname{Ran}C^{\mathsf T}$. We work first with the original response inequality
\begin{equation}
\int|s+\vec y\cdot\vec n|d\mu(\vec n)\ge N_{\mathrm P}(s,\vec y)=|\vec a s+C\vec y|.
\label{eq:analytic-original-response}
\end{equation}
Equality on a plane fixes the response there. Positivity of the gap on both sides also fixes its normal derivative wherever that derivative exists. These two data determine the spectrum; we first choose the plane so that its average bounds the total mass.

To bound the mass, average tests so that their kernels depend only on total weight and first moment. On $s=\vec q\cdot\vec y$, with $\vec q\in E$ and $|\vec q|<1$, write $(s,\vec y)=(\vec q\cdot\vec e,T_{\vec q}\vec e)$ for positive symmetric $T_{\vec q}$. Twice the spherical kernel average is $|\vec q+T_{\vec q}\vec n|$. Equating its square to $(1+\vec q\cdot\vec n)^2$ imposes $T_{\vec q}^2=(1-|\vec q|^2)\openone_3+\vec q\vec q^{\mathsf T}$ and $T_{\vec q}\vec q=\vec q$. The positive solution is
\[
\delta=1-|\vec q|^2,\qquad T_{\vec q}=\sqrt\delta\,\openone_3+ \frac{\vec q\vec q^{\mathsf T}}{1+\sqrt\delta}.
\]
Average over ambient $S^2$ at both ranks; rank-two hidden states remain on the circle in $E$. These tests lie on the plane, and $|\vec n|=1$ gives
\begin{equation}
2\int_{S^2}k_{\mathrm P}(\vec q\cdot\vec e,T_{\vec q}\vec e;\vec n)d\omega(\vec e) =|\vec q+T_{\vec q}\vec n|=1+\vec q\cdot\vec n. \label{eq:pvm-plane-balance}
\end{equation}
For a zero-moment spectrum, the integral of this affine kernel is exactly its mass. Thus the plane average realizes the mass and moment terms of the spectral balance by an explicit distribution of tests. Its source average is
\begin{equation}
\begin{aligned}
\Phi(\vec q)&=2\int_{S^2}N_{\mathrm P}(\vec q\cdot\vec e,T_{\vec q}\vec e)d\omega =2\int_{S^2}\sqrt{\vec e^{\mathsf T}M(\vec q)\vec e}\,d\omega,\\
M(\vec q)&=(\vec a+C\vec q)(\vec a+C\vec q)^{\mathsf T} +(1-|\vec q|^2)J,\qquad J=CC^{\mathsf T}-\vec a\vec a^{\mathsf T}.
\end{aligned}
\label{eq:pvm-semidefinite-integral}
\end{equation}
Indeed, $M=(CT_{\vec q}+\vec a\vec q^{\mathsf T})(CT_{\vec q}+\vec a\vec q^{\mathsf T})^{\mathsf T}$; rotation invariance allows the two Gram matrices to be interchanged under the square-root angular integral. Using the PVM kernel operator $\mathcal K$ from Eq.~\eqref{eq:pvm-support}, any positive spectrum $\mu$ of mass $t$ therefore obeys
\begin{equation}
\begin{aligned}
t+\vec q\cdot\int\vec n\,d\mu-\Phi(\vec q) =2\int_{S^2}\bigl[&(\mathcal K\mu)(\vec q\cdot\vec e,T_{\vec q}\vec e)-N_{\mathrm P}(\vec q\cdot\vec e,T_{\vec q}\vec e)\bigr]d\omega(\vec e).
\end{aligned}
\label{eq:pvm-plane-gap}
\end{equation}
Feasibility makes the right side nonnegative, yielding $t\ge\Phi(\vec q)$ for zero moment. Attainment forces the continuous gap to vanish on the entire plane. Wherever derivatives exist, nonnegativity on both sides fixes the normal derivative as well. Angular inversion will fix the even density, and positivity will then exclude mass on affine zero sets of the kernel, ensuring this differentiability.

Flattening the plane separates its data into even and odd weights. Under the condition below, their inverse is nonnegative and propagates above the source at every test. Transforming back gives first moment $\nabla_E\Phi(\vec q)$, so its zero determines the plane attaining the mass bound.
\begin{theorem}
\label{thm:pvm-finite-integral-root} \label{prop:pvm-unique-spectrum} Let $r=\rank C\in\{2,3\}$, and let $j_1\le j_2\le j_3$ be the eigenvalues of $J$. Suppose
\begin{equation}
\sum_{i=1}^{4-r}j_i\ge0. \label{eq:pvm-rank-two-root}
\end{equation}
Then $\Phi$ is strictly concave on the unit ball of $E$ and has a unique interior stationary point $\vec q_*$. The PVM scale and its unique minimizing spectrum are determined by this plane:
\begin{equation}
\nabla_E\Phi(\vec q_*)=0,\qquad c_{\mathrm P}(\vec a,C)=\Phi(\vec q_*). \label{eq:pvm-semidefinite-value}
\end{equation}
This plane varies continuously with the data throughout the domain, including at its rank-two boundary.
\end{theorem}
At full rank, Eq.~\eqref{eq:pvm-rank-two-root} is $J\succeq0$; on the effective circle, it is $\Tr J-\lambda_{\max}(J)\ge0$. A single angular inversion formula covers both ranks.

\begin{proof}
Fix an arbitrary $\vec q\in E$, $|\vec q|<1$, and make the Lorentz change
\[
s=\gamma(\tau+\vec q\cdot\vec x),\qquad \vec y=S_{\vec q}\vec x+\gamma\vec q\tau, \qquad \gamma=\delta^{-1/2},\quad S_{\vec q}=\gamma T_{\vec q}.
\]
It satisfies $s-\vec q\cdot\vec y=\tau/\gamma$ and brings the contact plane to $\tau=0$. In these coordinates the source is $|\vec d\tau+B\vec x|$, where
\begin{equation}
\vec d=\gamma(\vec a+C\vec q),\qquad B=CS_{\vec q}+\gamma\vec a\vec q^{\mathsf T},\qquad BB^{\mathsf T}-\vec d\vec d^{\mathsf T}=J, \qquad \delta BB^{\mathsf T}=M(\vec q). \label{eq:pvm-lorentz-source}
\end{equation}
On $E$, the matrix $B$ has rank $r$. At full rank this follows from $\det B=\gamma\det C(1+\vec q\cdot C^{-1}\vec a)$ and $|C^{-1}\vec a|\le1$. At rank two, a transverse component of $\vec a$ makes $B\vec x=0$ imply first $\vec q\cdot\vec x=0$ and then $C\vec x=0$. Without that component, Eq.~\eqref{eq:pvm-rank-two-root} reduces to $J\succeq0$ on the source plane, and the same determinant argument applies there.

At $\tau=0$, contact prescribes $|B\vec x|$ and normal derivative $\vec d\cdot B\vec x/|B\vec x|$. Value depends on the even measure, derivative on the odd measure. Both are reconstructed using the cosine identity on the source sphere in $\operatorname{Ran}B$:
\[
\int|\vec e\cdot\vec z|\,d\omega_{r-1}(\vec e) =\frac{\Gamma(r/2)}{\sqrt\pi\,\Gamma((r+1)/2)}|\vec z|
\]
and differentiation with respect to $\vec z$ yields
\[
\int\vec e\operatorname{sgn}(\vec e\cdot\vec z)d\omega_{r-1}(\vec e) =\frac{\Gamma(r/2)}{\sqrt\pi\,\Gamma((r+1)/2)}\frac{\vec z}{|\vec z|}.
\]
Weighting the first identity by $|B^{\mathsf T}\vec e|$ supplies the contact value. Contracting the second with $\vec d$ supplies its derivative through odd weight $\vec d\cdot\vec e$. Push these weights to $\vec n'=B^{\mathsf T}\vec e/|B^{\mathsf T}\vec e|$. The angular Jacobian $\sqrt{\det{}'(BB^{\mathsf T})}/|B^{\mathsf T}\vec e|^r$ gives density
\begin{equation}
f^{(\vec q)}(\vec n')= \frac{\sqrt\pi\,\Gamma((r+1)/2)}{\Gamma(r/2)} \frac{1+(B^+\vec d)\cdot\vec n'}{\sqrt{\det{}'(BB^{\mathsf T})}} [\vec n'^{\mathsf T}(B^{\mathsf T}B)^+\vec n']^{-(r+1)/2}. \label{eq:pvm-semidefinite-inverse-spectrum}
\end{equation}
Here $\det{}'$ is the product of positive eigenvalues. The angular factor is $2$ for $r=3$ and $\pi/2$ for $r=2$. Its kernel $K_{\vec q}$ obeys
\[
K_{\vec q}(0,\vec x)=|B\vec x|,\qquad \partial_\tau K_{\vec q}(0,\vec x) =\frac{\vec d\cdot B\vec x}{|B\vec x|}.
\]
Both contact data are therefore reproduced.

Nonnegativity reduces to $|B^+\vec d|\le1$, which follows from Eq.~\eqref{eq:pvm-rank-two-root}. At full rank, writing $\vec v=B^{-1}\vec d$ gives $J=B(\openone-\vec v\vec v^{\mathsf T})B^{\mathsf T}\succeq0$ and hence $|\vec v|\le1$. At rank two let $\vec v=B^+\vec d$. If $|\vec v|>1$, the $J$-form is negative on $B^{+\mathsf T}\vec v$ and nonpositive on a vector perpendicular to $\operatorname{Ran}B$. These directions are orthogonal, contradicting the nonnegative trace of $J$ on every two-dimensional subspace. Thus $|B^+\vec d|\le1$ in both dimensions.

Unit hidden directions imply $\Box K=0$, where $\Box=\partial_\tau^2-\Delta_{\vec x}$ on the effective space. Since value and derivative of the gap vanish on the plane, its propagation is driven by the source $N=|\vec d\tau+B\vec x|$ alone:
\begin{equation}
-\Box N=\Gamma(\tau,\vec x) =\frac{\Tr J-\vec e^{\mathsf T}J\vec e}{N}, \qquad \vec e=\frac{\vec d\tau+B\vec x}{N}. \label{eq:pvm-wave-source}
\end{equation}
Both ranks in the stated domain satisfy
\begin{equation}
(\Tr J)\openone_3-J\succeq0. \label{eq:pvm-propagation-region}
\end{equation}
The wave source is consequently nonnegative. Zero Cauchy data give, for $\tau>0$,
\begin{equation}
\begin{aligned}
r=3:\quad &(K_{\vec q}-N)(\tau,\vec x) =\int_0^\tau t\int_{S^2}\Gamma(\tau-t,\vec x+t\vec u)d\omega(\vec u)dt,\\
r=2:\quad &(K_{\vec q}-N)(\tau,\vec x) =\frac1{2\pi}\int_0^\tau\int_{|\vec z-\vec x|<\tau-t} \frac{\Gamma(t,\vec z)}{\sqrt{(\tau-t)^2-|\vec z-\vec x|^2}}d\vec z\,dt.
\end{aligned}
\label{eq:pvm-contact-propagation}
\end{equation}
Both integrals are nonnegative; reversing $\tau$ gives the other half-space. At source zeros, use $N_\eta=\sqrt{N^2+\eta^2}$. Its wave source has numerator
\[
(\vec d\tau+B\vec x)^{\mathsf T}[(\Tr J)\openone_3-J](\vec d\tau+B\vec x) +\eta^2\Tr J\ge0
\]
and denominator $(N^2+\eta^2)^{3/2}$. Let $K_\eta$ solve the homogeneous wave equation with the plane value and normal derivative of $N_\eta$. The propagation formula gives $K_\eta-N_\eta\ge0$. As $\eta\downarrow0$, these solutions converge to $K_{\vec q}$ and $N$, respectively, extending the inequality through source zeros.

Transforming back changes directions and weights to
\[
\vec n'\longmapsto \frac{S_{\vec q}\vec n'-\gamma\vec q}{\gamma(1-\vec q\cdot\vec n')}, \qquad d\mu\longmapsto\gamma(1-\vec q\cdot\vec n')d\mu.
\]
The weight factor is positive and each transformed direction is unit in $E$. Thus every $\vec q$ supplies a nonnegative spectrum satisfying all projective inequalities. Imposing its first moment will select the optimal one.

Write its mass and moment as $t_{\vec q}$ and $\vec m_{\vec q}$. Averaging its response on any other plane and using contact on its own plane yields
\[
t_{\vec q}+\vec r\cdot\vec m_{\vec q}\ge\Phi(\vec r),\qquad t_{\vec q}+\vec q\cdot\vec m_{\vec q}=\Phi(\vec q).
\]
For each $\vec q$, this affine function of $\vec r$ supports $\Phi$ from above and agrees with it at $\vec r=\vec q$. Hence $\Phi$ is concave; differentiating at contact identifies the spectral moment and mass:
\begin{equation}
\vec m_{\vec q}=\nabla_E\Phi(\vec q),\qquad t_{\vec q}=\Phi(\vec q)-\vec q\cdot\nabla_E\Phi(\vec q). \label{eq:pvm-plane-moments}
\end{equation}
Zero moment is therefore stationarity of $\Phi$. Any interior stationary plane gives a feasible measure of mass $\Phi(\vec q)$ and attains the bound. Strict concavity and inward growth at the boundary ensure a unique such plane.

Strict concavity follows by strengthening propagation away from contact. At full rank, $J\succeq0$ has rank at least two, so $\Tr J>\lambda_{\max}(J)$ and the wave integral is positive at nonzero height. At rank two, the nonnegative quadratic numerator of $\Gamma$ is not identically zero in $(\tau,\vec x)$. With transverse bias, $[\vec d\ B]$ has rank three and the middle matrix in Eq.~\eqref{eq:pvm-propagation-region} is nonzero. Without transverse bias, $J\succeq0$ has nonzero range in the source plane; the middle matrix cannot vanish on that plane, since its trace would force $J=0$ and then $\rank B\le1$. A nonzero quadratic polynomial is positive on an open subset of each propagation cone. The rank-two wave integral is therefore also strictly positive away from the contact plane. Integrating on a different plane makes the upper support strict and proves strict concavity.

To exclude a boundary maximum, fix $|\vec q_0|=1$ and write $\vec w_0=\vec a+C\vec q_0$. If $\vec w_0=0$, then $\Phi(\vec q_0)=0$ and this point cannot maximize. In the semidefinite case, $J$ has a positive direction perpendicular to any nonzero $\vec w_0$. At full rank this follows from $\rank J\ge2$. In a source plane the only exception could be a rank-one $J=\vec j\vec j^{\mathsf T}$ parallel to $\vec w_0$. Rotating right coordinates then gives $C=(\vec a\ \vec j)$ with independent columns. Parallelism forces $q_{0,1}=-1$ and $q_{0,2}=0$, which would give $\vec w_0=0$. Moving to $(1-\epsilon)\vec q_0$ thus opens a transverse squared singular value of order $\epsilon$. The resulting gain in the angular integral has the order
\[
\int_0^1(\sqrt{t^2+\epsilon}-t)dt =\frac\epsilon4\log\frac1\epsilon+O(\epsilon),
\]
which dominates the $O(\epsilon)$ longitudinal change.

With transverse bias at rank two, choose $\vec v\in E$ perpendicular to $\vec q_0$. The two source columns along this inward displacement are $\vec a(1-\epsilon)+C\vec q_0$ and $\sqrt{2\epsilon-\epsilon^2}C\vec v$. Their limiting directions are independent because the first has a transverse component. This again produces a positive angular gain.

More generally, the argument applies to any inward displacement $\vec q=\vec q_0+\epsilon\vec v$ with $\vec q_0\cdot\vec v<0$: the transverse opening is proportional to $1-|\vec q|^2=-2\epsilon\vec q_0\cdot\vec v+O(\epsilon^2)$, and the other changes are $O(\epsilon)$. This also excludes endpoint maxima when some coordinates are held fixed. Strict concavity gives the unique interior stationary point.

In the reduced coordinates of Section~S3, let $s_*$ denote the height at which a test meets the optimal plane. Contact and tangency give
\begin{equation}
F_{u_*}(s_*;\vec x)=\widetilde N_{\mathrm P}(s_*,\vec x),\qquad D_{u_*}(s_*;\vec x)=\frac{|\vec a_\perp|^2s_*}{\widetilde N_{\mathrm P}(s_*,\vec x)}. \label{eq:pvm-height-contact-equations}
\end{equation}
When $\vec a_\perp=0$, the second equation is $D_{u_*}=0$. Positivity of the density almost everywhere gives the unique height of Section~S3, and every eliminated source inequality is saturated:
\begin{equation}
(\mathcal K_{\mathrm P}u_*)(\vec x) =F_{u_*}(\zeta_{u_*}(\vec x);\vec x) =|C_r\vec x|=\mathcal S_{\mathrm P}(\vec x). \label{eq:pvm-reduced-saturation}
\end{equation}
At rank two with transverse bias, Eq.~\eqref{eq:pvm-height-contact-equations} is instead the tangency condition for the retained-height source. The strict propagation established above makes the gap positive at every other height. Thus the contact plane selects the active height in both forms of the PVM equation.

At full rank the contact data also determine the projective response functions. For a measurement direction $\vec e$, put $\vec y=(C+\vec a\vec q_*^{\mathsf T})^{-1}\vec e$ and take the test $(s,\vec y)=(\vec q_*\cdot\vec y,\vec y)$. The source vector on this plane is $\vec e$. Equality of the value and the first derivatives of the source and kernel gives
\[
f_{\vec e}(\vec n)=\operatorname{sgn}[(\vec n+\vec q_*)\cdot\vec y],\qquad \int f_{\vec e}(\vec n)\,d\mu=\vec a\cdot\vec e,\qquad \int f_{\vec e}(\vec n)\vec n\,d\mu=C^{\mathsf T}\vec e.
\]
Hence $p(\pm|\vec e,\vec n)=(1\pm f_{\vec e}(\vec n))/2$ divides the boundary measure into the required projective outcomes. Contact fixes both the weights and their divisions.

For uniqueness, consider any minimizing spectrum. Its gap has zero average on the optimal plane, so it agrees with the source throughout that plane. At rank two, the zero-source tests $\vec y\in\operatorname{Ker}C$ belong to the plane; equality there first confines the spectrum to $E$. After the Lorentz change, its even part has cosine transform $|B\vec x|$. Injectivity fixes this part to the even density in Eq.~\eqref{eq:pvm-semidefinite-inverse-spectrum}. Positivity bounds the whole measure by twice its even part, so it is absolutely continuous and assigns no mass to the kernel's zero sets at nonzero contact tests. Its normal derivative therefore exists, and tangency fixes the odd part's hemispherical transform to $\vec d\cdot B\vec x/|B\vec x|$.

Proposition~\ref{prop:spherical-cauchy-inversion} identifies the measure uniquely, on the circle at rank two and the sphere at full rank.

To include rank changes in the continuity statement, maximize $\Phi$ on the common closed three-dimensional ball. The displacement of the tests at $\vec q'$ from the optimal plane has coefficient $\vec q'-T_{\vec q'}\vec q_*$. For $|\vec q'|<1$ it vanishes only at $\vec q'=\vec q_*$; for $|\vec q'|=1$ it equals $\vec q'(1-\vec q'\cdot\vec q_*)\ne0$. Strict propagation therefore makes $\vec q_*$ the unique maximum even on this common ball. The functions $\Phi$ converge uniformly when the data converge. Compactness and uniqueness then imply convergence of their maximizing planes.
\end{proof}

To recover the scalar height of Section~S3, substitute $\vec y=V(\vec x-s\vec p_0)$ into $s=\vec q_*\cdot\vec y$. Positivity at $\vec q=0$ gives $|\vec p_0|\le1$, so the denominator below is positive. Solving for $s$ yields the contact height, which minimizes $F_{u_*}$ when $\vec a_\perp=0$:
\begin{align}
s_*(\vec x)&=\frac{\vec q_*\cdot V\vec x} {1+\vec q_*\cdot V\vec p_0}, \label{eq:pvm-explicit-contact-height}\\
\zeta_{u_*}(\vec x)&=s_*(\vec x)\qquad(\vec a_\perp=0). \label{eq:pvm-explicit-median}
\end{align}
Thus one plane vector resolves the entire family of heights in Eq.~\eqref{eq:pvm-median-integral}.

For the critical Werner state $\rho_{1/2}=\tfrac12\ketbra{\psi^-}+\tfrac12\openone_4/4$, the canonical data are $\vec a=0$, $C=-\openone_3/2$ and $J=\openone_3/4$. Its eigenvalues are $5/8,1/8,1/8,1/8$. Substitution in Theorem~\ref{thm:pvm-finite-integral-root} gives $\vec q_*=0$, $\Phi(0)=1$ and $d\mu_*=d\omega$. Hence uniform sphere measure is the unique minimizing pure-state spectrum.

For a mass-$t$ zero-moment measure, the binary outcome set is a zonoid~\cite{AubrunLancien2016,ChitambarZhangV1}, with support function
\begin{equation}
\begin{aligned}
Z_\mu&=\left\{\int f(\vec n)(1,\vec n)d\mu:0\le f\le1\right\},\\
2h_{Z_\mu}(s,\vec y)&=ts+\int|s+\vec y\cdot\vec n|d\mu=ts+K_\mu(s,\vec y).
\end{aligned}
\label{eq:zonoid-response-normalization}
\end{equation}
Indeed, maximizing $f$ pointwise gives $h_{Z_\mu}=\int[s+\vec y\cdot\vec n]_+d\mu$, and the first moment removes the linear vector term. This identifies the binary support function with the response used here. Multi-outcome divisions retain the maximum-response constraints of Section~S3 and their region moments below.

At full rank, source-sphere coordinates give a direct representation of the optimal spectrum and its POVM region moments. Evaluate $(B,\vec d)$ at $\vec q_*$ and set
\begin{equation}
S=(BB^{\mathsf T})^{1/2},\qquad r(\vec e)=|S\vec e|, \qquad R=\int_{S^2}r\,d\omega, \qquad \gamma=(1-|\vec q_*|^2)^{-1/2}. \label{eq:povm-direct-source-data}
\end{equation}
Pushforward of $2(r+\vec d\cdot\vec e)d\omega$ gives Eq.~\eqref{eq:pvm-semidefinite-inverse-spectrum}; transforming back yields
\begin{equation}
\begin{gathered}
\vec n'(\vec e)=\frac{B^{\mathsf T}\vec e}{r(\vec e)},\qquad \ell_*(\vec e)=\gamma[1-\vec q_*\cdot\vec n'(\vec e)],\\
\vec n_*(\vec e)=\frac{S_{\vec q_*}\vec n'(\vec e)-\gamma\vec q_*}{\ell_*(\vec e)}, \qquad w_*(\vec e)=2\ell_*(\vec e)[r(\vec e)+\vec d\cdot\vec e].
\end{gathered}
\label{eq:povm-direct-spectrum-coordinates}
\end{equation}
Thus, for any continuous $f$,
\begin{equation}
\int f(\vec n)u_*(\vec n)d\omega(\vec n) =\int w_*(\vec e)f(\vec n_*(\vec e))d\omega(\vec e). \label{eq:povm-direct-spectrum}
\end{equation}
Its mass is $2R/\gamma$. The density in the original coordinates follows from the two angular Jacobians:
\[
u_*(\vec n_*(\vec e))= \frac{2\ell_*(\vec e)^3r(\vec e)^3[r(\vec e)+\vec d\cdot\vec e]}{|\det B|}.
\]
These source coordinates will be used for the POVM region moments.

At both effective ranks in Theorem~\ref{thm:pvm-finite-integral-root}, the explicit spectrum is the unique weak limit of the densities determined in Section~S3:
\begin{equation}
u_{N_\varepsilon}\,d\omega_{r-1} \rightharpoonup u_*\,d\omega_{r-1},\qquad \lim_{\varepsilon\downarrow0}\int u_{N_\varepsilon}d\omega_{r-1}=\Phi(\vec q_*). \label{eq:pvm-explicit-spectral-limit}
\end{equation}
Bounded mass on the compact effective sphere gives weakly convergent subsequences. Their limits are feasible and minimizing by Section~S3; uniqueness identifies each with $u_*d\omega_{r-1}$. Hence both regularized densities and finite recurrences converge to this spectrum.

One matrix integral evaluates both the stationary equation and its mass:
\begin{equation}
\mathsf R(M)=\int_0^\infty \frac{(\openone_3+u^2M)^{-1}}{\sqrt{\det(\openone_3+u^2M)}}\,du. \label{eq:pvm-matrix-integral}
\end{equation}
Gaussian radial integration gives $\mathsf R(M)=2\int\vec e\vec e^{\mathsf T}/\sqrt{\vec e^{\mathsf T}M\vec e}\,d\omega$. Thus, with $\vec w=\vec a+C\vec q_*$ and $M=M(\vec q_*)$, Eq.~\eqref{eq:pvm-semidefinite-value} is
\begin{equation}
C^{\mathsf T}\mathsf R(M)\vec w=\vec q_*\Tr[\mathsf R(M)J], \qquad c_{\mathrm P}=\Tr[M\mathsf R(M)]. \label{eq:pvm-semidefinite-root}
\end{equation}
To solve the stationary equation by integrals alone, we apply the sign integration used for the heights in Section~S3 to successive plane coordinates. Choose orthonormal coordinates on $E$ and start with $\Phi^{(0)}=\Phi$. Since $r=\rank C=\dim E$, for a fixed tail $\vec v=(q_{j+1},\ldots,q_r)$ set $b(\vec v)=\sqrt{1-|\vec v|^2}$. Define successively
\begin{equation}
\begin{aligned}
\theta_j(\vec v)&=\frac12\int_{-b(\vec v)}^{b(\vec v)} \operatorname{sgn}[\partial_s\Phi^{(j-1)}(s,\vec v)]\,ds,\\
\Phi^{(j)}(\vec v)&=\Phi^{(j-1)}(\theta_j(\vec v),\vec v), \qquad j=1,\ldots,r.
\end{aligned}
\label{eq:pvm-nested-plane-integrals}
\end{equation}
Strict concavity and inward growth give a unique interior maximum of each profile. Its derivative is positive before the maximum and negative after it, so sign integration returns the maximizing coordinate. Substitution leaves the next partially maximized profile.

Comparing maximizing values at neighboring tails shows that the profile derivative is the corresponding partial derivative of $\Phi$ at the earlier maximizing coordinates. Evaluate it as $\partial_j\Phi=\tfrac12\Tr[\mathsf R(M)\partial_jM]$. At each trial $(s,\vec v)$, recover the earlier coordinates backwards through $\theta_{j-1},\ldots,\theta_1$, as in Eq.~\eqref{eq:povm-height-integral}. Finally, $q_r^*=\theta_r$ and $q_j^*=\theta_j(q_{j+1}^*,\ldots,q_r^*)$, so
\begin{equation}
c_{\mathrm P}=\Phi^{(r)}=\Phi(q_1^*,\ldots,q_r^*). \label{eq:pvm-nested-plane-value}
\end{equation}
Profile values at a boundary tail are defined by continuity.

Standard elliptic functions evaluate Eq.~\eqref{eq:pvm-matrix-integral}. For positive eigenvalues $\lambda_1,\lambda_2,\lambda_3$ of $M$, define Carlson's integrals
\[
R_F(a,b,c)=\frac12\int_0^\infty\frac{dx}{\sqrt{(x+a)(x+b)(x+c)}},\qquad R_D(a,b,c)=\frac32\int_0^\infty\frac{dx}{(x+c)\sqrt{(x+a)(x+b)(x+c)}}.
\]
The substitution $u=x^{-1/2}$ yields
\begin{equation}
\mathsf R_i=R_F(\lambda_1,\lambda_2,\lambda_3)-\frac{\lambda_i}{3}D_i, \qquad D_i=R_D(\lambda_j,\lambda_k,\lambda_i), \quad\{i,j,k\}=\{1,2,3\}. \label{eq:pvm-semidefinite-elliptic}
\end{equation}
Integration by parts gives $\sum_iD_i=3/\sqrt{\lambda_1\lambda_2\lambda_3}$ and $\sum_i\lambda_iD_i=3R_F$. These identities leave $R_F$ and one $D_i$ as independent evaluations. Coincident eigenvalues follow continuously; for $M=\lambda\openone_3$, $\mathsf R_i=2/(3\sqrt\lambda)$.

For rank-two $M$, let $\kappa_1\le\kappa_2$ be its positive eigenvalues and use
\[
\mathrm E(m)=\int_0^{\pi/2}\sqrt{1-m\sin^2\theta}\,d\theta,\qquad \mathrm K(m)=\int_0^{\pi/2}(1-m\sin^2\theta)^{-1/2}\,d\theta.
\]
Projection of uniform sphere area onto the source plane has mean radial length $\pi/4$, so the angular value becomes
\begin{equation}
\Phi=\sqrt{\kappa_2}\,\mathrm E(m),\qquad m=1-\kappa_1/\kappa_2. \label{eq:pvm-rank-two-elliptic-value}
\end{equation}
In the two positive directions and the null direction, respectively, the matrix integral is
\[
\mathsf R_1=\frac{\mathrm K-\mathrm E}{m\sqrt{\kappa_2}},\qquad \mathsf R_2=\frac{\mathrm E-(1-m)\mathrm K}{m\sqrt{\kappa_2}},\qquad \mathsf R_0=\frac{\mathrm K}{\sqrt{\kappa_2}}.
\]
At $m=0$ these are $\pi/(4\sqrt{\kappa_2})$, $\pi/(4\sqrt{\kappa_2})$, and $\pi/(2\sqrt{\kappa_2})$, respectively.

An isotropic transformed source makes the integral solution algebraic. This family can be specified directly by its original data.
\begin{theorem}
\label{thm:pvm-isotropic-algebraic} An invertible datum in the PVM domain has $S=\lambda\openone_3$ at its optimal plane if and only if
\begin{equation}
J=\lambda^2\openone_3- \frac{9\lambda^2}{4\lambda^2+|\vec a|^2}\vec a\vec a^{\mathsf T}, \qquad |\vec a|^2\le\frac{\lambda^2}{2},\qquad \lambda>0. \label{eq:pvm-isotropic-original-family}
\end{equation}
On this family the optimal plane and PVM scale are
\begin{equation}
\vec q_* =\frac{C^{\mathsf T}\vec a}{2\lambda^2-|\vec a|^2},\qquad c_{\mathrm P}=\frac{4\lambda^2}{\sqrt{4\lambda^2+|\vec a|^2}}. \label{eq:pvm-isotropic-algebraic-value}
\end{equation}
\end{theorem}
\begin{proof}
If $S=\lambda\openone_3$, the pre-Lorentz spectrum is $2(\lambda+\vec d\cdot\vec e)d\omega$. It has mass $2\lambda$ and normalized mean $B^{\mathsf T}\vec d/(3\lambda^2)$. Its inverse Lorentz moment is zero exactly when
\[
\vec q_* =\frac{B^{\mathsf T}\vec d}{3\lambda^2},\qquad |\vec q_*|^2=\frac{|\vec d|^2}{9\lambda^2}.
\]
Transforming the source back gives $\vec a=\gamma(\vec d-B\vec q_*)=2\gamma\vec d/3$ and $J=\lambda^2\openone_3-\vec d\vec d^{\mathsf T}$. With relative source bias $h=|\vec d|/\lambda$,
\begin{equation}
h^2=\frac{9|\vec a|^2}{4\lambda^2+|\vec a|^2}. \label{eq:pvm-isotropic-original-bias}
\end{equation}
Eliminating $h$ gives Eq.~\eqref{eq:pvm-isotropic-original-family}, with $h\le1$ equivalent to its bias bound.

Conversely, define $h$ by Eq.~\eqref{eq:pvm-isotropic-original-bias} and write $\vec u=\vec a/|\vec a|$ when $\vec a\ne0$. The stated identity implies
\[
CC^{\mathsf T}=\lambda^2(\openone_3-\vec u\vec u^{\mathsf T}) +\lambda^2\frac{(3-h^2)^2}{9-h^2}\vec u\vec u^{\mathsf T}.
\]
These eigenvalues are positive for $0\le h\le1$. The unit vector $\vec v=\sqrt{9-h^2}\,C^{\mathsf T}\vec u/[\lambda(3-h^2)]$ gives $\vec q_*=h\vec v/3$ and $\gamma=3/\sqrt{9-h^2}$. Substitution yields
\[
BB^{\mathsf T}=\lambda^2\openone_3,\qquad \vec d=\lambda h\vec u,\qquad \frac{B^{\mathsf T}\vec d}{3\lambda^2}=\vec q_*.
\]
This plane has zero spectral moment and is therefore the unique optimum. Its mass $2\lambda/\gamma$ gives the stated value. For $\vec a=0$ the same formulas reduce to $CC^{\mathsf T}=\lambda^2\openone_3$, $\vec q_*=0$, and $c_{\mathrm P}=2\lambda$.
\end{proof}
Here $\lambda^2$ is the repeated largest eigenvalue of $J$. In principal coordinates, the family has transverse correlations $\lambda,\lambda$, axial correlation magnitude $\lambda(3-h^2)/\sqrt{9-h^2}$, and axial bias $2\lambda h/\sqrt{9-h^2}$, for the entire interval $0\le h\le1$.

A pure conditional state places an extreme demand on a common ensemble. When correlations extend along at least two independent directions, that demand forces the mass above one. This will explain the limiting behavior of several families in Section~S5. Specifically, for $\rank C\ge2$, the spectral inequality implies
\begin{equation}
c_{\mathrm P}(\vec a,C)>\max_{|\vec n|=1}|\vec a+C\vec n|. \label{eq:pvm-strict-null-contact}
\end{equation}
To prove strictness, compare the response and source near a maximizing unit vector $\vec n_0$. Choose a tangent vector $\vec v\perp\vec n_0$ with $|\vec v|=1$ and $C\vec v\ne0$; rank at least two ensures that such a tangent exists. Set $\vec w=\vec a+C\vec n_0$. Stationarity on the sphere implies $\vec w\cdot C\vec v=0$, so the source at $(s,\vec y)=(1,\vec n_0+r\vec v)$ grows as $|\vec w|+|C\vec v|^2r^2/(2|\vec w|)+o(r^2)$. A zero-moment spectrum of mass $t$ has kernel
\[
t+2\int[-1-\vec n_0\cdot\vec n-r\vec v\cdot\vec n]_+\,d\mu.
\]
The positive part is supported in a shrinking cap about $-\vec n_0$, bounded by $r^2/2$, and zero at that point. Its integral is therefore $o(r^2)$, too small to supply the source's quadratic increase at $t=|\vec w|$. Attainment of the mass minimum yields the strict inequality \eqref{eq:pvm-strict-null-contact}. In a canonical state, a pure effect on Bob prepares Bloch vector $\vec a+C\vec n$ on Alice. If this conditional state is pure, the maximum equals one and $c_{\mathrm P}>1$.

\subsection{Outcome partitions and POVM equivalence} \label{subsec:povm-region-moments}

Winning regions of the POVM kernel must have the probabilities and Bloch moments of the requested effects. Height equations in Section~S3 balance their signed masses at fixed spectrum; derivatives with respect to all test components also fix their vector moments. At a most restrictive test, effect positivity $|\vec x_i|\le q_i$ constrains these region moments. On the circle, arcs saturate binary bounds. On the sphere, uniform cone-moment estimates give a full bias ball at each finite anisotropy.

A deficit of a PVM spectrum at a POVM test identifies a measurement requiring additional mass, because the algebraic source calculation in Section~S3 supplies an attaining POVM. Fix such a measurement and a positive probability density $\varpi$ on the effective circle or sphere, with mean $\vec b$ and source data $(B,\vec d)$. Denote by $T_M(\varpi)$ the mass needed for this density to supply its outcomes. After removing zero effects, write
\begin{equation}
\begin{aligned}
L_i(\vec n)&=\alpha_i+\vec\beta_i\cdot\vec n,\\
D_{q,\varpi}(Y)&=\int\max_iL_i\,d\varpi -\sum_iq_i(\alpha_i+\vec b\cdot\vec\beta_i),\\
N_M(Y)&=\sum_i\vec x_i\cdot(\vec d\alpha_i+B\vec\beta_i),\qquad T_M(\varpi)=\max_Y\frac{N_M(Y)}{D_{q,\varpi}(Y)}.
\end{aligned}
\label{eq:povm-circle-source-ratio}
\end{equation}
Common shifts and scaling allow centered branches with $\sum_i(\alpha_i^2+|\vec\beta_i|^2)=1$. The denominator is positive on this compact set: it could vanish only if every positive-weight branch coincided everywhere, which centering and normalization exclude. A maximizing test therefore exists. At a positive maximum, combine identical branches and their effects. The remaining ties have measure zero. On the winning regions $\Omega_i$, the derivatives of the same centered kernel are
\begin{equation}
\partial_{\alpha_i}D_{q,\varpi}=\int_{\Omega_i}d\varpi-q_i,\qquad \nabla_{\vec\beta_i}D_{q,\varpi}=\int_{\Omega_i}\vec n\,d\varpi-q_i\vec b. \label{eq:povm-region-derivatives}
\end{equation}
These derivatives measure the excess probability and Bloch moment of a region over the centered effect weights. To recover the height equation of Section~S3, take $(B,\vec d,\vec b)=(C_r,\vec a,\vec0)$ with $\vec a=C_r\vec p_0$ and $d\varpi=u\,d\omega/t$. The source-preserving derivative then gives
\[
t\left(\partial_{\alpha_i}D_{q,\varpi} -\vec p_0\cdot\nabla_{\vec\beta_i}D_{q,\varpi}\right) =\int_{\Omega_i}g u\,d\omega-q_it.
\]
The source derivative along this change is zero, recovering Eq.~\eqref{eq:povm-height-stationarity}. For all scalar and vector components, stationarity at $T_M$ equates derivatives of $D_{q,\varpi}$ with those of $N_M/T_M$, giving
\begin{equation}
\int_{\Omega_i}(1,\vec n)d\varpi =q_i(1,\vec b)+T_M^{-1}(\vec d\cdot\vec x_i,B^{\mathsf T}\vec x_i). \label{eq:povm-winning-cell-moments}
\end{equation}
Shift invariance and homogeneity remove the centering and normalization multipliers. Thus each winning region in Eq.~\eqref{eq:povm-winning-cell-moments} is linked to its own effect. Together with effect positivity, the region of a smallest-height branch bounds the mass.

On the effective circle, this region is an arc of length at most $\pi$, so its probability and Bloch moment saturate a binary response bound. Increasing the mass beyond the PVM requirement moves the effect's required moments strictly inside that bound. The region equation cannot then hold. This argument covers every bias and both nonzero correlation singular values.
\begin{theorem}
\label{thm:qubit-rank-two} At rank two, every positive zero-moment measure on the unit circle of $E$ satisfying the PVM inequalities also satisfies all POVM inequalities for the same datum and mass. Consequently, every qubit datum with $\rank C\le2$ satisfies
\begin{equation}
c_{\mathrm A}(\vec a,C)=c_{\mathrm P}(\vec a,C). \label{eq:qubit-rank-two-value}
\end{equation}
\end{theorem}
\begin{proof}
At rank at most one, let $t_0$ be the value in Eq.~\eqref{eq:qubit-rank-one-value}. A PVM-feasible spectrum on $\pm\vec v$ has equal weights and mass $t\ge t_0$. It realizes any POVM with responses
\[
f_i(\pm\vec v)=q_i+\frac{(\vec a\pm\vec b)\cdot\vec x_i}{t}.
\]
The effect bound $|\vec x_i|\le q_i$ gives nonnegativity, and the POVM normalization gives $\sum_i f_i=1$. Their mass and vector moment are the required ones. If $t_0=0$, then $\vec a=\vec b=0$ and both scales vanish.

For rank two, fix a positive continuous probability density $\varpi$ of zero mean on the effective circle, and let $t_\varpi$ be the least mass for which $t_\varpi\varpi$ satisfies the PVM inequalities. For an effect $M=q\openone_2+\vec x\cdot\vec\sigma$, its required mass and vector moment are
\[
F_t(M)=(q+\vec a\cdot\vec x/t,C_r^{\mathsf T}\vec x/t).
\]
Binary feasibility from Section~S3 gives, for every $(s,\vec y)$,
\[
(s,\vec y)\cdot F_{t_\varpi}(M) \le\int[s+\vec y\cdot\vec n]_+d\varpi.
\]
For a projection this is the PVM source inequality, since the right-hand side is $(s+\int|s+\vec y\cdot\vec n|d\varpi)/2$. Every effect is a convex combination of a projection, zero and the identity, so the same bound holds for $M$.

For $0<q<1$, the constant response $q$ realizes $(q,\vec0)$. Its gap from the binary bound is strictly positive at every nonzero test:
\[
(1-q)\int[s+\vec y\cdot\vec n]_+d\varpi +q\int[-s-\vec y\cdot\vec n]_+d\varpi>0.
\]
Full support ensures strictness. For $t>t_\varpi$, $F_t(M)$ is a strict convex combination of $F_{t_\varpi}(M)$ and $(q,\vec0)$. Hence every proper nonzero effect obeys
\begin{equation}
(s,\vec y)\cdot F_t(M) <\int[s+\vec y\cdot\vec n]_+d\varpi \qquad((s,\vec y)\ne0). \label{eq:povm-circle-strict-response}
\end{equation}
If a POVM required $T_M>t_\varpi$, Eq.~\eqref{eq:povm-winning-cell-moments} with $(B,\vec d,\vec b)=(C_r,\vec a,\vec0)$ would give
\begin{equation}
\int_{\Omega_i}(1,\vec n)d\varpi=F_{T_M}(M_i). \label{eq:povm-circle-contact}
\end{equation}
A nontrivial POVM with zero effects removed has $0<q_i<1$. Applying Eq.~\eqref{eq:povm-circle-strict-response} to the constant tests $(s,\vec y)=(\pm1,\vec0)$ shows that each region has probability strictly between zero and one, and hence positive length. Choose a branch of smallest height. Its inequalities are $(\vec\beta_i-\vec\beta_j)\cdot\vec n\ge\alpha_j-\alpha_i\ge0$, so its region on the circle is an arc of length at most $\pi$. If its midpoint is $\theta_0$ and half-length $\delta$, its indicator attains the binary test $\cos(\theta-\theta_0)-\cos\delta$. Equation~\eqref{eq:povm-circle-contact} therefore makes the inequality in Eq.~\eqref{eq:povm-circle-strict-response} an equality, a contradiction. No POVM requires more than the PVM mass of this fixed density.

Finally, let $\mu$ be any PVM spectrum on the effective circle. Circular heat convolution gives a positive smooth spectrum $\mu_\tau$ of the same mass and zero moment. The average rotation is $e^{-\tau}\openone_2$, so convexity of the source gives
\begin{equation}
\int|s+\vec y\cdot\vec n|d\mu_\tau \ge\int|\vec a s+C_rR_\varphi\vec y|d\eta_\tau(\varphi) \ge|\vec a s+e^{-\tau}C_r\vec y|. \label{eq:povm-circle-smoothing}
\end{equation}
The density result makes $\mu_\tau$ feasible for all POVMs at $(\vec a,e^{-\tau}C)$. Sending $\tau\downarrow0$ in each continuous test inequality recovers feasibility of $\mu$ at $(\vec a,C)$. Minimizing mass proves equality of the scales.
\end{proof}
This applies to every feasible circle measure, including those obtained by the general recurrence outside the analytic contact domain.

At full rank, we apply the same region equation and effect-positivity bound to the reconstructed PVM density $u_*$. In its source coordinates, the smallest-height region is a convex cone section on the sphere. A uniform bound on these cone moments will ensure, in the reduced notation of Section~S3, that
\[
\mathcal K(q,\vec y;u_*)\ge\mathcal S(q,\vec y) \qquad\text{for every retained POVM test}.
\]
Using the source data $S,r,R$ from Eq.~\eqref{eq:povm-direct-source-data}, define
\begin{equation}
H=\int\frac{\vec e\vec e^{\mathsf T}}{r(\vec e)}d\omega,\qquad \vec h=H\vec d,\qquad \kappa=R-\vec d\cdot\vec h,\qquad \vec p=S^{-1}\vec d. \label{eq:povm-biased-source-data}
\end{equation}
These quantities use the same matrix integral as the PVM value: $H=\mathsf R(S^2)/2$ and $R=\Tr[S^2\mathsf R(S^2)]/2$. Since $J=S(\openone-\vec p\vec p^{\mathsf T})S\succeq0$, one has $|\vec p|\le1$ and $\kappa=\int\vec e^{\mathsf T}J\vec e/r\,d\omega>0$. The source spectrum $2(r+\vec d\cdot\vec e)d\omega$ has mass $2R$ and mean hidden direction $B^{\mathsf T}\vec h/R$. The vanishing moment after the inverse Lorentz change is therefore
\begin{equation}
\vec q_* =B^{\mathsf T}\vec h/R,\qquad c_{\mathrm P}=2R\sqrt{1-|\vec q_*|^2}. \label{eq:povm-biased-spectral-value}
\end{equation}
Pushforward to $\vec n'=B^{\mathsf T}\vec e/r$ gives probability measure $\varpi$ of mean $\vec b=\vec q_*$. A source mass $T_M$ maps to original mass $T_M/\gamma$, with $\gamma=(1-|\vec q_*|^2)^{-1/2}$. Bounding outcome moments in terms of anisotropy and relative bias will exclude a POVM mass above the projective value.
\begin{theorem}
\label{thm:povm-finite-source-region} \label{thm:povm-isotropic-source-ball} Let $C$ be invertible and $J\succeq0$. At the optimal PVM plane, put
\begin{equation}
\chi=\frac{\lambda_{\max}(S)}{\lambda_{\min}(S)},\qquad s=|S^{-1}\vec d|,\qquad \eta=\frac{2(\chi^3-1)}{(\chi+1)(\chi^2+1)}. \label{eq:povm-source-anisotropy}
\end{equation}
Equality of scales holds whenever
\begin{equation}
\left(1-\frac54s\right)\left(1-\frac78s\right) \ge4\eta s(2\chi+s) +\frac23(\chi-1)s\left[4+(1+4\eta)(\chi+2)s^2\right]. \label{eq:povm-finite-source-region}
\end{equation}
For each $\chi\ge1$, this domain is the full bias ball $s\le s_\chi$, where $s_\chi$ is the unique root of equality in Eq.~\eqref{eq:povm-finite-source-region} on $(0,4/5]$. In particular $s_1=4/5$, and $s_\chi>0$ for every finite anisotropy. Throughout the domain,
\[
c_{\mathrm A}=c_{\mathrm P}=2R\sqrt{1-|\vec q_*|^2}.
\]
\end{theorem}
\begin{proof}
Since $0\le s\le1$, the condition forces $s\le4/5$: its left-hand side is negative for $s>4/5$, whereas its right-hand side is nonnegative.

The region equation reduces POVM equivalence to a bound on the probability and first moment of a convex cone section. We derive this bound from the cone geometry for isotropic sources, then control its change under anisotropy. Normalize the source spectrum to $(r+\vec d\cdot\vec e)d\omega/R$ and write, for source region $D_i$,
\[
R_i=\int_{D_i}r\,d\omega,\qquad H_i=\int_{D_i}\frac{\vec e\vec e^{\mathsf T}}r\,d\omega,\qquad \vec m_i=\int_{D_i}\vec e\,d\omega.
\]
Its probability is $(R_i+\vec d\cdot\vec m_i)/R$ and its first moment is $B^{\mathsf T}(\vec m_i+H_i\vec d)/R$. In Eq.~\eqref{eq:povm-winning-cell-moments}, solve the vector equation for $\vec x_i/T_M$ and substitute into the scalar equation. The terms $\vec d\cdot\vec m_i$ cancel, leaving
\begin{equation}
q_i=Q_i:=\frac{R_i-\vec d^{\mathsf T}H_i\vec d}{\kappa},\qquad \vec x_i=\frac{T_M}{R}\vec V_i,\qquad \vec V_i=\vec m_i+(H_i-Q_iH)\vec d. \label{eq:povm-biased-cell-moments}
\end{equation}
Effect positivity gives $(T_M/R)|\vec V_i|\le Q_i$. A smallest-height branch has region
\[
(B\vec\beta_i-B\vec\beta_j)\cdot\vec e\ge(\alpha_j-\alpha_i)|S\vec e|.
\]
All coefficients on the right are nonnegative. Homogeneous extension therefore gives a proper convex cone, reducing the required bound to
\begin{equation}
Q_D\le2|\vec V_D| \quad\text{for every positive-area proper closed convex cone section }D. \label{eq:povm-cone-condition}
\end{equation}
Indeed, if a POVM needed $T_M>2R$, its selected region would obey $0<Q_i=q_i$ and
\[
\frac{T_M}{R}|\vec V_i|=|\vec x_i|\le Q_i\le2|\vec V_i|,
\]
which is impossible. A source spectrum realizing all POVMs maps back to mass $2R/\gamma$ by the inverse Lorentz change.

Compare directional absolute moments with the length of the cone's first moment. Their difference will express the weighted region moments through a nonnegative scalar function. Put $A=\int_Dd\omega$ and $\vec m=\int_D\vec e\,d\omega=M\vec u$, where $M>0$ and $|\vec u|=1$, and define
\begin{equation}
E(\vec v)=M|\vec v|-\int_D|\vec v\cdot\vec e|d\omega \label{eq:povm-centroid}
\end{equation}
and split the cone section at $\vec v\cdot\vec e=0$, with first moments $\vec m_+$ and $\vec m_-$. To show $E(\vec v)\ge0$, group direction pairs in the double integral for $\vec m_+\cdot\vec m_-$ by their common plane through the origin. Convexity makes the cone section in each plane an arc of length at most $\pi$, cut into adjacent arcs of lengths $a$ and $b$. If $s$ and $t$ measure distance from their shared endpoint, the angle between the paired directions is $s+t$. The decomposition has a positive plane-measure factor and angular Jacobian $|\sin(s+t)|=\sin(s+t)$, since $a+b\le\pi$. Each plane therefore contributes with the sign of
\[
\int_0^a\int_0^b\cos(s+t)\sin(s+t)dsdt =\tfrac12\sin a\sin b\sin(a+b)\ge0.
\]
Thus $\vec m_+\cdot\vec m_-\ge0$ and $|\vec m_+-\vec m_-|\le M$. Its component in direction $\vec v$ proves $E(\vec v)\ge0$.

Its source-weighted mass and normalized second moment are
\[
\delta=2\int E(S\vec n)d\omega=2MR-R_D,\qquad Z=\frac{2\int\vec n\vec n^{\mathsf T}E(S\vec n)d\omega}{\delta}.
\]
For $\delta>0$, $Z\succeq0$ and $\Tr Z=1$. Set
\begin{equation}
\vec v=(\openone-4Z)\vec p,\qquad L_0=\kappa S^{-1}+\vec h\vec p^{\mathsf T}. \label{eq:povm-deficit-affine}
\end{equation}
Spherical integration by parts for the degree-one function $E(S\vec n)$ gives
\[
\int(\vec p\cdot\vec n)\nabla E(S\vec n)d\omega =MS\vec h-\tfrac12SH_D\vec d =\tfrac\delta2(4Z-\openone)\vec p.
\]
Here the gradient is with respect to $\vec n$, and $H$ commutes with $S$. It follows that $H_D\vec d=2M\vec h+\delta S^{-1}\vec v$. Combining this with $R_D=2MR-\delta$ yields the direct cell formulas
\begin{equation}
Q_D=2M-\frac\delta\kappa(1+\vec p\cdot\vec v),\qquad \vec V_D=M\vec u+\frac\delta\kappa(\vec h+L_0\vec v). \label{eq:povm-deficit-identity}
\end{equation}
Applying the supporting inequality for the norm gives
\begin{equation}
2|\vec V_D|-Q_D\ge\frac\delta\kappa\mathcal L, \qquad \mathcal L=1+\vec p\cdot\vec v+2\vec u\cdot(\vec h+L_0\vec v). \label{eq:povm-source-gap}
\end{equation}
If $\delta=0$, continuity and nonnegativity imply $E=0$, and the same identities give $Q_D=2M$, $\vec V_D=\vec m$. For positive deficit, it therefore suffices to bound $\mathcal L$ uniformly from below.

Before the source change, the normalized deficit has moments
\[
s_0=\int E\,d\omega=M-A/2,\qquad K_D=\int_D\vec e\vec e^{\mathsf T}d\omega,\qquad Z_0=\frac{\int\vec n\vec n^{\mathsf T}E(\vec n)d\omega}{s_0}.
\]
Positive $\delta$ and invertibility of $S$ give $s_0>0$. The spherical identity $\int\vec n\vec n^{\mathsf T}|\vec n\cdot\vec e|d\omega=(\openone+\vec e\vec e^{\mathsf T})/8$ gives
\[
Z_0=\frac{M\openone/3-(A\openone+K_D)/8}{s_0}.
\]
For $q_0=2s_0/M\in(0,1)$ and $P_{\vec u}=\openone-\vec u\vec u^{\mathsf T}$, its moments satisfy
\begin{equation}
\begin{aligned}
&\Tr Z_0=1,\qquad Z_0\succeq\frac{\openone}6+\frac{P_{\vec u}}{6(1+q_0)},\\
&P_{\vec u}Z_0P_{\vec u}\succeq\frac{P_{\vec u}}3,\qquad \vec u^{\mathsf T}Z_0\vec u\le\frac14-\frac{q_0}{12(2-q_0)}.
\end{aligned}
\label{eq:povm-moment-domain}
\end{equation}
These bounds retain the coupling between the cone axis and moment matrix needed to estimate $\mathcal L$.

For their derivation, let $C_D$ be the radial cone. Translation along $\vec x\in C_D$ gives $C_D\subset C_D-t\vec x$ for $t\ge0$. The Gaussian integral $\int_{C_D}e^{-|\vec z-t\vec x|^2/2}d\vec z$ is therefore nondecreasing at zero. Its derivative shows $\vec m\cdot\vec x\ge0$, so $\vec u\in C_D^*$. The centroid is also interior to $C_D$. In polar coordinates about $\vec u$, each radial section thus ends at a cosine $c(\varphi)\in[0,1]$, and
\[
4\pi A=\int(1-c)d\varphi,\quad 8\pi M=\int(1-c^2)d\varphi,\quad 12\pi\vec u^{\mathsf T}K_D\vec u=\int(1-c^3)d\varphi.
\]
It follows that $\vec u^{\mathsf T}Z_0\vec u=1/6+\int c(1-c)^2/[12\int c(1-c)]$. Under the weight $1-c$, the mean of $c$ is $q_0/(2-q_0)$. Nonnegative variance then gives the longitudinal upper bound in Eq.~\eqref{eq:povm-moment-domain}. Pr\'ekopa log-concavity~\cite{Prekopa1973} of the translated Gaussian integral gives, through its Hessian,
\[
3K_D/A-(8/\pi)(M/A)^2\vec u\vec u^{\mathsf T}\preceq\openone.
\]
Compression to $\vec u^\perp$ and substitution into $Z_0$ give its transverse lower bound.

For the remaining matrix bound, take $\vec y\in C_D^*$ and weight the translated Gaussian integral by $\vec y\cdot\vec z$. Shifting the integration variable changes the domain to $C_D-t\vec x$ and the weight to $\vec y\cdot\vec z+t\vec x\cdot\vec y$. The domain enlarges and the added weight is nonnegative. Differentiation therefore gives $\vec x^{\mathsf T}K_D\vec y\ge0$. By duality and symmetry, $K_D$ maps both $C_D$ and $C_D^*$ into themselves.

Suppose first that $C_D$ contains no line. The powers of $K_D/\lambda_{\max}(K_D)$ applied to $\vec m$ converge to a nonzero top eigenvector. If the limit were zero, applying those powers to $\vec m\pm\epsilon\vec w\in C_D$ for a top eigenvector $\vec w$ would put both $\vec w$ and $-\vec w$ in $C_D$. Normalize the nonzero limit to $\vec v_*\in C_D\cap C_D^*$. Radial integration about this vector yields
\[
4\vec v_*\cdot\vec m-A-3\vec v_*^{\mathsf T}K_D\vec v_* =\frac1{4\pi}\int c(1-c)^2d\varphi\ge0.
\]
Since $\vec v_*$ is a top eigenvector, $A\openone+3K_D\preceq4M\openone$ and $Z_0\succeq\openone/6$. The transverse bound gives $\lambda_2(Z_0)\ge1/3$, while the longitudinal value bounds the smallest eigenvalue by $1/4$. This eigenvalue is simple; write it as $1/6+e$ for $0\le e\le1/12$. Its eigenvector is $\vec v_*$, and radial integration gives $k=\vec u\cdot\vec v_*\ge1-3q_0e$.

Let $h=[6(1+q_0)]^{-1}$ and write $\vec u=k\vec v_*+\sqrt{1-k^2}\vec w$, with $\vec w\perp\vec v_*$. Since the other two eigenvalues are at least $1/3$,
\[
Z_0-\frac{\openone}{6}-hP_{\vec u} \succeq e\vec v_*\vec v_*^{\mathsf T}+\frac16P_{\vec v_*}-hP_{\vec u}.
\]
On the span of $\vec v_*,\vec w$, the right-hand matrix is
\[
\begin{pmatrix}
e-h(1-k^2)&hk\sqrt{1-k^2}\\
hk\sqrt{1-k^2}&1/6-hk^2
\end{pmatrix}
.
\]
The lower-right entry and perpendicular eigenvalue $1/6-h$ are positive. Its determinant is $h/6$ times
\[
(1-6e)k^2-1+e/h \ge e^2(36q_0+9q_0^2-54q_0^2e)\ge0.
\]
This gives the matrix bound for cones containing no line. Intersecting a general cone with $\{\vec x:\vec u\cdot\vec x\ge\epsilon|\vec x|\}$ and taking $\epsilon\downarrow0$ extends it by continuity of angular moments.

For an isotropic source, scale $S$ to $\openone$ and rotate the bias to $p\vec e_3$. Then $Z=Z_0$, $\vec h=p\vec e_3/3$ and $\kappa=1-p^2/3$. At $p=4/5$, we need the endpoint estimate
\begin{equation}
\mathcal L(\openone,p\vec e_3;\vec u,Z_0)>1/2000. \label{eq:povm-rational-margin}
\end{equation}
To prove it, reduce the allowed moment matrices to a two-variable polynomial inequality. Put $x=p^2$ and $z=u_3$, and define
\[
\begin{aligned}
A_z&=xz^2+2pxz^3/3+2p(1-x/3)z,\\
B_z&=x(1-z^2)(1+2pz/3),\\
k&=xz+2pxz^2/3+p(1-x/3),\qquad L_z=1+x+8pz/3.
\end{aligned}
\]
Reflection in the plane of $\vec u$ and $\vec e_3$ preserves the moment constraints and the linear gap. Averaging the reflected matrices removes their coupling to the normal direction. In the basis $(\vec u,(\vec e_3-z\vec u)/\sqrt{1-z^2})$, write the remaining block as $\left(
\begin{smallmatrix}
Z_{11}&Z_{12}\\
Z_{12}&Z_{22}
\end{smallmatrix}
\right)$ and set $v=12Z_{11}-2$. The moment bounds give
\[
0\le v\le1,\qquad q_0\le\frac{2(1-v)}{2-v},\qquad \frac1{6(1+q_0)}\ge\frac{2-v}{6(4-3v)},\qquad Z_{22}\le\frac{6-v}{12},
\]
and
\[
Z_{12}^2\le(Z_{11}-1/6)\left(Z_{22}-1/6-\frac1{6(1+q_0)}\right) \le\frac{v(3v^2-14v+12)}{144(4-3v)}.
\]
Continuity supplies $|z|=1$. In these coordinates,
\[
\mathcal L=L_z-4[Z_{11}A_z+Z_{22}B_z+2Z_{12}\sqrt{1-z^2}\,k].
\]
Here $B_z\ge0$, and $k>0$ because its quadratic discriminant is $p^4(8p^2-15)/9<0$. Increasing $Z_{22}$ increases both the subtracted expression and the bound on $|Z_{12}|$. Therefore
\[
\mathcal L\ge L_z-\frac{(2+v)A_z+(6-v)B_z}{3} -\frac23\sqrt{\frac{v(3v^2-14v+12)}{4-3v}}\sqrt{1-z^2}\,k.
\]
To bound this expression on $[0,1]\times[-1,1]$, define the polynomials
\[
\begin{aligned}
F(v,z)&=L_z-\frac{(2+v)A_z+(6-v)B_z}{3}-\frac1{2000},\\
P(v,z)&=(4-3v)F(v,z)^2 -\frac49v(3v^2-14v+12)(1-z^2)k^2.
\end{aligned}
\]
To bound the radical it suffices that both $F$ and $P$ be positive. Expand them in the Bernstein basis on the rectangles
\[
v=\frac{i+\xi_1}{4},\qquad z=-1+\frac{j+\xi_2}{4},\qquad i=0,\ldots,3,\quad j=0,\ldots,7,\quad 0\le\xi_1,\xi_2\le1.
\]
For a polynomial $\sum_{a,b}c_{ab}\xi_1^a\xi_2^b$ of bidegree $(m,n)$, the Bernstein coefficients are
\[
b_{k\ell}=\sum_{a=0}^k\sum_{b=0}^{\ell} c_{ab}\frac{\binom ka\binom{\ell}b}{\binom ma\binom nb}.
\]
Their bidegrees are $(1,3)$ for $F$ and $(3,6)$ for $P$. The smallest coefficients on these rectangles are
\[
\min b_{k\ell}(F)=\frac{877}{6000},\qquad \min b_{k\ell}(P)=\frac{65158379}{3888000000}.
\]
Because the Bernstein basis is nonnegative and sums to one, these positive rational coefficients imply $F,P>0$ throughout $[0,1]\times[-1,1]$. With $4-3v>0$, the definition of $P$ then bounds the radical strictly by $F$. Substitution proves the cone-gap margin in Eq.~\eqref{eq:povm-rational-margin}.

Cubic interpolation extends this endpoint margin to a uniform lower bound throughout the isotropic bias ball. Fix a unit bias direction $\vec e$ and write the gap as
\[
\mathcal L_0(s)=1+c_1s+c_2s^2+c_3s^3.
\]
Equation~\eqref{eq:povm-moment-domain} gives $\operatorname{spec}Z_0\subset[1/6,1/2]$. Expanding the gap gives
\[
c_2=1-4\vec e^{\mathsf T}Z_0\vec e\le\frac13,\qquad c_3=\frac83[\vec u^{\mathsf T}Z_0\vec e -(\vec u\cdot\vec e)\vec e^{\mathsf T}Z_0\vec e],\qquad |c_3|\le\frac49.
\]
A matrix element between orthogonal unit vectors is bounded by half the spectral width, giving the last inequality. With $a=4/5$, cubic interpolation yields
\[
\mathcal L_0(s)-\left[1-\frac sa+\frac sa\mathcal L_0(a)\right] =s(s-a)[c_2+(s+a)c_3].
\]
For $0\le s\le a$, the endpoint estimate and the coefficient bounds imply
\[
\mathcal L_0(s)\ge \left(1-\frac54s\right) \left(1-\frac{124}{225}s-\frac{16}{45}s^2\right) \ge\left(1-\frac54s\right)\left(1-\frac78s\right).
\]
The difference in the last inequality is $s(1-5s/4)[583/1800-16s/45]\ge0$ on this interval.

To transfer the isotropic bound to general $S$, we control the two effects of anisotropy: the change of angular directions and the reweighting of the deficit. Let $d\mu=E(\vec w)d\omega/s_0$, so $Z_0=\int\vec w\vec w^{\mathsf T}d\mu$. With
\[
q_S(\vec w)=|S^{-1}\vec w|,\qquad \vec n(\vec w)=S^{-1}\vec w/q_S(\vec w),
\]
the source-weighted moment is
\[
Z=\frac{\int q_S^{-4}\vec n\vec n^{\mathsf T}d\mu} {\int q_S^{-4}d\mu}.
\]
For a positive matrix with eigenvalues in $[a,b]$, a unit vector $\vec w$ satisfies
\[
\vec w^{\mathsf T}A^2\vec w\le(a+b)\vec w^{\mathsf T}A\vec w-ab \le\frac{(a+b)^2}{4ab}(\vec w^{\mathsf T}A\vec w)^2.
\]
Applying this to $S^{-1}$ bounds the sine of the angle between $\vec w$ and $\vec n$ by $(\chi-1)/(\chi+1)$, and hence bounds the difference of their rank-one projectors by the same number.

Reweighting a probability by a function in $[m,M]$ changes it in total variation by at most $(\sqrt M-\sqrt m)/(\sqrt M+\sqrt m)$. Indeed, if $t$ is its mean weight, convexity of the positive part bounds this variation by $(M-t)(t-m)/[(M-m)t]$, whose maximum occurs at $t=\sqrt{mM}$. The ratio of the extreme weights $q_S^{-4}$ is at most $\chi^4$. Since a unit-direction quadratic form of a rank-one projector lies in $[0,1]$, the direction and weight estimates together give
\[
\|Z-Z_0\|_{\mathrm{op}} \le\frac{\chi-1}{\chi+1}+\frac{\chi^2-1}{\chi^2+1}=\eta.
\]
This uniform cone estimate controls the matrix in $\mathcal L$. For its other coefficients, compare anisotropic and isotropic values at the same $\vec p$:
\[
\vec h_0=\vec p/3,\qquad L_{00}=(1-s^2/3)\openone+\vec p\vec p^{\mathsf T}/3, \qquad\|L_{00}\|_{\mathrm{op}}=1.
\]
Because $H$ commutes with $S$,
\[
\|HS-\openone/3\|_{\mathrm{op}}\le(\chi-1)/3,\qquad \|RS^{-1}-\openone\|_{\mathrm{op}}\le\chi-1.
\]
These bounds follow by placing $r(\vec e)$ between the extreme eigenvalues of $S$ and using $\int e_i^2d\omega=1/3$. The same bounds put every eigenvalue of $(\vec p^{\mathsf T}SHS\vec p)S^{-1}$ between $s^2/(3\chi^2)$ and $\chi^2s^2/3$. Substitution into $L_0$ therefore gives
\[
|\vec h-\vec h_0|\le\frac{\chi-1}{3}s,\qquad \|L_0-L_{00}\|_{\mathrm{op}} \le D_\chi(s):=(\chi-1)\left[1+\frac{\chi+2}{3}s^2\right].
\]
Let $\vec v_0=(\openone-4Z_0)\vec p$. The spectrum of $Z_0$ gives $|\vec v_0|\le s$, and the moment perturbation gives $|\vec v-\vec v_0|\le4\eta s$. Comparing the anisotropic and isotropic gaps in Eq.~\eqref{eq:povm-source-gap} now yields
\[
\begin{aligned}
|\mathcal L-\mathcal L_0| &\le4\eta s^2+\tfrac23(\chi-1)s +2sD_\chi(s)+8\eta s[1+D_\chi(s)]\\
&=4\eta s(2\chi+s) +\tfrac23(\chi-1)s[4+(1+4\eta)(\chi+2)s^2].
\end{aligned}
\]
This is the anisotropy allowance in Eq.~\eqref{eq:povm-finite-source-region}. Under that condition $\mathcal L\ge0$ for every cone, proving Eq.~\eqref{eq:povm-cone-condition}. Stationary region moments and effect positivity then exclude every POVM mass above the projective value.

Finally, the left-hand side of Eq.~\eqref{eq:povm-finite-source-region} is strictly decreasing from one to zero on $[0,4/5]$, and the right-hand side is a polynomial with nonnegative coefficients. Equality has a unique root there, equal to $4/5$ at $\chi=1$ and positive for every finite $\chi$. Both sides depend continuously on $\chi$, and the right-hand side increases with $\chi$, so the root is continuous and decreasing.
\end{proof}
At $\chi=1$, the condition is the entire isotropic source ball
\begin{equation}
S=\lambda\openone_3,\qquad |\vec d|\le4\lambda/5,\qquad\lambda>0. \label{eq:povm-isotropic-source-ball}
\end{equation}
Here $R=\lambda$ and $H=\openone/(3\lambda)$, so its common scale is
\begin{equation}
c_{\mathrm A}=c_{\mathrm P} =2\lambda\sqrt{1-|\vec d|^2/(9\lambda^2)}. \label{eq:povm-isotropic-value}
\end{equation}
In the original parameters, Eq.~\eqref{eq:pvm-isotropic-original-bias} identifies this ball with $|\vec a|^2\le64\lambda^2/209$ inside the algebraically solvable family \eqref{eq:pvm-isotropic-original-family}. At $s=0$, the same anisotropy condition holds for every positive definite $S$. In particular $\vec a=0$ gives $\vec q_*=0$, $B=C$, and $\vec d=0$, recovering~\cite{ZhangPOVM2026}
\begin{equation}
c_{\mathrm A}(0,C)=c_{\mathrm P}(0,C) =2\int_{S^2}|C^{\mathsf T}\vec e|d\omega. \label{eq:s3-unbiased-povm-value}
\end{equation}
Continuity gives this formula for singular $C$ as well.

Uniqueness also gives an exact equality criterion across the entire full-rank projective domain. Using the notation of Section~S3, define
\begin{equation}
\mathfrak D(\vec a,C)=\int_{\mathcal Q} [\mathcal S(q,\vec y)-\mathcal K(q,\vec y;u_*)]_+^2d\nu. \label{eq:povm-biased-integral-region}
\end{equation}
Then
\begin{equation}
\mathfrak D(\vec a,C)=0 \quad\Longleftrightarrow\quad c_{\mathrm A}(\vec a,C)=c_{\mathrm P}(\vec a,C)=\Phi(\vec q_*). \label{eq:povm-biased-exact-value}
\end{equation}
A zero deficit integral makes the continuous deficit zero at every interior test, since $\nu$ has full support; continuity then supplies boundary tests. Thus $u_*$ is a POVM spectrum of the PVM mass. Conversely, weak compactness gives a minimum-mass POVM spectrum. If its mass equals $c_{\mathrm P}$, uniqueness of the PVM spectrum identifies it with $u_*d\omega$, and the integral vanishes.

At equality, every minimum-mass POVM measure also minimizes the PVM mass and thus equals $u_*d\omega$. The POVM recurrence consequently has the weak limit in Eq.~\eqref{eq:pvm-explicit-spectral-limit}. Its region moments provide every required outcome division.

\section{S5. Two-qubit state families} \label{sec:analytical-examples}

The contact solutions let us follow how physical changes to a state alter its required distribution of local preparations. We begin with a biased family whose weights and mass are elementary, making their separate dependence on polarization and noise explicit. Its unit-mass spectrum constructs a channel parent measurement, and one member also supplies the reverse boundary of the Bowles family. Solving the forward contact equation then determines the complete arbitrary-measurement one-way interval.

We next vary the mixture and entanglement within established state families, obtaining the projective boundary throughout the generalized Werner and noisy MEMS parameter ranges and the stated analytic segments of the other examples. Comparing their spectra separates the effects of global mixedness, conditional purity and the trusted marginal. The final group follows filtering and damping, where a change of marginal support can make the limiting preparation problem differ from that at the endpoint.

We take Alice as the first tensor factor and use the basis $\ket{00},\ket{01},\ket{10},\ket{11}$. The superscripts $A\to B$ and $B\to A$ specify the steering direction, while $\mathrm P$ and $\mathrm A$ specify PVMs and POVMs. The states below have the Bloch form
\begin{equation}
\rho=\frac14\left[\openone_4+a_0\sigma_3\otimes\openone_2+b_0\openone_2\otimes\sigma_3+\sum_{j=1}^3t_j\sigma_j\otimes\sigma_j\right]. \label{eq:example-X-bloch}
\end{equation}
Whitening makes the role of the trusted marginal explicit. For $A\to B$, put $D=1-b_0^2$. When $D>0$, the canonical data are
\begin{equation}
\vec a=A\vec e_3,\qquad A=\frac{a_0-b_0t_3}{D},\qquad C=\operatorname{diag}\left(\frac{t_1}{\sqrt D},\frac{t_2}{\sqrt D},L\right),\qquad L=\frac{t_3-a_0b_0}{D}. \label{eq:example-X-whitening}
\end{equation}
Interchanging $a_0$ and $b_0$ gives the reverse data. If the trusted marginal is pure, the state is a product and the support restriction in Section~S1 gives the length of the other party's Bloch vector. For axial correlations we write $T=|t_1|/\sqrt D=|t_2|/\sqrt D$.

For axial data, $J\succeq0$ is equivalent to $L^2-A^2=(t_3^2-a_0^2)/D\ge0$. Rotational symmetry reduces the unique plane parameter of Theorem~\ref{thm:pvm-finite-integral-root} to $\vec q=q\vec e_3$. Define
\begin{equation}
X(q)=T^2(1-q^2),\qquad Y(q)=(L+Aq)^2-X(q),\qquad I(X,Y)=\int_0^1\sqrt{X+Yx^2}\,dx. \label{eq:example-axial-integral}
\end{equation}
Stationarity and mass become
\begin{equation}
\begin{gathered}
\int_0^1\frac{-T^2q(1-x^2)+A(L+Aq)x^2}{\sqrt{X(q)+Y(q)x^2}}\,dx=0,\qquad -1<q<1,\\
c_{\mathrm P}=2I(X(q),Y(q)),
\end{gathered}
\label{eq:pvm-axial-root}
\end{equation}
The mass integral is elementary:
\begin{equation}
I(X,Y)=
\begin{cases}
\tfrac12[\sqrt{X+Y}+X\operatorname{arsinh}\sqrt{Y/X}/\sqrt Y],&Y>0,\\
[2pt] \sqrt X,&Y=0,\\
[2pt] \tfrac12[\sqrt{X+Y}+X\arcsin\sqrt{-Y/X}/\sqrt{-Y}],&Y<0.
\end{cases}
\label{eq:example-axial-primitive}
\end{equation}
Use continuous limits for degenerate arguments. Within the analytic PVM domain the curves follow the scalar contact equation; elsewhere they use Eqs.~\eqref{eq:pvm-reduced-source-integral} and \eqref{eq:povm-source-integral} for PVMs and POVMs. Axial symmetry leaves a density in $n_3$, with azimuthal integrals split into elementary pieces by affine-branch crossings. Steering begins at $c=1$. In paired figures, POVMs appear on the left and PVMs on the right.
\subsection{Bias and measurement compatibility}

\begin{figure}[H]
\centering\includegraphics[width=.96\textwidth]{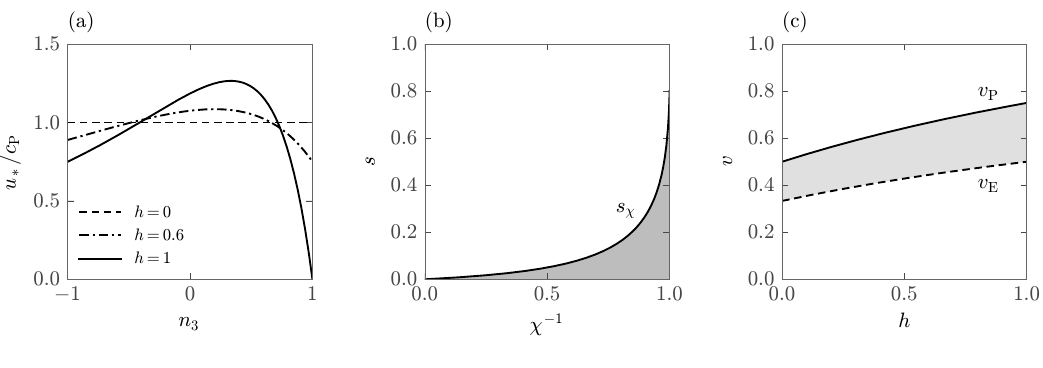}
\caption{Bias, angular weights and measurement equivalence. (a) Normalized optimal spectra of Eq.~\eqref{eq:example-elementary-spectrum} at $h=0,0.6,1$. (b) The shaded bias ball $s\le s_\chi$ from Theorem~\ref{thm:povm-finite-source-region} ensures equality of PVM and POVM scales at each source anisotropy $\chi$. (c) Separability and projective-steering boundaries of the same biased family. The shaded interval is entangled and projectively unsteerable; for $h\le4/5$, the steering boundary also holds for arbitrary POVMs.}\label{fig:example-steering-domains}
\end{figure}

\begin{example}[Biased states with elementary thresholds]
\label{ex:elementary-biased} An isotropic source lets us separate a change in ensemble mass from a redistribution of its weight. Take the source in Section~S4 with $B=-\lambda\openone_3$ and $\vec d=\lambda h\vec e_3$, where $0\le h\le1$. Its zero-moment plane is $\vec q_*=-h\vec e_3/3$. The inverse Lorentz transformation returns the canonical state parameters
\[
\vec a=\frac{2\lambda h}{\sqrt{9-h^2}}\vec e_3,\qquad C=-\lambda\operatorname{diag}\left(1,1,\frac{3-h^2}{\sqrt{9-h^2}}\right).
\]
Bias and longitudinal correlation vary together in this family. For $0\le v,h\le1$, put $\lambda=v\sqrt{9-h^2}/[(3-h)(1+h)]$. Its states are
\begin{equation}
\begin{aligned}
\rho(v,h)={}&\frac{\openone_4}{4}+\frac{vh}{2(3-h)(1+h)}\sigma_3\otimes\openone_2-\frac{v}{4(3-h)(1+h)}\left[\sqrt{9-h^2}\sum_{j=1}^2\sigma_j\otimes\sigma_j+(3-h^2)\sigma_3\otimes\sigma_3\right].
\end{aligned}
\label{eq:example-elementary-state}
\end{equation}
Here $h$ is the relative source bias and $v$ is white-noise visibility: $\rho(v,h)=v\rho(1,h)+(1-v)\openone_4/4$.

Its four eigenvalues give the physical range:
\[
\begin{gathered}
\frac14\left[1-\frac{v(3+h)(1-h)}{(3-h)(1+h)}\right],\qquad \frac{1-v}{4},\qquad
\frac14\left[1+\frac{v(3+h)}{1+h}\right],\qquad \frac14\left[1-\frac{v(3+h^2)}{(3-h)(1+h)}\right].
\end{gathered}
\]
They are nonnegative on the displayed square. In particular, $(3-h)(1+h)-(3+h^2)=2h(1-h)\ge0$ controls the last eigenvalue, while the second requires $v\le1$. At $v=1$ the states have rank three for $0<h<1$, with rank-one and rank-two endpoints at $h=0$ and $h=1$.

At fixed $h$, decreasing visibility weakens all conditional-state demands by the same factor. It rescales the spectral mass without moving the contact plane:
\begin{equation}
\vec q_*=-\frac h3\vec e_3,\qquad c_{\mathrm P}(v,h)=\frac{2v(3+h)}{3(1+h)},\qquad v_{\mathrm P}(h)=\frac{3(1+h)}{2(3+h)}. \label{eq:example-elementary-boundaries}
\end{equation}
Only $h$ changes the normalized angular weight. On the original Bloch sphere with normalized area $d\omega$, write $d\mu=u_*d\omega$ with
\begin{equation}
u_*(\vec n)=c_{\mathrm P} \frac{(1-h^2/9)[1+h^2/3-(4h/3)n_3]}{(1-hn_3/3)^4}. \label{eq:example-elementary-spectrum}
\end{equation}
This expression follows by transforming the source density $2\lambda(1-hn'_3)$ back through the plane $\vec q_*$. Its numerator is nonnegative because its least value is $(1-h)(1-h/3)$. Integration over $n_3$ gives
\[
\frac12\int_{-1}^1u_*(n_3)\,dn_3=c_{\mathrm P},\qquad \frac12\int_{-1}^1n_3u_*(n_3)\,dn_3=0.
\]
Axial symmetry makes the transverse moments zero. Thus the angular weight can change with local bias while its first moment still represents the same maximally mixed trusted marginal.

Contact derivatives give the projective response along $\vec e$. With $\gamma=(1-h^2/9)^{-1/2}$,
\begin{equation}
f_{\vec e}(\vec n)=-\operatorname{sgn} \left[e_1n_1+e_2n_2+\gamma e_3(n_3-h/3)\right],\qquad p(\pm|\vec e,\vec n)=\frac{1\pm f_{\vec e}(\vec n)}2. \label{eq:example-elementary-response}
\end{equation}
At $v=v_{\mathrm P}$ the measure has mass one, and these shifted spherical caps divide it into the required conditional preparations. Bias therefore changes both the distribution and its outcome divisions. For every positive visibility the same normalized spectrum models the map divided by $c_{\mathrm P}$.

Partial transposition gives eigenvalues
\[
\begin{gathered}
\frac{1+v}{4},\qquad \frac14\left[1+\frac{v(3+h)(1-h)}{(3-h)(1+h)}\right],\qquad
\frac14\left[1+\frac{v(3+h^2)}{(3-h)(1+h)}\right],\qquad \frac14\left[1-\frac{v(3+h)}{1+h}\right].
\end{gathered}
\]
The final eigenvalue is smallest, giving separability threshold and its relation to steering:
\begin{equation}
v_{\mathrm E}(h)=\frac{1+h}{3+h},\qquad v_{\mathrm P}(h)=\frac32v_{\mathrm E}(h),\qquad c_{\mathrm P}=\frac23\left[1-4\lambda_{\min}(\rho^{T_B})\right]. \label{eq:example-elementary-visibility}
\end{equation}
Figure~\ref{fig:example-steering-domains} compares these thresholds with the changing angular weights and the POVM equivalence region. Both thresholds increase strictly with $h$. The negativity $\mathcal N=(\|\rho^{T_B}\|_1-1)/2$ is $[v(3+h)/(1+h)-1]_+/4$, so throughout the entangled part,
\begin{equation}
c_{\mathrm P}=\frac23(1+4\mathcal N),\qquad \rho(v,h)\text{ is PVM-steerable}\quad\Longleftrightarrow\quad \mathcal N>\frac18. \label{eq:example-elementary-negativity}
\end{equation}
At fixed negativity, the entangled members of this family require the same PVM mass even though their angular distributions and outcome divisions differ. In particular, the entire steering boundary carries the same amount of entanglement as bias changes the visibility needed to reach it.

The isotropic source condition admits $h\le4/5$, so
\begin{equation}
c_{\mathrm A}=c_{\mathrm P}=\frac{2v(3+h)}{3(1+h)},\qquad v_{\mathrm A}=v_{\mathrm P},\qquad 0\le v\le1,\quad 0\le h\le\frac45. \label{eq:example-elementary-equality}
\end{equation}
In this rectangle, $v\le v_{\mathrm E}$ is separable, $v_{\mathrm E}<v\le v_{\mathrm P}$ has an all-POVM LHS model, and $v>v_{\mathrm P}$ is steerable with either class. The same condition $\mathcal N>1/8$ locates its all-POVM boundary.

Since Bob is maximally mixed, $\rho(v,h)$ is also a normalized Choi state. Define $\mathcal E_h$ by $\rho(1,h)=(\mathcal E_h\otimes\mathrm{id})(\ketbra{\Phi^+})$, with $\ket{\Phi^+}=(\ket{00}+\ket{11})/\sqrt2$ and output on Alice. Positivity and the Bob marginal give complete positivity and trace preservation. At $v=1$, the channel's Bloch action has drift $\vec a$ and matrix $CF$, $F=\operatorname{diag}(1,-1,1)$. Nonzero drift makes it nonunital for $h>0$. Output depolarization $\mathcal D_v(X)=vX+(1-v)\Tr(X)\openone_2/2$ produces $\rho(v,h)$.

For any effect $M_i$, the Choi identity reads
\begin{equation}
2\Tr_A[(M_i\otimes\openone_2)\rho(v,h)]=[(\mathcal D_v\circ\mathcal E_h)^*(M_i)]^{\mathsf T}. \label{eq:example-biased-choi}
\end{equation}
Thus a normalized LHS model is equivalent to a common parent for the Heisenberg images of the measurements~\cite{Uola2015}. At $v=v_{\mathrm P}(h)$, let $\mu_h$ be the unit-mass measure in Eq.~\eqref{eq:example-elementary-spectrum}. The operator measure
\begin{equation}
G_h(d\vec n)=2\tau_{\vec n}^{\mathsf T}d\mu_h(\vec n),\qquad \int G_h=\openone_2, \label{eq:example-biased-parent}
\end{equation}
is a parent POVM, and the ensemble's outcome probabilities give its classical postprocessing. For $0\le h\le4/5$, Eq.~\eqref{eq:example-elementary-equality} supplies these probabilities for every POVM. If $v\le v_{\mathrm P}$, put $\lambda=v/v_{\mathrm P}$ and add a separate parent outcome of effect $(1-\lambda)\openone_2$ to the scaled measure $\lambda G_h$. Assign that outcome to $i$ with probability $\Tr M_i/2$. This realizes the depolarizing part and constructs a common parent throughout the interval. For $v>v_{\mathrm P}$ the Choi state is projectively steerable, so its measurement images are incompatible. Consequently
\begin{equation}
\mathcal D_v\circ\mathcal E_h\text{ is incompatibility breaking}\quad\Longleftrightarrow\quad v\le\frac{3(1+h)}{2(3+h)},\qquad 0\le h\le\frac45. \label{eq:example-biased-channel-threshold}
\end{equation}
Comparing with the separability threshold gives an exact interval $v_{\mathrm E}<v\le v_{\mathrm P}=3v_{\mathrm E}/2$ for $0\le h\le4/5$: these channels preserve entanglement for some inputs while making all output measurements jointly measurable. Their common parent is specified by the same spectrum.

These states exhaust the potentially entangled canonical family with an isotropic optimal source. Equation~\eqref{eq:pvm-isotropic-original-family} requires two equal transverse singular values, the longitudinal value displayed above, and bias along the distinguished axis. Proper local rotations bring the negative-determinant branch to Eq.~\eqref{eq:example-elementary-state}. The positive-determinant branch is locally equivalent to its partial transpose; it is physical precisely for $v\le v_{\mathrm E}$ and is PPT there, so it contains only separable states.

An invertible filter on Bob preserves these directional scales. Its normalization is $\Tr(K^\dagger K)/2$, independent of $v$ and $h$, because Bob's original marginal is $\openone_2/2$. The same thresholds therefore hold along the filtered lines whose zero-visibility member is the product of Alice's maximally mixed state and Bob's filtered marginal.
\end{example}

\begin{example}[Bowles one-way states]
\label{ex:bowles}

In the Bowles family~\cite{BowlesOneWay}, both steering directions share the same isotropic correlations, but the two marginals have unequal local polarizations. Reversing trust therefore changes the normalization of the conditional states and the preparations that an ensemble must supply. Consider
\begin{equation}
\rho_{\mathrm B}(p)=p\ketbra{\Psi_-}+\frac{1-p}{5}\left(2\ketbra0\otimes\frac{\openone_2}{2}+3\frac{\openone_2}{2}\otimes\ketbra1\right),\qquad 0\leq p\leq1. \label{eq:example-bowles-state}
\end{equation}
These demands give distinct directional thresholds, shared by PVMs and POVMs:
\begin{equation}
\begin{aligned}
p_{\mathrm A}^{A\to B}=p_{\mathrm P}^{A\to B}&=0.488862118569\ldots,\\
p_{\mathrm A}^{B\to A}=p_{\mathrm P}^{B\to A}&=\frac12.
\end{aligned}
\label{eq:example-bowles-thresholds}
\end{equation}
Thus $0.488862118569\ldots<p\le1/2$ is the exact Alice-to-Bob one-way interval for arbitrary measurements. At both boundaries, the projective ensembles supply all POVM outcomes. Mixing with the separable state at $p=0$ covers lower singlet weights, while increasing conditional demands proves steering above each crossing. The reverse endpoint belongs to the elementary family. In the chosen basis,
\begin{equation}
\rho_{\mathrm B}=
\begin{pmatrix}
(1-p)/5&0&0&0\\
0&1/2&-p/2&0\\
0&-p/2&p/2&0\\
0&0&0&3(1-p)/10
\end{pmatrix}
. \label{eq:example-bowles-matrix}
\end{equation}
Trusted normalization uses $D_B=1-9(1-p)^2/25$ forward and $D_A=1-4(1-p)^2/25$ in reverse. Put $H=-p+6(1-p)^2/25$. Whitening gives forward data and primed reverse data
\begin{equation}
\begin{aligned}
A&=\frac{(1-p)(2-3p)}{5D_B},&T&=\frac p{\sqrt{D_B}},&L&=\frac H{D_B},\\
A'&=-\frac{(1-p)(3-2p)}{5D_A},&T'&=\frac p{\sqrt{D_A}},&L'&=\frac H{D_A}.
\end{aligned}
\label{eq:example-bowles-data}
\end{equation}
Their longitudinal $J$ entries are
\begin{equation}
L^2-A^2=\frac{7p-2}{8-3p},\qquad {L'}^2-{A'}^2=\frac{8p-3}{7-2p}. \label{eq:example-bowles-J}
\end{equation}
At the reverse threshold $p=1/2$, the stationary plane is $q=1/5$, and
\[
S=\frac5{4\sqrt6}\openone_3,\qquad \vec d=-\frac3{4\sqrt6}\vec e_3,\qquad \vec p=-\frac35\vec e_3.
\]
A simultaneous local rotation reversing the third axis changes the bias sign and leaves the diagonal correlations fixed. The resulting canonical state is exactly $\rho(2/3,3/5)$ in Eq.~\eqref{eq:example-elementary-state}. Since $2/3=v_{\mathrm P}(3/5)$ and $3/5<4/5$, Eq.~\eqref{eq:example-elementary-equality} gives common scale one. In the original reverse canonical coordinates, Eq.~\eqref{eq:example-elementary-spectrum} becomes the explicit probability density
\[
u_{B\to A}(\vec n)=\frac{96(7+5n_3)}{(5+n_3)^4}.
\]
Its zero first moment supplies Alice's canonical marginal, while the spherical partitions reproduce the conditional states at the reverse boundary.

For the forward direction, the scalar stationary equation locates the crossing. Both crossings lie in the interval
\begin{equation}
\frac{61}{125}\le p\le\frac{51}{100}. \label{eq:example-bowles-common-interval}
\end{equation}
Throughout this interval, the stationary transformed sources have sufficiently small anisotropy and bias for their PVM ensembles to supply all POVM outcomes. Specifically,
\begin{equation}
A\to B:\quad\chi<\frac{11}{10},\quad s<\frac15; \qquad B\to A:\quad\chi\le\frac{1007}{1000},\quad s<\frac{16}{25}, \label{eq:example-bowles-source-bounds}
\end{equation}
where $\chi,s$ are defined in Eq.~\eqref{eq:povm-source-anisotropy}. At the two upper corners, the margin in Eq.~\eqref{eq:povm-finite-source-region} exceeds $1/4$ and $1/1000$, respectively. Monotonicity in $\chi,s$ puts every $p$ in Eq.~\eqref{eq:example-bowles-common-interval} inside the equivalence region.

To prove the bounds, we compare trial source axis ratios with the stationary one. For either direction put $a=|A|/T$, $l=|L|/T$, $g=a^2$, and $j=l^2-a^2$. The stationary plane has the sign of $AL$. With $u=|q_*|$, its source axis ratio and relative bias are
\[
r=\frac{l+au}{\sqrt{1-u^2}},\qquad s=\frac{a+lu}{l+au},\qquad j=r^2(1-s^2).
\]
Define the positive integrals
\[
I_z(r)=\int_0^1\frac{x^2\,dx}{\sqrt{1+(r^2-1)x^2}},\qquad I_t(r)=\int_0^1\frac{1-x^2}{\sqrt{1+(r^2-1)x^2}}\,dx, \qquad K(r)=\frac{rI_z(r)}{I_t(r)}.
\]
Bounding the common denominator between $\min(1,r)$ and $\max(1,r)$ gives
\[
\frac{\min(1,r^2)}2\le K(r)\le\frac{\max(1,r^2)}2.
\]
Differentiating the plane value with respect to $u$ gives the sign of $aK(r)-u/\sqrt{1-u^2}$. For $w=u/\sqrt{1-u^2}$, $r=l\sqrt{1+w^2}+aw$ increases strictly. At a trial ratio $r>l$ with $r-gK>0$, comparison with stationarity reduces to
\begin{equation}
P(r,K;p)=r^2-j-g\bigl(1+2rK+jK^2\bigr). \label{eq:example-bowles-ratio-comparison}
\end{equation}
Positive $P$ places the stationary ratio below the trial ratio; negative $P$ places it above. This follows by squaring the comparison between $r-gK$ and $l\sqrt{1+gK^2}$, and using strict concavity of the plane value. The function $P$ decreases with $K\ge0$.

In the forward direction,
\[
g=\frac{(1-p)^2(2-3p)^2}{p^2(8-3p)(2+3p)},\qquad j=\frac{(7p-2)(2+3p)}{25p^2}.
\]
On the interval in Eq.~\eqref{eq:example-bowles-common-interval}, $g$ decreases and $j$ increases. Their endpoint bounds give $41/50\le j\le43/50$ and $1/125\le g\le3/200$, so $l<1$ and $1-g/2>0$. At $r=1$, where $K=1/2$, Eq.~\eqref{eq:example-bowles-ratio-comparison} is positive because
\[
1-j-g(2+j/4)\ge1-\frac{43}{50} -\frac3{200}\left(2+\frac{43}{200}\right)>0.
\]
Hence $r<1$, while $r\ge l\ge\sqrt{207/250}>10/11$. This proves the forward anisotropy bound. The zero-moment condition reads
\[
u=s\frac{r^2I_z}{I_t+r^2I_z}=s\,m(r), \qquad m(r)\le\frac13\quad(r\le1).
\]
Transforming the source back gives $a/l=s(1-m)/(1-s^2m)\ge2s/3$. The ratio
\[
\frac al=\frac{5(1-p)(2-3p)}{25p-6(1-p)^2}
\]
decreases with $p$ and equals $21440/166049$ at $p=61/125$. Thus $s\le32160/166049<1/5$.

For the reverse direction,
\[
g=\frac{(1-p)^2(3-2p)^2}{p^2(7-2p)(3+2p)},\qquad j=\frac{(3+2p)(8p-3)}{25p^2}.
\]
Take $r_+=1007/1000$ and $r_-=1000/1007$. On this interval $g<1/5$ and $j<7/10$, giving $l^2<9/10<r_-^2$. Both trials have $K<1$, and hence $r-gK>r_--1/5>0$. The integral bounds on $K$ reduce the upper and lower comparisons to the degree-six polynomials
\[
\begin{aligned}
F_+(p)&=p^4(7-2p)(3+2p)P(r_+,r_+^2/2;p),\\
F_-(p)&=-p^4(7-2p)(3+2p)P(r_-,r_-^2/2;p).
\end{aligned}
\]
Under $p=61/125+(11/500)x$, each polynomial has the form $\sum_{k=0}^6b_k\binom6k x^k(1-x)^{6-k}$. Substitution in the displayed expressions gives $b_k(F_+)\ge1/1300$ and $b_k(F_-)\ge3/500$ for all seven coefficients. The basis functions are nonnegative and sum to one on $0\le x\le1$, proving $F_\pm>0$ throughout the interval. The scalar comparison yields $r_-\le r\le r_+$ and hence $\chi\le1007/1000$. Finally, $j'(p)=18(1-p)/(25p^3)>0$, so
\[
s^2=1-\frac j{r^2}\le1-\frac{j(61/125)}{r_+^2} <\left(\frac{16}{25}\right)^2.
\]
Both directional sources therefore obey Eq.~\eqref{eq:example-bowles-source-bounds} throughout the interval, giving POVM equivalence there.

To locate the forward unit-mass crossing, apply Jensen inequality to obtain $\Phi\le2\sqrt{\Tr M/3}$ and maximize the quadratic trace over $q$. This gives
\[
c_{\mathrm P}^2\le\frac83T^2\left(1+\frac{L^2}{2T^2-A^2}\right)<1 \qquad\left(p=\frac{61}{125}\right).
\]
At the upper endpoint, take $q=0$ and bound the square root below by its chord to obtain
\[
c_{\mathrm P}\ge\frac23(2T-L)>1 \qquad\left(p=\frac{51}{100}\right).
\]
Both endpoint inequalities follow from Eq.~\eqref{eq:example-bowles-data}. Above each crossing, the conditional demands continue to grow. To see this, write the longitudinal combinations as
\begin{equation}
\begin{aligned}
L+A&=\frac{2-7p}{2+3p},& L-A&=-\frac{2+3p}{8-3p},\\
L'+A'&=-\frac{3+2p}{7-2p},& L'-A'&=\frac{3-8p}{3+2p}.
\end{aligned}
\label{eq:example-bowles-longitudinal}
\end{equation}
All four functions are strictly decreasing. Both transverse correlations also increase, since
\[
\frac{d}{dp}\frac p{\sqrt{1-\kappa_0(1-p)^2}} =\frac{1-\kappa_0(1-p)}{[1-\kappa_0(1-p)^2]^{3/2}}>0 \quad\text{for }\kappa_0=9/25,\ 4/25.
\]
For a fixed $|q|<1$, the plane value is
\[
\Phi_p(q)=2\int_0^1\sqrt{T^2(1-q^2)(1-x^2)+(L+Aq)^2x^2}\,dx.
\]
The longitudinal factor combines $L+A$ and $L-A$ convexly. It is negative and strictly decreasing for $p\ge2/7$ forward and $p\ge3/8$ in reverse. Meanwhile the transverse factor increases. Hence every fixed-plane value, and its maximum, increases strictly on the corresponding interval. Once above one, the scale stays above one.

Below each threshold $p_*$, the affine identity
\[
\rho_{\mathrm B}(p)=\frac p{p_*}\rho_{\mathrm B}(p_*) +\left(1-\frac p{p_*}\right)\rho_{\mathrm B}(0),\qquad 0\le p\le p_*,
\]
mixes an all-POVM boundary parent with the separable state at zero. It supplies every lower $p$. Thus neither direction is steerable up to the forward threshold, only Alice steers between it and $1/2$, and both parties steer for $p>1/2$. Both scales reach two at the singlet.
\end{example}
\begin{figure}[H]
\centering\includegraphics[width=.88\textwidth]{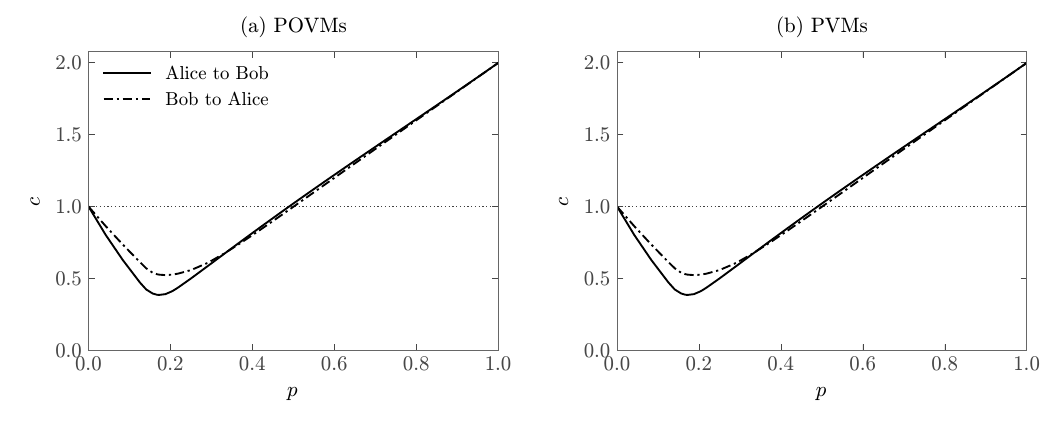}
\caption{Common measurement thresholds with one-way steering in the Bowles family. Solid and dash-dotted curves show Alice-to-Bob and Bob-to-Alice scales; the dotted line is $c=1$. Both measurement classes have the forward threshold $0.488862\ldots$ and reverse threshold $1/2$, giving the same nonzero one-way interval. Panels (a,b) correspond to POVMs and PVMs.} \label{fig:example-bowles}
\end{figure}
\subsection{Mixed-state boundaries}
\begin{example}[Bell-diagonal states]
\label{ex:bell-diagonal}

With both marginals maximally mixed, the correlations alone fix the ensemble mass and angular distribution. The Bell-diagonal family is
\begin{equation}
\rho_{\mathrm T}(t_1,t_2,t_3)=\frac14\left(\openone_4+\sum_{j=1}^3t_j\sigma_j\otimes\sigma_j\right) =\frac14
\begin{pmatrix}
1+t_3&0&0&t_1-t_2\\
0&1-t_3&t_1+t_2&0\\
0&t_1+t_2&1-t_3&0\\
t_1-t_2&0&0&1+t_3
\end{pmatrix}
. \label{eq:example-bell-matrix}
\end{equation}
Their physical region is the tetrahedron with vertices $(1,-1,1)$, $(-1,1,1)$, $(1,1,-1)$, and $(-1,-1,-1)$. The canonical data are $\vec a=0$ and $C=\operatorname{diag}(t_1,t_2,t_3)$, and Eq.~\eqref{eq:s3-unbiased-povm-value} gives
\begin{equation}
c_{\mathrm A}=c_{\mathrm P} =2\int_{S^2}\sqrt{t_1^2n_1^2+t_2^2n_2^2+t_3^2n_3^2}\,d\omega(\vec n). \label{eq:example-bell-scale}
\end{equation}
For PVMs this is the critical-radius formula for $T$ states~\cite{NguyenVuCritical,ZhangZhangBell}. Its level set $c=1$ gives the common PVM and POVM steering boundary. Partial transposition places the separable states in the octahedron $|t_1|+|t_2|+|t_3|\le1$. Between this octahedron and the curved surface, entanglement is present but one ensemble still supplies every POVM preparation.

Along the Werner line~\cite{Werner1989},
\begin{equation}
\rho_{\mathrm W}(p)=p\ketbra{\Psi_-}+(1-p)\frac{\openone_4}{4},\qquad \ket{\Psi_-}=\frac{\ket{01}-\ket{10}}{\sqrt2},\qquad -\frac13\leq p\leq1, \label{eq:example-werner-state}
\end{equation}
the scale reduces to $2|p|$, with common threshold $p=1/2$ on the nonnegative branch. Other symmetry and rank limits are also elementary. Axial data have $q=0$ and common value $2I(T^2,L^2-T^2)$, with $I$ from Eq.~\eqref{eq:example-axial-primitive}. A single nonzero correlation has scale equal to its magnitude. With two nonzero eigenvalues $0<\lambda_1\le\lambda_2$ of $CC^{\mathsf T}$, the common value is $\sqrt{\lambda_2}\,\mathrm E(1-\lambda_1/\lambda_2)$.
\end{example}

\begin{figure}[H]
\centering\includegraphics[width=.82\textwidth]{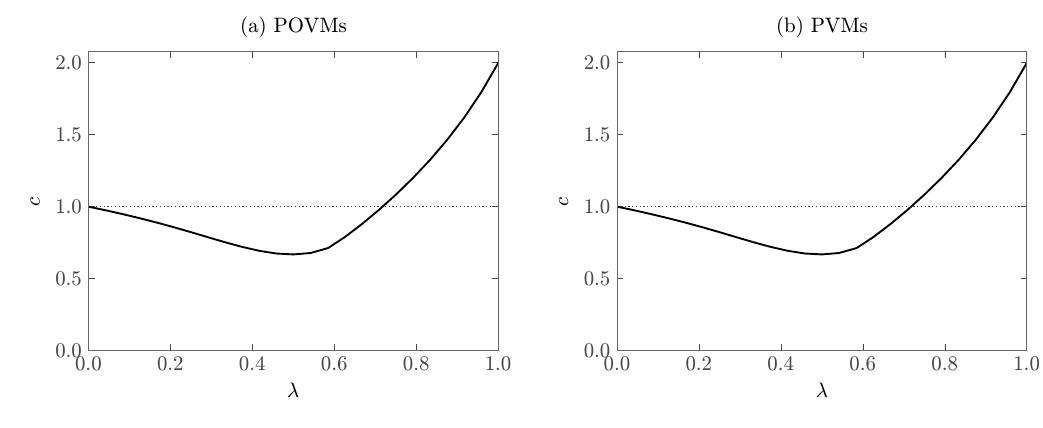}
\caption{Common directional scales in Gisin states at $\theta=\pi/8$. The weight $\lambda$ interpolates between a classical mixture and a pure entangled state. At the entanglement boundary $\lambda=2-\sqrt2$, the rank-two formula gives the exact common PVM and POVM value $0.714853\ldots$. The dotted line is the steering threshold $c=1$. Panels (a,b) correspond to POVMs and PVMs.} \label{fig:example-gisin}
\end{figure}

\begin{example}[Gisin states]
\label{ex:gisin}

Gisin states~\cite{Gisin1996} give an exact example of entanglement beginning below unit ensemble mass:
\begin{equation}
\rho_{\mathrm G}(\lambda,\theta)=\lambda\ketbra{\phi_\theta}+\frac{1-\lambda}{2}(\ketbra{00}+\ketbra{11}),\qquad \ket{\phi_\theta}=\sin\theta\ket{01}+\cos\theta\ket{10}, \label{eq:example-gisin-state}
\end{equation}
with $0\le\lambda\le1$ and $0\le\theta\le\pi/4$. The entangled component occupies the odd-parity subspace, while the remaining weight is a classical mixture of $\ket{00}$ and $\ket{11}$. Explicitly,
\begin{equation}
\rho_{\mathrm G}=
\begin{pmatrix}
(1-\lambda)/2&0&0&0\\
0&\lambda\sin^2\theta&\lambda\sin\theta\cos\theta&0\\
0&\lambda\sin\theta\cos\theta&\lambda\cos^2\theta&0\\
0&0&0&(1-\lambda)/2
\end{pmatrix}
. \label{eq:example-gisin-matrix}
\end{equation}
Party exchange followed by $\sigma_1\otimes\sigma_1$ fixes the state, so each measurement class has equal directional scales. At $\theta=\pi/4$ this is the Bell-diagonal line $(\lambda,\lambda,1-2\lambda)$. For $k=\cos2\theta$, $s=\sin2\theta$, and $D=1-\lambda^2k^2$, the canonical data are
\begin{equation}
A=-\frac{2\lambda(1-\lambda)k}{D},\qquad T=\frac{\lambda s}{\sqrt D},\qquad L=\frac{1-2\lambda+\lambda^2k^2}{D}. \label{eq:example-gisin-data}
\end{equation}
For $\theta>0$, entanglement begins at $\lambda\sin2\theta=1-\lambda$, exactly where $L=0$. The correlation then has rank two, and the POVM and PVM scales coincide by Section~S4. At $\theta=\pi/8$, this occurs at $\lambda=2-\sqrt2$ with $A^2<T^2$. Symmetry gives $q_*=0$, and hence
\[
c_{\mathrm A}=c_{\mathrm P} =\frac\pi2\sqrt{\frac{\sqrt2-1}{2}} =0.714853481333\ldots.
\]
Entanglement thus appears while the required mass is below one. Figure~\ref{fig:example-gisin} follows the family from the classical mixture of mass one through entanglement onset to a pure entangled state of mass two. The axial PVM equation applies for $|1-2\lambda|\ge\lambda\cos2\theta$. At the displayed angle, the two analytic intervals end at $(2+1/\sqrt2)^{-1}$ and start at $(2-1/\sqrt2)^{-1}$; general spectral evolution evaluates the intervening curves.
\end{example}

\begin{figure}[H]
\centering\includegraphics[width=.82\textwidth]{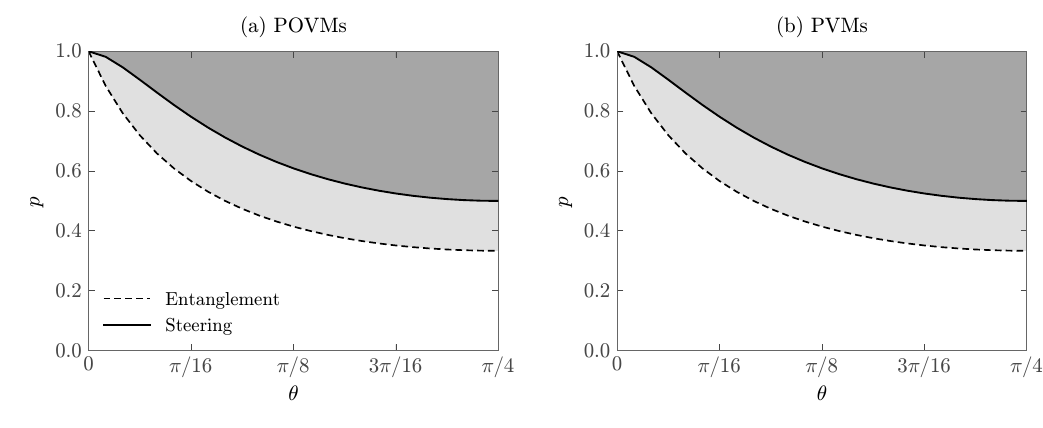}
\caption{Entanglement and steering in generalized Werner states. The dashed and solid curves are the entanglement and steering thresholds. A horizontal line fixes the visibility $p$ and hence the full global spectrum while the Schmidt angle changes. White, light-gray, and dark-gray regions denote separable states, entangled unsteerable states, and steerable states. The right edge is the Werner line. Panels (a,b) correspond to POVMs and PVMs.} \label{fig:example-generalized-werner}
\end{figure}

\begin{figure}[H]
\centering\includegraphics[width=.82\textwidth]{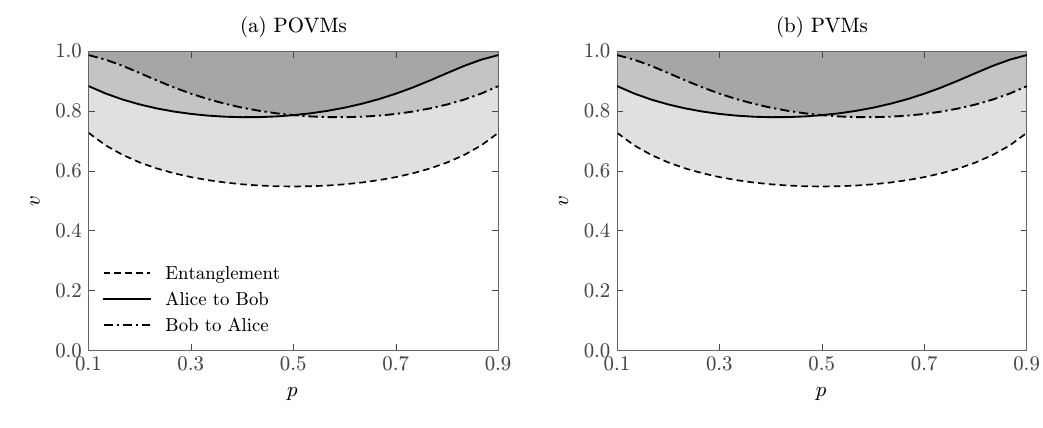}
\caption{Directional steering at a fixed global spectrum in the noisy rank-two family. The Schmidt-state weight is $2/3$. At fixed visibility $v$, changing $p$ leaves all eigenvalues unchanged. Dashed, solid, and dash-dotted curves give the entanglement, Alice-to-Bob, and Bob-to-Alice thresholds. White through dark gray denote separability, entanglement without steering, one-way steering, and two-way steering. Reflection about $p=1/2$ exchanges the directions. Panels (a,b) correspond to POVMs and PVMs.} \label{fig:example-noisy-rank-two}
\end{figure}

\begin{figure}[H]
\centering\includegraphics[width=.82\textwidth]{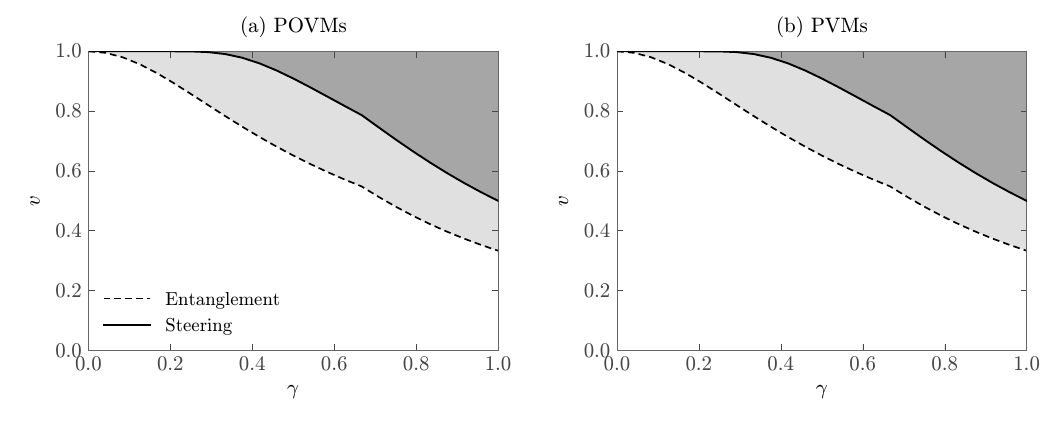}
\caption{Noise thresholds across the two MEMS branches. The dashed curve is the entanglement threshold and the solid curve the steering threshold. The two directions agree by symmetry. White, light-gray, and dark-gray regions denote separable, entangled unsteerable, and steerable states. The junction at $\gamma=2/3$ is the symmetric member of the noisy rank-two family. Panels (a,b) correspond to POVMs and PVMs.} \label{fig:example-noisy-mems}
\end{figure}

\begin{example}[Generalized Werner states]
\label{ex:generalized-werner}

Generalized Werner states~\cite{JirakovaGeneralizedWerner2021} have a common global spectrum at fixed visibility and different conditional preparations as the Schmidt angle varies:
\begin{equation}
\rho_{\mathrm{GW}}(p,\theta)=p\ketbra{\psi_\theta}+(1-p)\frac{\openone_4}{4},\qquad \ket{\psi_\theta}=\cos\theta\ket{00}+\sin\theta\ket{11}, \label{eq:example-GW-state}
\end{equation}
For $-1/3\le p\le1$ and $0\le\theta\le\pi/4$, their eigenvalues are $(1+3p)/4,(1-p)/4,(1-p)/4,(1-p)/4$, independently of $\theta$. Writing $d=(1-p)/4$, the matrix is
\begin{equation}
\rho_{\mathrm{GW}}=
\begin{pmatrix}
d+p\cos^2\theta&0&0&\tfrac p2\sin2\theta\\
0&d&0&0\\
0&0&d&0\\
\tfrac p2\sin2\theta&0&0&d+p\sin^2\theta
\end{pmatrix}
. \label{eq:example-GW-matrix}
\end{equation}
Exchange symmetry makes the two steering directions equivalent, but varying $\theta$ changes their common noise tolerance. For nonnegative visibility, entanglement begins at $p_{\mathrm E}=(1+2\sin2\theta)^{-1}$; at $\theta=\pi/8$, the PVM steering boundary lies at $p=0.608841950835\ldots$, above $p_{\mathrm E}=\sqrt2-1$. A horizontal line in Figure~\ref{fig:example-generalized-werner} fixes $p$ and hence every global eigenvalue, yet crosses the entanglement and steering regions as $\theta$ varies. This dependence enters the canonical parameters through $k=\cos2\theta$ and $D=1-p^2k^2$:
\begin{equation}
A=\frac{pk(1-p)}D,\qquad T=\frac{|p|\sin2\theta}{\sqrt D},\qquad L=\frac{p(1-pk^2)}D. \label{eq:example-GW-data}
\end{equation}
Since $L^2-A^2=T^2$, the axial PVM equation covers all $T>0$. At $\theta=\pi/4$, local unitary equivalence to Werner states reduces both scales to $2|p|$. At $\theta=0$, the rank-one formula instead yields $2|p|/(1+p)$. For a pure entangled state, $p=1$ and $\theta>0$, whitening recovers Bell data and scale two. At the product endpoint $(1,0)$, the scale is one.
\end{example}
\begin{example}[Noisy rank-two states]
\label{ex:noisy-rank-two}

A fixed global spectrum also permits opposite steering directions. To exhibit this, take the rank-two family of Ref.~\cite{McCloskeySteering2017} with fixed weights $2/3$ and $1/3$ and add white noise:
\begin{equation}
\rho_{\mathrm R}(v,p)=v\left(\frac23\ketbra{\psi_p}+\frac13\ketbra{01}\right)+(1-v)\frac{\openone_4}{4},\qquad \ket{\psi_p}=\sqrt p\ket{00}+\sqrt{1-p}\ket{11},\qquad 0\leq v,p\leq1. \label{eq:example-rank-two-state}
\end{equation}
Orthogonality of the two components fixes the eigenvalues to $d+2v/3,d+v/3,d,d$, where $d=(1-v)/4$, independently of $p$. In matrix form,
\begin{equation}
\rho_{\mathrm R}=
\begin{pmatrix}
d+2vp/3&0&0&2v\sqrt{p(1-p)}/3\\
0&d+v/3&0&0\\
0&0&d&0\\
2v\sqrt{p(1-p)}/3&0&0&d+2v(1-p)/3
\end{pmatrix}
,\qquad d=\frac{1-v}{4}. \label{eq:example-rank-two-matrix}
\end{equation}
Changing $p$ to $1-p$ exchanges the two parties up to $\sigma_1\otimes\sigma_1$, and hence
\[
c_{\mathfrak M}^{B\to A}(v,p) =c_{\mathfrak M}^{A\to B}(v,1-p), \qquad \mathfrak M=\mathrm A,\mathrm P.
\]
States at $p$ and $1-p$ also share entanglement threshold $v_{\mathrm E}=3/[1+2\sqrt{1+16p(1-p)}]$, but their local populations exchange the steering demands. At $p=0.2$, the numerical thresholds are about $0.82227$ forward and $0.92724$ in reverse, with PVM and POVM curves coincident at the displayed resolution [Fig.~\ref{fig:example-noisy-rank-two}]. They exchange roles at $p=0.8$. For $v=0.87$, the numerical diagram thus places two states of common spectrum $(0.6125,0.3225,0.0325,0.0325)$ in opposite one-way regions.

For the forward canonical data, put $D=1-v^2(4p-3)^2/9$:
\begin{equation}
A=\frac{v[3(4p-1)-v(4p-3)]}{9D},\qquad T=\frac{4v\sqrt{p(1-p)}}{3\sqrt D},\qquad L=\frac{3v-v^2(4p-1)(4p-3)}{9D}. \label{eq:example-rank-two-data}
\end{equation}
Since $L^2-A^2=8v^2p(1-2p)/(9D)$, $p\le1/2$ lies in the axial PVM domain; spectral evolution evaluates $p>1/2$. Direction exchange gives the reverse values, meeting the forward ones at $p=1/2$.
\end{example}
\begin{example}[Noisy maximally entangled mixed states]
\label{ex:noisy-mems}

Munro, James, White and Kwiat's states~\cite{MunroMEMS} maximize concurrence at fixed linear entropy. Their two branches include the symmetric point of the preceding rank-two family:
\begin{equation}
\rho_{\mathrm M}(\gamma)=
\begin{pmatrix}
g(\gamma)&0&0&\gamma/2\\
0&1-2g(\gamma)&0&0\\
0&0&0&0\\
\gamma/2&0&0&g(\gamma)
\end{pmatrix}
,\qquad g(\gamma)=
\begin{cases}
1/3,&0\leq\gamma\leq2/3,\\
\gamma/2,&2/3\leq\gamma\leq1.
\end{cases}
\label{eq:example-MEMS-state}
\end{equation}
Their white-noise descendants are
\begin{equation}
\rho_{\mathrm M}(\gamma,v)=v\rho_{\mathrm M}(\gamma)+(1-v)\frac{\openone_4}{4},\qquad 0\leq v\leq1. \label{eq:example-MEMS-noise}
\end{equation}
At their junction, $\rho_{\mathrm R}(v,1/2)=\rho_{\mathrm M}(2/3,v)$. Exchange followed by $\sigma_1\otimes\sigma_1$ fixes every MEMS state, making the directional scales equal. Set $g=g(\gamma)$, $h=1-2g$ and $D=1-v^2h^2$; the canonical data are
\begin{equation}
A=\frac{vh[1+v(4g-1)]}{D},\qquad T=\frac{v\gamma}{\sqrt D},\qquad L=\frac{v(4g-1)+v^2h^2}{D}. \label{eq:example-MEMS-data}
\end{equation}
Since $L^2-A^2=4v^2g(3g-1)/D\ge0$, the axial PVM formula applies on both branches. The endpoint values are $2v/(3-v)$ at $\gamma=0$ and $2v$ at $\gamma=1$, for both measurement classes.

Partial transposition gives the entanglement boundary
\begin{equation}
4v^2\gamma^2=(1-v)\bigl[1+v(3-8g(\gamma))\bigr]. \label{eq:example-MEMS-entanglement}
\end{equation}
Before white noise is added, every state with $v=1$ and $\gamma>0$ has a product vector in its kernel and hence a pure conditional state. The strict inequality \eqref{eq:pvm-strict-null-contact}, proved in Section~S4, then implies PVM steering. Figure~\ref{fig:example-noisy-mems} shows how much white noise removes this steering, alongside the distinct loss of entanglement, as the state passes through the branch junction.
\end{example}
\subsection{Filtering, damping and marginal support}
\begin{figure}[H]
\centering\includegraphics[width=.82\textwidth]{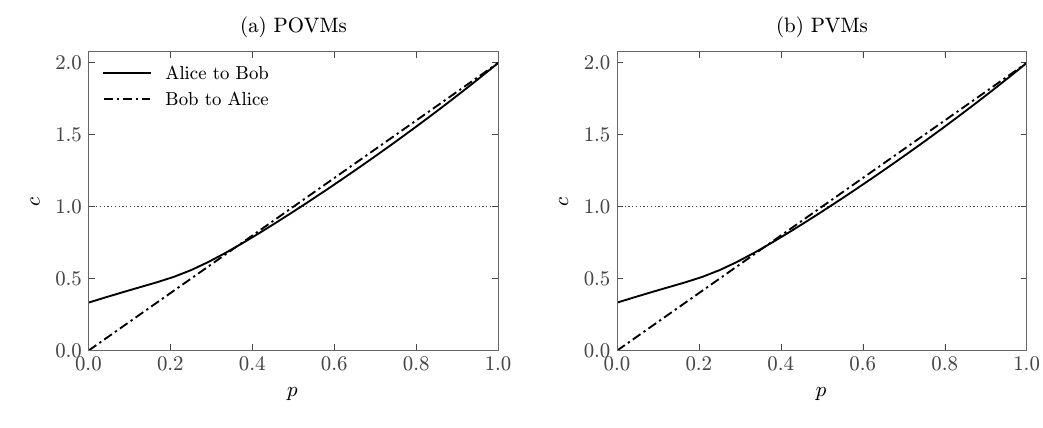}
\caption{Directional effect of filtering Alice in a Werner state, with $r=1/2$. The reverse scale (dash-dotted) remains exactly $2p$ because whitening the trusted party removes the invertible filter. The forward scale (solid) gives a PVM threshold $0.518366\ldots$, above the reverse threshold $1/2$. The dotted line is $c=1$. Panels (a,b) correspond to POVMs and PVMs.} \label{fig:example-filtered-werner}
\end{figure}

\begin{example}[Alice-filtered Werner states]
\label{ex:filtered-werner}

A local filter changes steering according to which party is trusted. Acting on Alice's Werner state with $K_r=\operatorname{diag}(\sqrt r,1)$, $r\ge0$, gives
\begin{equation}
\rho_{\mathrm{FW}}(p,r)=\frac{(K_r\otimes\openone_2)\rho_{\mathrm W}(p)(K_r\otimes\openone_2)}{(1+r)/2} =\frac1{2(1+r)}
\begin{pmatrix}
r(1-p)&0&0&0\\
0&r(1+p)&-2p\sqrt r&0\\
0&-2p\sqrt r&1+p&0\\
0&0&0&1-p
\end{pmatrix}
, \label{eq:example-filtered-werner-matrix}
\end{equation}
When Alice is trusted, whitening removes her invertible filter: for $r>0$, the reverse canonical data remain $A'=0$, $C'=-p\openone_3$. When she performs the measurements, the filter changes the conditional preparations required on Bob. With $0\le p\le1$, $\delta=(r-1)/(r+1)$, and $D=1-p^2\delta^2$, the forward data become
\begin{equation}
A=\frac{\delta(1-p^2)}D,\qquad T=\frac{p\sqrt{1-\delta^2}}{\sqrt D},\qquad L=-\frac{p(1-\delta^2)}D. \label{eq:example-filtered-werner-data}
\end{equation}
Forward projective contact applies at $p\ge|\delta|$, because $L^2-A^2=(p^2-\delta^2)/(1-p^2\delta^2)$. Reverse data give
\begin{equation}
c_{\mathrm A}^{B\to A}(p,r)=c_{\mathrm P}^{B\to A}(p,r)=2p. \label{eq:example-filtered-werner-reverse}
\end{equation}
Bob can therefore steer Alice above the same boundary $p=1/2$ for every invertible filter. Entanglement likewise begins at $p=1/3$, since invertible local filtering preserves separability. Alice's steering requirement changes: at $r=1/2$, her PVM threshold rises to $0.518365561644\ldots$. This opens the projective one-way interval $1/2<p\le0.518365561644\ldots$ from Bob to Alice [Fig.~\ref{fig:example-filtered-werner}].

At $r=1$ the two directions have the Werner value. A singular filter changes the support: $r=0$ makes Alice's marginal pure and gives $c^{A\to B}=1$ and $c^{B\to A}=p$. This endpoint anticipates the support change under complete damping below.
\end{example}

\begin{figure}[H]
\centering\includegraphics[width=.88\textwidth]{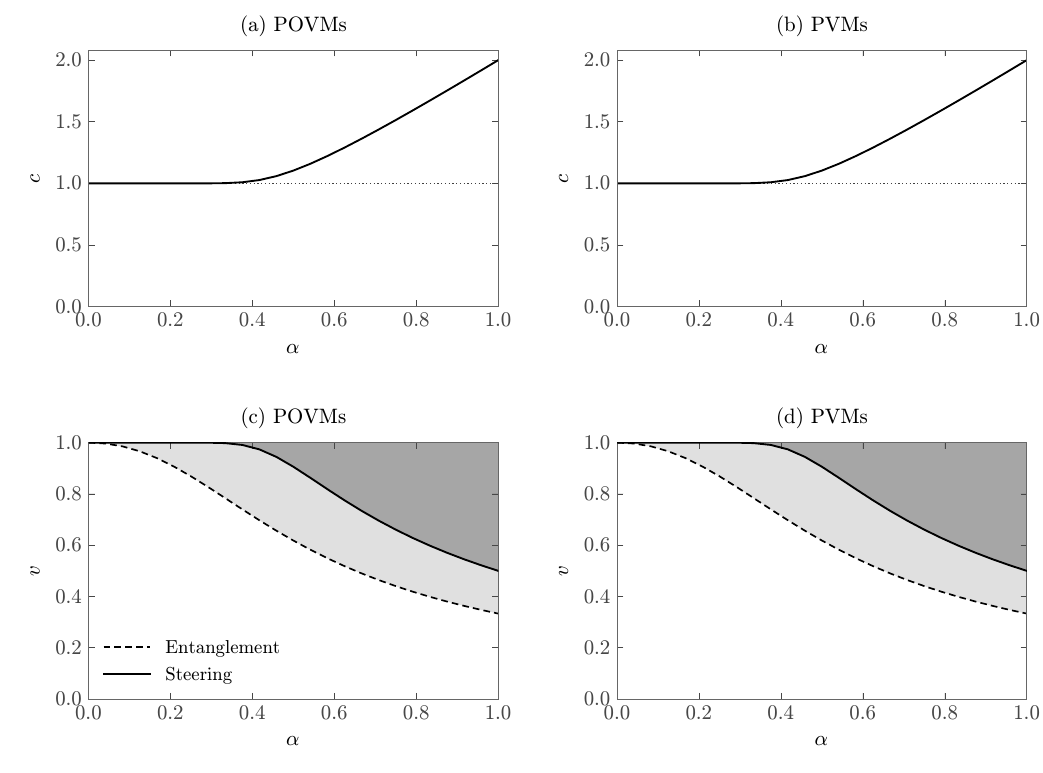}
\caption{Steering and noise tolerance near the polarized product endpoint of the Hirsch family. Panels (a,b) show the forward POVM and PVM scales; the dotted line is $c=1$. A pure conditional state ensures $c_{\mathrm P}>1$ for every $\alpha>0$. Panels (c,d) give the white-noise thresholds, with steering at $v=1/c$ (solid) and entanglement at the dashed curve. White, light-gray, and dark-gray regions denote separable, entangled unsteerable, and steerable states.} \label{fig:example-hirsch}
\end{figure}

\begin{example}[Hirsch states and their white-noise descendants]
\label{ex:hirsch}

Even an arbitrarily small singlet contribution in the Hirsch family~\cite{HirschHiddenNonlocality} leaves a pure conditional state and thus preserves steering:
\begin{equation}
\rho_{\mathrm H}(\alpha)=\alpha\ketbra{\Psi_-}+(1-\alpha)\ketbra0\otimes\frac{\openone_2}{2} =\frac12
\begin{pmatrix}
1-\alpha&0&0&0\\
0&1&-\alpha&0\\
0&-\alpha&\alpha&0\\
0&0&0&0
\end{pmatrix}
,\qquad 0\leq\alpha\leq1. \label{eq:example-hirsch-state}
\end{equation}
Bob is maximally mixed, with canonical data $\vec a=(1-\alpha)\vec e_3$ and $C=-\alpha\openone_3$. For every $\alpha>0$, the correlation has full rank and $|\vec a-C\vec e_3|=1$, so Eq.~\eqref{eq:pvm-strict-null-contact} implies $c_{\mathrm P}^{A\to B}>1$. As the curve approaches one near $\alpha=0$, the tolerance to added noise vanishes, but steering survives at every nonzero singlet weight. At the product endpoint the scale is one; at the singlet it is two.

Add white noise to quantify that tolerance:
\begin{equation}
\rho_{\mathrm H}(\alpha,v)=v\rho_{\mathrm H}(\alpha)+(1-v)\frac{\openone_4}{4},\qquad 0\leq v\leq1. \label{eq:example-hirsch-noise}
\end{equation}
Now $A=v(1-\alpha)$, $T=v\alpha$ and $L=-v\alpha$. Uniform scaling by $v$ gives
\begin{equation}
c_{\mathfrak M}^{A\to B}(\alpha,v)=v\,c_{\mathfrak M}^{A\to B}(\alpha,1),\qquad v_{\mathfrak M}(\alpha)=\frac1{c_{\mathfrak M}^{A\to B}(\alpha,1)},\qquad \mathfrak M=\mathrm A,\mathrm P. \label{eq:example-hirsch-radial}
\end{equation}
Thus a single curve at unit visibility contains the full forward noise tolerance. Figure~\ref{fig:example-hirsch} compares the resulting steering boundary with $v_{\mathrm E}=[\alpha+\sqrt{(1-\alpha)^2+4\alpha^2}]^{-1}$. For $\alpha\ge1/2$, the PVM curve follows the axial equation; stronger polarization requires the general spectral formula.
\end{example}
\begin{example}[Thermal amplitude-damping Choi states]
\label{ex:thermal-damping}

\label{ex:amplitude-damped-bell}

Damping weakens correlations and can polarize the output, so the added noise needed for a common-ensemble description can differ between steering directions. Let $\mathcal E_{\gamma,r}$ be the generalized amplitude-damping channel~\cite{KhatriGADC2020}, with damping strength $\gamma$ and equilibrium Bloch polarization $r$. Place its output on Alice and add white noise:
\begin{equation}
\rho_{\mathrm{th}}(\gamma,r;v) =v(\mathcal E_{\gamma,r}\otimes\operatorname{id})(\ketbra{\Phi_+})+(1-v)\frac{\openone_4}{4},\qquad \ket{\Phi_+}=\frac{\ket{00}+\ket{11}}{\sqrt2},\qquad 0\leq\gamma,r,v\leq1. \label{eq:example-thermal-state}
\end{equation}
In the computational basis this state reads
\begin{equation}
\rho_{\mathrm{th}}=\frac14
\begin{pmatrix}
1+v[1-\gamma(1-r)]&0&0&2v\sqrt{1-\gamma}\\
0&1-v+v\gamma(1+r)&0&0\\
0&0&1-v+v\gamma(1-r)&0\\
2v\sqrt{1-\gamma}&0&0&1+v[1-\gamma(1+r)]
\end{pmatrix}
. \label{eq:example-thermal-matrix}
\end{equation}
Bob's marginal stays maximally mixed, so the forward data are
\begin{equation}
A=v\gamma r,\qquad T=v\sqrt{1-\gamma},\qquad L=v(1-\gamma). \label{eq:example-thermal-forward}
\end{equation}
White noise is radial in this direction: $c_{\mathfrak M}^{A\to B}(\gamma,r;v)=v\,c_{\mathfrak M}^{A\to B}(\gamma,r;1)$ for both measurement classes. In the reverse direction, whitening depends on the noisy polarization of Alice. With $D'=1-v^2\gamma^2r^2$, it gives
\begin{equation}
A'=-\frac{v^2\gamma r(1-\gamma)}{D'},\qquad T'=\frac{v\sqrt{1-\gamma}}{\sqrt{D'}},\qquad L'=\frac{v(1-\gamma)}{D'}. \label{eq:example-thermal-reverse}
\end{equation}
Thus damping changes the two normalized preparation demands differently, as shown for $r=4/5$ in Figure~\ref{fig:example-thermal-damping}. Forward projective contact applies at $\gamma\le1/(1+r)$, while reverse data satisfy
\[
(L')^2-(A')^2=\frac{v^2(1-\gamma)^2}{D'}\ge0\qquad(D'>0).
\]
At $r=0$, polarization vanishes and the Bell-diagonal formula gives the common directional scale $2vI(1-\gamma,-\gamma(1-\gamma))$, decreasing from $2v$ to zero.

Partial transposition gives the entanglement threshold
\begin{equation}
v_{\mathrm E}(\gamma,r)=\frac1{1-\gamma+\sqrt{4(1-\gamma)+\gamma^2r^2}}. \label{eq:example-thermal-entanglement}
\end{equation}
When $v_{\mathrm E}>1$, all physical visibilities are separable. At unit visibility, the entanglement boundary is
\[
\gamma_{\mathrm E}(r)=\frac{2}{1+\sqrt{2-r^2}}.
\]
Complete damping produces $(\openone_2+vr\sigma_3)/2\otimes\openone_2/2$, with forward scale $vr$ and reverse scale zero.
\begin{figure}[H]
\centering\includegraphics[width=.90\textwidth]{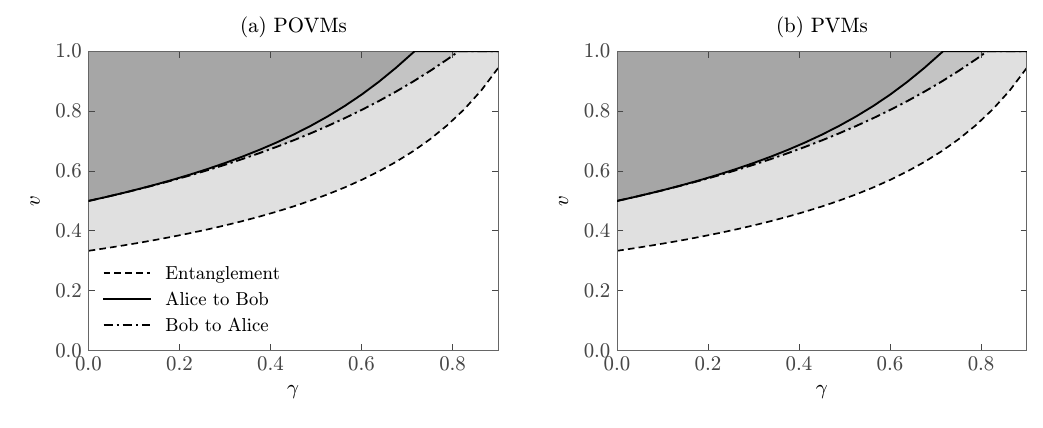}
\caption{Directional visibility thresholds under thermal amplitude damping at $r=4/5$. The dashed curve marks entanglement; solid and dash-dotted curves mark Alice-to-Bob and Bob-to-Alice steering. White through dark gray denote separability, entanglement without steering, one-way steering, and two-way steering. White noise rescales the forward canonical data radially, while the reverse data also change through whitening. Panels (a,b) correspond to POVMs and PVMs.} \label{fig:example-thermal-damping}
\end{figure}
At zero temperature and unit visibility, a pure conditional state survives in each direction at every partial damping strength. Steering then persists until the endpoint, where the trusted marginal's support becomes essential. Set $r=v=1$ and exchange the parties so that Bob is damped. The Kraus operators are
\[
E_0=\ketbra0+\sqrt{1-\gamma}\ketbra1,\qquad E_1=\sqrt\gamma\ket0\bra1.
\]
The state is
\begin{equation}
\rho_{\mathrm{AD}}(\gamma)=(\operatorname{id}\otimes\mathcal E_{\gamma,1})(\ketbra{\Phi_+}) =\frac12
\begin{pmatrix}
1&0&0&\sqrt{1-\gamma}\\
0&0&0&0\\
0&0&\gamma&0\\
\sqrt{1-\gamma}&0&0&1-\gamma
\end{pmatrix}
,\qquad 0\leq\gamma\leq1. \label{eq:example-AD-matrix}
\end{equation}
For $\gamma<1$, its whitened forward data are
\begin{equation}
A=-\frac\gamma{1+\gamma},\qquad T=\frac1{\sqrt{1+\gamma}},\qquad L=\frac1{1+\gamma}. \label{eq:example-AD-forward}
\end{equation}
Here $L-A=1$ and $L^2-A^2=(1-\gamma)/(1+\gamma)\ge0$. In reverse, the canonical data are $A'=\gamma$, $T'=\sqrt{1-\gamma}$, and $L'=1-\gamma$, with $A'+L'=1$. Each direction retains both full-rank correlations and a pure conditional state. Equation~\eqref{eq:pvm-strict-null-contact} then forces $c_{\mathrm P}>1$, and hence $c_{\mathrm A}>1$, at every partial damping strength.

Forward projective contact holds throughout $\gamma<1$. In reverse it holds up to $\gamma=1/2$; spectral evolution evaluates stronger damping.

Whitening keeps the forward canonical correlations finite even as the physical state approaches complete damping. As $\gamma\uparrow1$, the data approach $(A,T,L)=(-1/2,1/\sqrt2,1/2)$, with $J=\operatorname{diag}(1/2,1/2,0)$ and $|\vec a|^2=1/4$. This is the isotropic-source family in Eq.~\eqref{eq:pvm-isotropic-original-family} with $\lambda^2=1/2$, at its boundary $|\vec a|^2=\lambda^2/2$. Equation~\eqref{eq:pvm-isotropic-algebraic-value} then fixes $\vec q_*=-\vec e_3/3$ and
\begin{equation}
\lim_{\gamma\uparrow1}c_{\mathrm P}^{A\to B}(\gamma)=\frac43. \label{eq:example-AD-limit}
\end{equation}
At complete damping the state is $\openone_2/2\otimes\ketbra0$. Restricting to the trusted support gives scale zero forward and one in reverse for both measurement classes. The limit $4/3$ and the forward endpoint zero therefore correspond to different supports. Open and filled points in Figure~\ref{fig:example-amplitude-damping} distinguish them. Before support collapses, normalization retains nontrivial conditional-preparation geometry even as the physical state tends to a product.
\end{example}
\begin{figure}[H]
\centering\includegraphics[width=.90\textwidth]{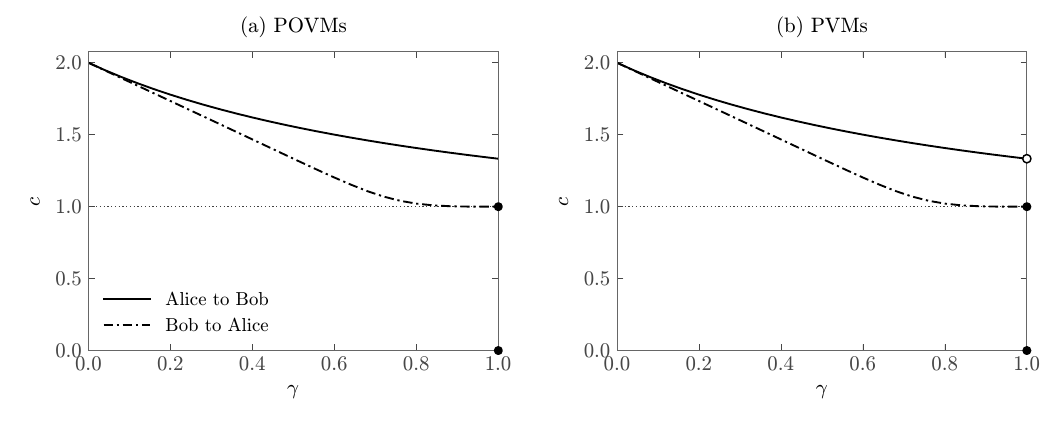}
\caption{Partial and complete damping of Bob in a Bell state. Solid and dash-dotted curves show the forward and reverse scales; the dotted line is $c=1$. Both directions remain steerable for every $\gamma<1$. The open point in (b) is the exact forward PVM limit $4/3$. At $\gamma=1$, the filled points give the product-state values zero and one after the trusted-support restriction. Panels (a,b) correspond to POVMs and PVMs.} \label{fig:example-amplitude-damping}
\end{figure}